\documentclass[final,5p,times,twocolumn,numbers]{elsarticle}

\usepackage[table,xcdraw]{xcolor}
\usepackage{lineno,hyperref}
\usepackage[british]{babel}
\usepackage[textsize=tiny,
	]{todonotes}
\modulolinenumbers[5]
\usepackage{amsmath,amssymb,amsfonts}
\usepackage[utf8]{inputenc}
\usepackage{multicol}
\usepackage{multirow}
\usepackage{subcaption}
\usepackage{comment}
\usepackage{booktabs}  
\usepackage[scale=2]{ccicons}
\usepackage{tikz}

\usepackage{listings,lstlang0}

\usepackage{csquotes}
\usepackage{mathrsfs}
\usepackage{algorithm}
\usepackage{caption}
\usepackage{algpseudocode}
\usepackage{amssymb}
\usepackage{amsthm}
\usepackage{enumitem}

\usepackage[
  separate-uncertainty = true,
  multi-part-units = repeat
]{siunitx}

\setcitestyle{square}

\setenumerate{noitemsep} 
\algnewcommand\algorithmicinput{\textbf{Input:}}
\algnewcommand\algorithmicoutput{\textbf{Output:}}
\algnewcommand\Input{\item[\algorithmicinput]}%
\algnewcommand\Output{\item[\algorithmicoutput]}%

\newtheorem{proposition}{Proposition}

\newtheorem{example}{Example}

\newcommand{\attr}[1]{#1}

\usepackage[normalem]{ulem}

\journal{Information Systems}

\begin{document}

\begin{frontmatter}


\title{LOG.io: Unified Rollback Recovery and Data Lineage Capture \\ 
for Distributed Data Pipelines  \tnoteref{t1}}




\author[1]{Eric Simon\corref{cor1}}
\ead{eric.el.simon@gmail.com}
\cortext[cor1]{Corresponding author}

\author[2]{Renato B. Hoffmann}
\ead{renato.hoffmann@edu.pucrs.br}

\author[2]{Lucas Alf}
\ead{lucasmachadoalf@gmail.com}

\author[2]{Dalvan Griebler}
\ead{dalvan.griebler@pucrs.br}

\address[1]{SAP Labs, Paris, France} 
\address[2]{School of Technology, Pontifical Catholic University of Rio Grande do Sul, Brazil}

\tnotetext[t1]{This work was supported by a research contract granted by SAP and a research contract granted by the Brazilian state;}




\begin{abstract}
This paper introduces LOG.io, a comprehensive solution designed for correct rollback recovery and fine-grain data lineage capture in distributed data pipelines. It is tailored for serverless scalable architectures and uses a log-based rollback recovery protocol.
LOG.io supports a general programming model, accommodating non-deterministic operators, interactions with external systems, and arbitrary custom code. It is non-blocking, allowing failed operators to recover independently without interrupting other active operators, thereby leveraging data parallelization, and it facilitates dynamic scaling of operators during pipeline execution.
Performance evaluations, conducted within the SAP Data Intelligence system, compare LOG.io with the Asynchronous Barrier Snapshotting (ABS) protocol, originally implemented in Flink. Our experiments show that when there are straggler operators in a data pipeline and the throughput of events is moderate (e.g., 1 event every 100 ms), LOG.io performs as well as ABS during normal processing and outperforms ABS during recovery. Otherwise, ABS performs better than LOG.io for both normal processing and recovery. However, we show that in these cases, data parallelization can largely reduce the overhead of LOG.io while ABS does not improve. Finally, we show that the overhead of data lineage capture, at the granularity of the event and between any two operators in a pipeline, is marginal, with less than 1.5\% in all our experiments.  

\end{abstract}


\end{frontmatter}

\section{Introduction}

\subsection{Data pipelines}
\label{sec:intro-pipelines}

The context considered in this paper is a data platform that enables the specification of applications in the form of a dataflow \cite{dataflow-programming:2004}, herein called \emph{data pipelines}, and their execution in a serverless scalable distributed architecture. We describe the programming model of our data pipelines using the concepts of flow-based programming \cite{morrison1994flow}.
A data pipeline is a directed graph of black box components, henceforth called \emph{operators}, that exchange information packets (also called messages or events interchangeably) through connections relating an operator’s output port to an operator’s input port. An operator represents an asynchronous process that is executed in a data-driven mode, that is, whenever its necessary inputs are available in its input ports. This  is similar to the model adopted by many distributed stream processing engines, such as \cite{FLINK:24, MURRAY:ACM:13, QIAN:ACM:13, AKIDAU:VLDB:13, AKIDAU:VLDB:15, WU:IEEE:15, LIN:ACM:16}. 
We assume that messages carry data that are tabular in format and have a granularity that varies from a single record to a collection of records, also called a \emph{batch}. Input data can be bounded (i.e., have a fixed size) or unbounded (i.e., infinite) and can be ordered or not.  
In our model, any operator can interact with the outside world, i.e., with external systems, to receive input data and send data. The \emph{source} (resp. \emph{sink}) operators have no input (resp. output) port; they interact with external systems to receive (resp. send) data. 

At the execution level, operators can be grouped to execute together in one execution environment (for instance, within the same containerized image), and each group can be executed on one or more processing nodes of the distributed system (i.e., its execution can be parallelized). 
A processing node can be as general as a physical machine, a virtual machine, a processor of a machine, or an encapsulation of a set of containerized applications (e.g., a Kubernetes pod \cite{Kubernetes-pod:25}). 


\subsection{Correct rollback recovery}
\label{sec:intro-rollback}

When a data pipeline is started, source operators start getting data from external systems and produce output events on their output ports. We sometimes say that source operators ingest data in the data pipeline. The \emph{global} state of a running data pipeline includes the state of the communication channels between operators (i.e., all messages sent and not yet consumed) and the state of each operator. A \emph{consistent} global state is such that if an operator's state reflects that an input event was received, or if the state of a communication channel reflects that an event was sent, then the state of the sender operator reflects sending that event. 

A global state can become inconsistent due to \emph{system failures} that occur during the communication between operators (e.g., a sent event is lost because the receiver operator has failed) or during the execution of an operator (e.g., the execution state of the operator is lost). 

When a failure occurs, the goal of a \emph{correct} rollback recovery protocol \cite{ELNOZAHY:02} is to restore a consistent global state of the data pipeline such that the result of the execution of the recovered data pipeline on external systems is equivalent to a failure-free execution of the failed pipeline, possibly scheduled at a point in time that follows the time at which it was originally scheduled. With the latter condition, it is possible that the data ingested in the recovered data pipeline by the source operators correspond to a state of the external systems observed at time T', which differs from the state observed when the data pipeline was originally started at an earlier time T. However, if this is the case, the recovered pipeline should produce the same result on external systems as a failure-free execution of the pipeline started at time T'. 
The notion of correct rollback recovery is similar to the \emph{end-to-end exactly once} processing semantics defined in the literature on distributed streaming systems \cite{AKIDAU:VLDB:15, CARBONE:VLDB:17}. 

A rollback recovery protocol must ensure that the developer of a data pipeline or the application does not require any intervention to recover a failed data pipeline. Instead, the system automatically maintains the state of pipeline execution, according to some predefined policy to which each operator in the data pipeline must adhere, and recovers automatically from failures when they occur.

A popular rollback recovery protocol is the coordinated asynchronous checkpointing of the state of a data pipeline, which is implemented in several commercial and open source systems, such as Flink \cite{FLINK:24}, Spark Structured Streaming \cite{SPARK:structured-streaming:23}, and SAP Data Intelligence \cite{SAP:DI:24}. With the asynchronous barrier snapshotting protocol (ABS) \cite{CARBONE:VLDB:17, FLINK:24}, the data ingested in the pipeline is divided into time intervals called \emph{epochs}. At the end of each epoch, source operators inject a "marker" event with a unique identifier in the data pipeline. When an operator receives the markers of a given epoch on all its inputs, it asynchronously checkpoints its state and sends the marker to its downstream operators. An epoch is \emph{complete} when all operators checkpoint their state for that epoch. In case of a failure, the data pipeline is restarted, and its execution resumes from the state of its last complete epoch. 

The main virtues of ABS are its ease of implementation and its small time overhead during the normal processing of a data pipeline. However, a first issue is that ABS is \emph{blocking} in the sense that a failure of one operator stops the processing and triggers a restart of the pipeline. So, when an operator is executed with parallel tasks running on different processes and one task fails, the whole pipeline is restarted without leveraging the possibility for the sibling tasks to take over the load and pursue the processing while the failed task recovers. A second issue is that ABS does not allow for dynamic scaling (up and down) of the processing of an operator when data is ingested at different frequencies or operators have very different processing times. To rescale an operator, it is necessary to take a savepoint (i.e., get a complete epoch), stop the computation, increase the maximum parallelism value for that operator, redeploy the data pipeline with the new configuration, and resume its execution from the complete epoch \cite{FLINK:elastic:25}. Finally, operators that perform updates on an external system can prepare their updates, but must wait for a complete epoch before committing their changes (following a two-step commit policy). The issue is that, apart from the latency on the external systems incurred by this synchronization, an update performed by an operator $OP1$ to an external system cannot be seen by an operator $OP2$ that executes \emph{after} $OP1$ in the data pipeline because $OP1$'s updates can only be committed when a complete epoch is reached. 

Log-based rollback recovery techniques address the first and third issues above. They rely on the assumption that they can identify all the non-deterministic events executed by each operator, and for each such event, log all information necessary to replay the events in case of recovery \cite{ELNOZAHY:02}. Using this assumption, it is possible to recover a failed operator and replay its execution as it occurred before the failure. 

MillWheel and Google Dataflow \cite{AKIDAU:VLDB:13, DATAFLOW:Overview:2024} log the output events of each operator synchronously, while ChronoStream \cite{WU:IEEE:15} and StreamS \cite{LIN:ACM:16} log them asynchronously, but require that the operators be deterministic. Output events are logged in addition to the checkpoints of operator states (in case of stateful operators). In case of failure, only the failed operators are restarted without disrupting the execution of the other operators. 
Instead of storing events, Spark Streaming \cite{SPARK:streaming:24} stores the graph of operations (called lineage graph) used to build the resilient distributed datasets (RDD) partitions resulting from the execution of any intermediate operator in a data pipeline, so that in case of failure, the partitions can be reconstructed from the source data. To ensure the correctness of the reconstruction, the operators must be deterministic and the source data must be logged. However, to minimize the amount of recomputation, intermediate RDD partitions (corresponding to output events) can be periodically checkpointed. In addition, Spark streaming uses a "micro-batch" processing approach in which the size of batches is statically defined, causing possible latency issues with recovery \cite{adapted-batching-survey:2022, Das:2014, Zhang:2016}. 
Lineage stashing \cite{lineage-stashing:2019}, a decentralized causal logging technique, enables asynchronous logging of a lineage graph of tasks while allowing for an event-driven execution model with dynamically sized batches of data. However, again, it requires that operators be deterministic, although the order in which events are combined from multiple inputs can be non-deterministic. 
Finally, some log-based methods like MillWheel, Google Dataflow, and ChronoStream address the issue of dynamic scaling. However, ChronoStream requires that operators be such that their computation function and the order in which events coming from different input streams are consumed be deterministic. 

\subsection{Fine-grain data lineage capture}
\label{sec:intro-data-lineage}
Another problem that is paramount for the debugging of data pipelines is the ability to capture the \textit{ data lineage}, also called data provenance, during the execution of a data pipeline \cite{provenance:2009, lineage-scientific:2005, Ikeda:2013}. In our context, 
given an operator OP, there exists a \emph{data lineage relationship} $(a, b)$ between an output event $b$ and an input event $a$ if and only if there exists a record in event $b$ and a record in event $a$ such that $a$ was used by OP to compute $b$. 

Given an output event produced by an operator, a \textit{backward lineage} query asks: 'What input events were used by the operator to produce this event?' Suppose this querying process is applied recursively along a path of connections in the data pipeline. In that case, it is possible to identify the events ingested by a source operator that were used to compute any intermediate output event. Similarly, given an input event received by an operator, a \textit{forward lineage} query asks 'what output events were produced from this event?' 


The existing log-based recovery methods that log the lineage graph of any intermediate result are insufficient to capture the data lineage as defined above. First, systems like Spark Streaming track the lineage at the fixed coarse-grain granularity of RDD partitions, each of which may span multiple batches. 
Thus, for a stateful operator OP, the lineage may indicate that an output partition $p_3$ of OP depends on input partitions $p_1$ and $p_2$. However, only some records in $p_1$ were used to calculate the records in $p_3$. This is because the intention of the lineage graph is only to reconstruct the past state of an operator, not to track the data lineage relationships defined earlier. Second, although they use finer granularity of events, the progress methods of the StreamS and ChronoStream systems, as well as the lineage stash method of \cite{lineage-stashing:2019}, suffer from the same problem. 

Dedicated data lineage techniques have therefore been developed for dataflow processing systems. Several methods exist to capture the data lineage in data pipelines in which bounded source data is ingested \cite{newT:2013, subzero:2013, Titian:2018, hippo:2017, Glavic-ICDE:2009, Glavic-ICDE:2017, Smoke:2018}. They generally rely on the instrumentation of operators to write the data lineage relationships to a lineage subsystem through an API provided by the subsystem. They provide a record-level granularity of data lineage to the detriment of a relatively high data capture overhead. In \cite{Titian:2018}, the reported overhead for Spark jobs is at least 20\% and most of the time below 30\%, which is better than the methods proposed in \cite{newT:2013, subzero:2013}. In \cite{Smoke:2018}, data lineage capture is limited to a dataflow of relational database queries, with a reported overhead for TPC-H queries between 6 and 22\% which is better than the methods of \cite{Glavic-ICDE:2009, Glavic-ICDE:2017}. 

Fewer methods have been proposed for the processing of unbounded data. The most recent methods also consist of instrumenting the operators of a data pipeline. The principle is to enrich the metadata of each event output by an operator with an annotation that describes the input events that were used to produce the output event. In \cite{glavic:2014}, the annotation consists of a set of identifiers of source events, which is propagated across the operators of the data pipeline. That is, each operator adds the metadata of the contributing input events to each output event. The method assumes that all source events are stored as well as all sink events (i.e., the final events produced by the pipeline). To answer a backward data lineage query, the identifiers in the sink events are matched with the identifiers of the source events, and the corresponding source data are retrieved. However, this method suffers from the following drawbacks: (1) the list of event identifiers carried by each new output event can grow arbitrarily, and (2) all source events must be logged until we can distinguish between those that actually contribute to sink events and those that do not. 

The two previous problems are addressed in \cite{genealog:2018, ananke:2020} using another annotation-based method that consists of annotating each output event with the identifiers of the input events that \emph{directly} contributed to it, which requires that the generated identifiers of each new output event must be stored. This limits the amount of metadata that must be carried by events flowing from source operators to sink operators. In addition, the source events that never contributed to an output event, in the operators that immediately consume the output of the source operators, can be discarded earlier. The experimental evaluations reported in \cite{genealog:2018} show that this technique results in better performance with respect to \cite{glavic:2014}.

However, all previous methods still have the following issues. First, they only support “full” backward data lineage queries, i.e., queries that start from sink events and return source events. Since intermediate results are not logged (only output identifiers are stored), the only way to reconstruct the output events of an intermediate operator is to replay the upstream operators up to these output events. This is a second issue, since it requires that each operator be deterministic as well as the order in which events coming from multiple streams are consumed by an operator. 
A third issue is that the previous methods instrument predefined deterministic operators such as source, map, filter, union, join, and aggregate, but do not provide a general data provenance mechanism for any arbitrary black box or non-deterministic operator. Finally, previous methods \cite{glavic:2014, genealog:2018} strictly add a data lineage capture overhead to the overhead of the rollback-recovery method that a data pipeline engine must implement, which includes the explicit storage of source data. For example, when the method of \cite{genealog:2018} runs on top of Flink, it is reported that the latency for a simple data pipeline involving two aggregate operators is between 5 and 6\%.      

\subsection{Research contributions}
\label{sec:intro-contribs}

Our key contribution in this paper is to provide a unified solution for both the correct rollback recovery and the data lineage capture of data pipeline executions in a distributed and scalable serverless architecture. 

We propose a log-based rollback recovery protocol, named LOG.io, which works as follows. When an operator receives an event, it updates its internal state (for a stateful operator) and acknowledges the event by durably assigning it to an Input Set. When the computation of a set of output events (Output Set) is triggered, the operator starts an atomic transaction to log each output event with a status "undone" and set to "done" the logged events of the Input Set used to compute the output event. In addition, when data lineage capture is enabled, the association between each event of an Output Set and the identifier of the corresponding Input Set is logged within the same atomic transaction. Then, output events are sent to the downstream operators (pessimistic logging). When an operator fails, it first recovers its "undone" and unacknowledged output events from the log and sends them to the downstream operators. Then, it recovers its acknowledged "undone" input events (i.e., input events already assigned to an Input Set) from the log and uses them to reconstruct its internal state. Afterwards, it returns to its normal processing. 

We show that LOG.io has the following features. First, it accepts a \textit{general programming model}, as in Google Dataflow \cite{DATAFLOW:Overview:2024} or Flink \cite{FLINK:24}, in which (1) operators and the order of consumption of events coming from multiple input streams, can be non-deterministic, (2) interactions with an external system are allowed in any operator, and (3) operators can have arbitrary custom code. In addition, records are dynamically batched into events of varying size, which can be decided by each operator. 

As a second feature, LOG.io is \textit{not blocking}, as in \cite{DATAFLOW:Overview:2024}: in case of a failure, each failed operator recovers independently without interrupting the processing of other alive operators. Thus, unlike ABS, this enables LOG.io to pursue the processing of an operator when it is executed in parallel on different processing nodes, and one of the nodes fails. To achieve this, as in previous log-based methods (e.g. \cite{AKIDAU:VLDB:13, DATAFLOW:Overview:2024, WU:IEEE:15, LIN:ACM:16}), LOG.io logs output events, but it does not require one to log the internal state of operators. The only checkpointing done by LOG.io, within the same atomic transaction that logs output events, is the state of global variables such as timers or counters that are used to decide when to trigger the computation of output events. This has a tiny footprint as compared to the size of the input data that a stateful operator must manage (e.g., windows of events). 

Third, from the algorithmic perspective, LOG.io can \textit{dynamically scale up and down operators} that can be executed in parallel, without interrupting the processing of a data pipeline. 

Fourth, we present a version of LOG.io in which only the identifiers of output events are logged, but not the actual payload of output events. This version is only applicable to deterministic operators in a data pipeline. When an upstream operator of a failed operator did not log the data of its output events, it is asked, during the recovery phase, to recompute its output events that are still "undone" and send them to the failed operator. This can happen recursively along the path of upstream operators. We show that this technique can be used, similarly to systems like \cite{WU:IEEE:15, LIN:ACM:16, lineage-stashing:2019}, to asynchronously (i.e., optimistically) log output events in the case of deterministic operators. 

Finally, during the execution of a data pipeline, LOG.io \textit{captures the data lineage relationships} of operators at the granularity of events in such a way that backward and forward lineage queries can be supported between any two arbitrary operators of a data pipeline. 

We describe the implementation of LOG.io in the data pipeline engine of the SAP Data Intelligence system \cite{SAP:DI:24}, which already implements a variant of the ABS snapshotting recovery protocol. We also present the LOG.io API that must be used by the implementer of a custom operator to conform to the LOG.io protocol. We then evaluate the performance of the pessimistic logging version of LOG.io against ABS with different use cases to assess the impact of the non-blocking property of LOG.io. We demonstrate that during normal processing, there are cases where LOG.io outperforms ABS when straggler operators exist, that is, operators that execute relatively slower than other operators located on the same path of a data pipeline. In our experience, we regularly see this case because stateful and custom-coded operators execute more slowly than stateless operators (e.g., filters). When there is a little difference between the processing time of operators, and the traffic of events is moderate and regular, the additional overhead of LOG.io with respect to ABS remains small (below 5\%). In the extreme case, when the traffic of events is high, all operators have similar processing times, stateful operators have a small state, and there is a single sink operator, LOG.io overhead with respect to ABS can grow up to 40\%. We then show that LOG.io always has a lower overhead than ABS during recovery, and the difference is more significant when operators can be executed in parallel. Finally, we show that the overhead of data lineage capture is marginal in all scenarios (less than 1.2\% ). 


\subsection{Paper outline}

Section \ref{sec:premininaries} introduces useful definitions and describes the assumptions made by our protocol. Section \ref{sec:logio-normal-protocol} describes the behavior of LOG.io with pessimistic logging during the normal processing of a data pipeline. Section \ref{sec:logio-recovery} presents the behavior of LOG.io to correctly recover the execution of a data pipeline when one or more failures occurred. Section \ref{sec:replay-operators} presents a version of LOG.io that only stores the identifiers of output events without storing their data. Section \ref{sec:logioImplementation} describes the implementation of LOG.io in the SAP Data Intelligence system, a commercial cloud offering for data pipelines in a distributed serverless architecture based on Kubernetes. Section \ref{sec:logio-advantages} provides a qualitative analysis in which we discuss the main benefits of LOG.io. Section \ref{sec:related_work} gives a detailed description of previous work on rollback recovery and compares it with LOG.io. Section \ref{sec:logio-evaluation} compares LOG.io and ABS in the context of SAP Data Intelligence. Finally, Section \ref{sec:conclusions} concludes.

\section{Preliminary definitions and assumptions}
\label{sec:premininaries}

\subsection{Connections and events}
\label{sec:connections}

We assume that all communications between operators are performed using an asynchronous message passing framework, all connections between processes are reliable, and messages are delivered in the order they are sent (FIFO ordering). 
A \emph{back-pressure} mechanism is used to block a connection if the buffer of events associated with the connection is full. When a connection is blocked, an operator cannot send an event on that connection. 

Data pipeline operators deployed on the same execution environment on a processing node communicate using local inter-process or inter-thread communication, and communications between operators located on different processing nodes happen through remote process communications. Each processing node has access to a durable storage, 


Finally, events are batches of records of variable size, which can be dynamically determined by each operator. Each event sent on an output port by an operator is identified by an \textit{event ID}, which is a System-generated Sequential Number (SSN), that is unique for each output port of an operator. 


\subsection{Read and write actions}
\label{sec:atomic-actions}

A data pipeline interacts with the outside world to receive input data from external systems or to use the outcome of a computation to perform actions on external systems. If a failure occurs, in general, the outside world cannot be relied on to roll back and cannot participate in our rollback recovery protocol. Examples of external systems include file systems, databases, publish-subscribe systems, or terminal consoles. We model the interaction of operators with external systems through read and write actions.

\paragraph{Read and write actions} 
An operator can send read or write action to an external system through a connection. 
A read action observes a state of an external system at a given point in time, while a write action performs a change from an observable state of the external system to another observable state. By observable state, we mean a state of the data in an external system that can be observed by an application process interacting with the external system. 

For instance, in a database system, a read action is a database query that returns a result set, while a write action is an insert, delete, or update to a database table. The state observed or changed by the read or write action is defined by the version of each database table that is read or written by the action, which in turn depends on the isolation degree of the embedding transaction. For a publish-subscribe system, a read action returns one or more events from a queue to which a reader operator has subscribed while a write action publishes one or more events into a queue. 

The \emph{effect} of a read action A on the observable state S of an external system is the sequence of events $[a_1,…,a_n ]$, denoted $r(A,S)$, that results from applying $A$ on $S$. 
For example, the effect of a database query is the set of records returned by the query in successive batches, each of which represents an input event $a_i$.

A read action $A$ is \emph{replayable} if the data returned by $A$ at time $T+\Delta T$ always contain the data returned by $A$ at a previous time $T$, and in the same order. So, if $S$ is the observable state at time $T$ and $S'$ is the state observed at time $T+\Delta T$, $r(A,S) \preceq r(A,S')$, where $\preceq$ denotes the subsequence relationship. 

\begin{example}
A database query "SELECT * FROM SALESORDER WHERE ORDERDATE $\leq$ ’10-02-2010’ ORDER BY ORDERDATE”, in which table SALESORDER is append-only and new records have an increasing order date, is immutable and therefore replayable. A read action that reads sorted events from a queue starting from a given offset, and a database query “SELECT * FROM SALESORDER ORDER BY ORDERDATE”, are replayable read actions because they both return sorted data and new data at time $T+\Delta T$ occur at the end of the data returned at time $T$.
\end{example}

A write action is considered as an output event produced by an operator that must be sent on a connection to an external system. A write action $A$ is said to be \emph{idempotent} if for any observable state $S$ of an external system, the state resulting from two consecutive executions of $A$ on $S$ is the same as the state resulting from a single execution of $A$ on $S$.

We assume that write actions are \emph{durable}, that is, when the external system returns a success response to an execution request of a write action, it is eventually executed by the external system, even if the external system failed after acknowledging the reception and handling of the execution request. We therefore rely on the fault-tolerance capability of the external system.


\subsection{Data pipeline operators}
\label{sec: operators}
 
\paragraph{Types of operators} We distinguish the following types of operators in a data pipeline. \emph{Source} operators ingest data into the data pipeline through output ports and have no input port. A source operator can either read data from one or more external systems using read actions or can be a self-contained data generation process.
\emph{Middle} operators take intermediate results received on their input ports and produce other intermediate results on their output ports.
\emph{Reader} operators read data from one or more external systems using read actions while \emph{Writer} operators write data into one or more external systems using write actions. Finally, \emph{Sink} operators have no output port. 

A Source operator is generally a Reader operator. In our model, a Middle or Sink operator can also be categorized as a Reader (resp. a Writer) operator when it performs read (resp. write) actions during the processing of the events received on its input ports. 
We sometimes call read and write actions performed by an operator \emph{side effect} actions.  

An operator is said to be \emph{deterministic} if for the same arbitrary sequence of input events received on each of its input ports, it always returns the same sequence of output events on each of its output ports and each of its connections to an external system. 
We consider the following sources of nondeterminism: the order of arrival of events at the input ports, which may influence the computation of output events, and the use of a nondeterministic function to compute output events (e.g., the use of a non-replayable read action within a middle operator). 

\paragraph{Local state} 
An operator can maintain a local volatile state built from the data contained in the events that the operator receives at its input ports. We shall call this state \emph{event state}. In addition, an operator can maintain a state, called \emph{global state}, that contains contextual information used to decide when to trigger the computation of output events, such as a counter that counts the total number of input events received, or a local clock timer. 
The distinction made between an event state and a global state is illustrated in Example \ref{ex:single-input-set}.

\begin{figure}
    \centering
    \includegraphics[width=0.70\linewidth]{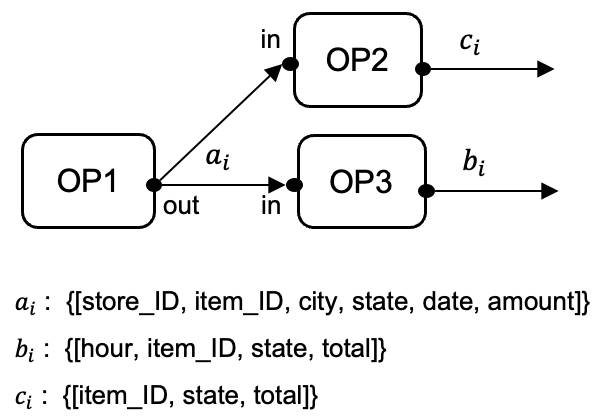}
    \caption{Sales data pipeline}
    \label{fig:Example4}
\end{figure}

\begin{example}
\label{ex:single-input-set}
Consider the Middle operator OP2 in Figure \ref{fig:Example4}.  It receives events from the Source operator OP1, each of which represents a batch of sales records of the form $[store\_ID, item\_ID, city, state, date, amount]$. 
Each input event is used to update an \textbf{event state} (initialized as an empty set) consisting of an incrementally maintained sum (amount) grouped by item\_ID and state over the input events received so far. 
The \textbf{global state} counts the total number of events received.
When 100 new events have been received (triggering condition) 
OP2 generates a single output event from the event state which is a batch of records of the form $[item\_ID, state, total]$. The event state is then re-initialized.  
\end{example}

\paragraph{Stateless and stateful operator}
An operator is said to be \emph{stateless} if it computes a set of output events $O$ (possibly empty) from \emph{each} input event received on one of its input ports without using any global state. Otherwise, the operator is said to be \emph{stateful}. 

To better explain our protocol, it is convenient to model the behavior of a stateful operator as a succession of the following phases:
\begin{itemize}
    \item \textbf{State Update}: each input event received on an input port is used to update the event state and the global state. 
    \item \textbf{Triggering}: a condition is evaluated against the event state and the global state to determine if the generation of output events can be triggered
    \item \textbf{Generation}: a set of output events is computed using a generative function and published on zero or more output ports and zero or more connections to external systems, depending on the type of operator
\end{itemize}

\paragraph{Input and Output Sets} When a set of output events $O$ is generated, we call the set of input events $I$ that was used to generate $O$ the \emph{Input Set} for $O$ while we refer to $O$ as the \emph{Output Set} for $I$. During the Generation phase, input events include both those received on input ports and those obtained from the effect of possible read actions. Several generative functions can be used to generate different Output Sets using not necessarily distinct sets of input events. 
Each Input Set used by a generative function to produce an Output Set has a unique identifier, henceforth noted InSet\_ID. 


When a new input event $e$ is processed during the State Update phase, it is assigned an InSet\_ID that corresponds to a part of the event state that is updated by some records of $e$ and that can be used by a generative function to produce an Output Set. This assignment is determined by both the records of $e$ and the global state. Multiple InSet\_ID can be assigned to an event $e$ because the records in batched in $e$ can be used to generate different Output Sets using different generative functions. 
Once an Input Set with some InSet\_ID has been used to generate an Output Set during the Generation phase, no new event can be assigned that InSet\_ID. These notions are illustrated in Examples \ref{ex:OP2} and \ref{ex:two-input-sets}. 

Note that for a stateless operator, an input event is immediately used to generate output events (there are no State Update and Triggering phases). Therefore, each Input Set identifier refers to a singleton set of input events.  

\begin{example}
\label{ex:OP2}
Consider the operator OP2 in Figure \ref{fig:Example4}. 
As explained before, OP2 uses a single generative function which, given an event state consisting of the incrementally maintained sum(amount) grouped by item\_ID and state of the records contained in the last 100 input events, produces a single output event. Thus, the InSet\_ID assigned to an input event can be defined as the multiple of 100 events received by OP2. Hence, the first 100 events are assigned InSet\_ID "1", the next 100 events are assigned InSet\_ID "2", and so on so forth. Now, suppose that OP2 uses another generative function which, given the last 100 input events, produces a batch of records of the form [item\_ID, minAmount, maxAmount] on another output port of OP2. Then, two InSet\_IDs are needed for each set of 100 input events to distinguish the two incrementally maintained events states used by the two generative functions.
\end{example}

\begin{example}
\label{ex:two-input-sets}
Consider now the operator OP3 in Figure \ref{fig:Example4}. 
During the State Update phase, each input event of OP3 is used to update one or more windows, each of which representing an Input Set that contains sales records for the same hour of the day identified by the record field "date". Thus, an InSet\_ID can be defined as an hour of the day. An input event can be assigned multiple InSet\_ID because it may contain records covering different hours. For simplicity, let us assume that input events carry sales records that are totally ordered with respect to their field "date", i.e., there is no event reaching OP3 which contains (late) records. 
Then, the Triggering phase checks if a window for hour $H$ is "complete", that is, if at least one record for hour $H+1$ has already been received.  
In the Generation phase, for each complete window, a generative function produces a single output event that is a batch of records of the form $[hour, item\_ID, state, total]$, where the field "hour" contains the hour associated with the complete window and the field "total" is the sum (amount) grouped by item\_ID and state over the window. 
\end{example}

\section{LOG.io Normal Processing}
\label{sec:logio-normal-protocol}

In Section \ref{sec:data-lineage-config}, we describe how a user configures the data lineage capture enabled for a data pipeline. Section \ref{sec:logs} describes the schema of the database tables used by LOG.io. Then, Sections \ref{sec:normal-source} to \ref{sec:normal-generate} describe the behavior of LOG.io during the normal execution of a data pipeline operator. Finally, Section  \ref{sec:garbage-collect} describes the garbage collection of LOG.io database tables. 

\subsection{Configuration of data lineage capture}
\label{sec:data-lineage-config}

Given an operator OP, there exists a \emph{data lineage relationship} $(a, b)$ between an output event $b$ and an input event $a$ if and only if there exists a record in event $b$ and a record in event $a$ such that $a$ was used by OP to compute $b$. The purpose of a \emph{data lineage capture} for an operator OP is to build all data lineage relationships for OP during the execution of a data pipeline. 

It is however the user's responsibility to determine the operators of a data pipeline for which data lineage capture is enabled. More specifically, the user must define a \emph{data lineage scope} (or lineage scope, for short) as a couple $(start, target)$ of operator's port IDs, which define the operator's output port IDs from which data lineage capture is \textit{started} and \textit{ended}. This implicitly defines a set of data lineage paths, each consisting of a sequence of operator's port IDs, that connect the two ports of a data lineage scope $(start, target)$. If there exists a subsequence (OP.in, OP.out) in a data lineage path, for an operator OP, then data lineage capture is \emph{enabled} for these two ports of OP. 

Suppose that a data pipeline is configured in such a way that an operator OP has its data lineage capture enabled for a set of input ports IN and a set of output ports OUT. During the execution of the pipeline, suppose that an Output Set $O$ is generated from an Input Set $I$. 
A data lineage relationship is recorded between an output event $o \in O$ and an input event $e \in I$ if and only if $o$ is sent to a port $q \in $ OUT and $e$ was received on a port $p \in$ IN. 

We finally extend data lineage capture to the input events coming from the read actions issued during the generation of an Output Set. If there exists an output event $o \in O$ that is sent to a port $q \in $ OUT and input events coming from a read action $r$ are used to compute $O$, then a data lineage relationship is recorded between $r$ and $o$. 

\begin{figure}
    \centering
    \includegraphics[width=0.95\linewidth]{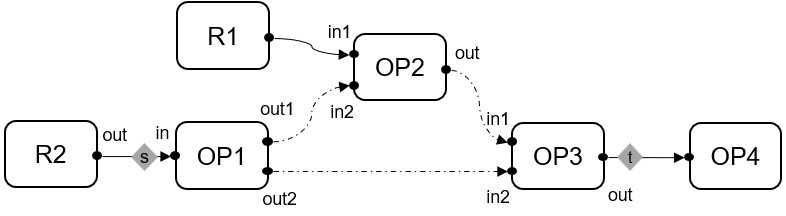}
    \caption{Data lineage paths}
    \label{fig:data-lineage-paths}
\end{figure}

\begin{example}
Consider the example on Figure \ref{fig:data-lineage-paths}, where ports have names like "in" and "out". For a configured data lineage scope $(R_2.out, OP_3.out)$, the set $\{\Phi_1, \Phi_2\}$ of data lineage paths (having dotted lines in the Figure) is: $\Phi_1= (R_2.out, OP_1.in, OP_1.out1, OP_2.in2, OP_2.out, OP_3.in1, OP_3.out)$ and $\Phi_2=(R_2.out, OP_1.in, OP_1.out2, OP_3.in2, OP_3.out)$. Thus, for operator OP2, data lineage is enabled for IN = \{in2\} and OUT = \{out\}, while for operator OP3, data lineage is enabled for IN = \{in1, in2\} and OUT = \{out\}.  

Now, suppose that an event $c_1$ is generated on OP2.out using a generative function whose Input Set only contains input events received on OP2.in1, then no data lineage relationship is recorded for event $c_1$. However, if event $c_2$ is generated on OP2.out using a function whose Input Set contains an input event $e$ received on OP2.in2 then a data lineage relationship $(e, c_2)$ will be recorded. 
\end{example}

\subsection{Schema of LOG.io log tables}
\label{sec:logs}
LOG.io persistently logs the events of each Output Set generated by a generative function together with the marking of its corresponding Input Set identifier. LOG.io also logs the serialized global state that was used to generate that Output Set. More specifically, LOG.io uses the following five persistent tables whose presentation is simplified (details corresponding to the multi-tenancy model are not shown). 

The EVENT\_LOG table stores the output events generated by an operator. Each event is identified by: an event\_ID, the operator\_ID and port\_ID of the sender and receiver operators for the event, an event status (which is either "done" or "undone") and an InSet\_ID to which the event was assigned during the State Update phase. 

The EVENT\_DATA table stores the body (payload) of output events. Each event is identified by: an event\_ID, the operator\_ID and port\_ID on which the event is output, and two BLOB fields respectively for the header and the body of the event. 

The READ\_ACTION table stores the read actions issued by an operator. Each read action is identified by: an action\_ID, a status ("complete" or "incomplete"), an operator\_ID, a connection\_ID (on which the read action is issued to an external system), and a textual description of the action to be executed. An action\_ID is unique for a connection\_ID and an operator\_ID.

The STATE table stores a global state of an operator, which is identified by a state\_ID, an operator\_ID, and a BLOB field containing the serialized value of the global state. 

The EVENT\_LINEAGE table stores the data lineage relationships between an output event $e$ and the InSet\_ID of the Input Set that was used to generate it. A lineage is therefore identified by the event\_ID of $e$, a port\_ID and operator\_ID on which event $e$ is sent, and the InSet\_ID of the Input Set used to generate $e$.

\subsection{Processing of a Source operator}
\label{sec:normal-source}
We describe the case of a Source operator OP that performs a read action $A$ on an external system, via a connection $Cx$, to generate data on an output port “out” connected to the port "in" of an operator OP'. In that case, the steps in Algorithm \ref{alg:normal-source} must be implemented by OP. 

All transactions in the different steps of Algorithm \ref{alg:normal-source} are atomic. 
In Step 2.a, read action $A$ can be retried an arbitrary number of times before raising a fatal error. Since $A$ is not replayable, the effect of $r$ is stored, although LOG.io does not impose any technical solution to the developer of a Source operator to store the effect of $r$. In Step 2.b, the read action is marked as "complete" and logged in EVENT\_LOG as an "undone" event with a \textit{null} value for the receiver port, which is what distinguishes a read action event. The data part of the event in EVENT\_DATA is a reference to the stored effect of $r$. In Step 2.c, OP decides how the batching of output events should be done (dynamic batching). Also, since the effect of $r$ is already stored, it is sufficient to keep offsets that indicate the end of each batch in that store. If OP fails, the generation of events is restarted from the last batch sent (identified with the offset). However, this requires adding the offset to the global state of OP and therefore storing in STATE the last value of the offset (see part (2) of the atomic transaction in Step 2.c). 

In Step 3, since the consumption of the effect of $r$ and the generation of output events are pipelined, there is no need to manage an event in EVENT\_LOG for $r$ as in Step 2.b. In Step 3.a, since $A$ is replayable, an index to the (reproducible) effect of $r$ can be kept in the global state to indicate from where new output events must be generated during a replay of $r$. The management of offsets is specific to the external system and the type of read actions created by OP. 

\begin{example}
Tables \ref{tab:event-log}, \ref{tab:read-action}, \ref{tab:event-data} show the contents of three log tables after a non-replayable read action with action\_ID $r_1$ has been executed by a Source operator OP. Steps 2.a and 2.b of Algorithm \ref{alg:normal-source} have passed: the effect of $r_1$ is stored, action $r_1$ is "complete" in READ\_ACTION and an event is added to the EVENT\_LOG. Step 2.c is on-going and two output events $a_1$ and $a_2$ have been generated, logged into EVENT\_LOG, and sent to "OP.out". However, the event $r_1$ in EVENT\_LOG is still "undone" because Step 2.c is not finished. When it will be, by Step 2.d, event $r_1$ in EVENT\_LOG will be set to "done", the store of the effect of $r_1$ will be deleted, and event $r_1$ in EVENT\_DATA will be deleted.  

\begin{table}[htbp]
\centering
  \caption{Table READ\_ACTION}
  \label{tab:read-action}
    \setlength{\tabcolsep}{4pt}
\resizebox{0.37\textwidth}{!}
{%
\begin{tabular}{l|l|l|l|l}
\toprule
\attr{Action\_ID}&	\attr{Status}& \attr{OP\_ID}&	\attr{Conn\_ID}& \attr{Action\_Desc}\\
\midrule
    $r_1$&	complete& OP&	Cx&	text of action $A$\\
\bottomrule
\end{tabular}
}
\end{table}

\begin{table}[htbp]
\centering
  \caption{Table EVENT\_LOG}
  \label{tab:event-log}
    \setlength{\tabcolsep}{4pt}
\resizebox{0.45\textwidth}{!}
{%
  \begin{tabular}{l|l|l|l|l|l|l}
\toprule
\attr{Event\_ID}&	\attr{Status}& \attr{Send\_Op}&	\attr{Send\_p}& \attr{Rec\_Op}&	\attr{Rec\_p}& \attr{InSet\_ID}\\
\midrule
    $r_1$&	undone& OP&	Cx&	OP&	null & null\\
    $a_1$&	undone& OP&	out&	OP'&	in& null\\
    $a_2$&	undone& OP&	out&	OP'&	in& null\\
\bottomrule
\end{tabular}
}
\end{table}
\end{example}

\begin{table}[htbp]
\centering
  \caption{Table EVENT\_DATA}
  \label{tab:event-data}
    \setlength{\tabcolsep}{4pt}
\resizebox{0.36\textwidth}{!}
{%
\begin{tabular}{l|l|l|l|l}
\toprule
\attr{Event\_ID}&	\attr{Send\_Op}& \attr{Send\_p}&	\attr{Header}& \attr{Body}\\
\midrule
    $r_1$&	OP & null &	null & ref to $r_1$ store \\
\bottomrule
\end{tabular}
}
\end{table}

\begin{algorithm}[tb]
\caption{Processing of a Source operator OP}
\label{alg:normal-source}
\begin{tabular*}{\linewidth}{rp{0.75\linewidth}}
\textbf{input:} & read action $A$ on connection $Cx$\\
\textbf{output:} & output events on "OP.out" connected to "OP'.in"; \\
& update LOG.io logs \\
\end{tabular*}
\begin{enumerate}
    \item \textit{Execution:} Get a new action\_ID $r$ for $A$; get new state\_ID $s$ for global state; use a transaction to add $r$ in READ\_ACTION with status "incomplete"; execute action $A$ on $Cx$. 
    
    \item \textit{If $A$ is not replayable:} 
    \begin{itemize}
        \item[a.] Store the effect of $r$. If $r$ fails, discard the store for $r$ and replay $r$. 
        
        \item[b.] When $r$ is completed: use a transaction to (1) update $r$ in READ\_ACTION with status "complete" and (2) add an event in EVENT\_LOG and EVENT\_DATA with event\_ID $r$, status "undone", "OP.Cx" as sender port, and "null" as receiver port

        \item[c.] Iterate over the effect of $r$: generate an output event $o$ with a new SSN; use a transaction to (1) add $o$ in EVENT\_LOG and EVENT\_DATA with status “undone”, "OP.out” as sender, "OP'.in" as receiver, (2) store in STATE the offset in $r$'s store of the batch end and (3) if $o$ is last event, set status of event\_ID $r$ in EVENT\_LOG to “done”. Send $o$ on "OP.out" 

        \item[d.] Garbage collect: Discard the store for $r$, and delete event $r$ from EVENT\_DATA
    \end{itemize}
            
    \item \textit{If $A$ is  replayable:} OP continuously consumes the effect of $r$: It generates output events as in Step 2.c above, except for the last event where the status of $r$ is set to “complete” in READ\_ACTION. If action $r$ fails, it can be replayed and the generation of events is resumed from the last event logged. 
\end{enumerate}
\end{algorithm}

\subsection{State Update phase of a Middle or Sink operator}
\label{sec:normal-state-update}
When a stateful operator OP consumes an event on one of its input ports from an operator OP', Algorithm \ref{alg:input-events} describes the processing steps of the State Update phase (see Section \ref{sec: operators}) that OP must implement. LOG.io internal in-memory data structures (e.g., InSet\_ID and State\_ID variables), later called \emph{LOG.io context}, are initialized when OP is deployed and started. 

\begin{algorithm}[tb]
\caption{State Update phase of a stateful operator OP}
\label{alg:input-events}
\begin{tabular*}{\linewidth}{rp{0.75\linewidth}}
\textbf{input:} & event $e$ received on input port "OP.in" from OP'\\
\textbf{output:} & Updates of OP state and LOG.io logs \\
\end{tabular*}
\begin{enumerate}

    \item \textit{Filtering:} Discard $e$ if it is obsolete, i.e., its event\_ID is smaller than the greatest event\_ID received on "OP.in" with InSet\_ID $\neq$ null 
                
    \item \textit{State update:} Use $e$ to update the global state and LOG.io context, if not already done; Update the event state; Accordingly, assign to $e$ one or more InSet\_IDs; Use a transaction to update the InSet\_ID(s) of $e$ in EVENT\_LOG; 
    
    \item \textit{Triggering:} if a triggering condition evaluates to $true$ then the corresponding generation phase is triggered by using the processing steps of Algorithm \ref{alg:generate-general}
        
\end{enumerate}
\end{algorithm}

The filtering condition in Step 1 detects "obsolete" events.  
In Step 2, an array of the latest event\_ID from each input port used to update the global state is maintained within the LOG.io context. If the event\_ID of the input event is less than or equal to the latest event\_ID of the corresponding port in the array, the global state was already updated and the update is skipped. Otherwise, the global state and the array are updated.  
The rationale for the filtering condition and the check of the array of event\_IDs will be explained later when we present our recovery method. 

Then, the event state is updated. The developer of an operator must manage the Input Sets required by the processing logic of the operator, while LOG.io provides an API (described in Section \ref{sec:4.Protocol-api}) to create a new InSet\_ID and assign one or more InSet\_IDs to an input event. 
When an event $e$ is assigned more than one InSet\_ID, as many entries for $e$ are created in EVENT\_LOG to represent the different assignments of $e$ (each entry represents a single InSet\_ID assignment), so that the Input Set ID is unique for each generated Output Set.
When Steps 1 and 2 have been completed, the input event $e$ is said to be \emph{acknowledged} and $e$ is removed from the connection to OP. 
Finally, Step 3 evaluates the triggering condition to start the generation of an Output Set. 

\subsection{Generation Phase of a Middle or Sink operator}
\label{sec:normal-generate}

We consider the general case of a stateful Middle or Sink operator that can execute read or write actions. Read actions can be executed to obtain additional input events used to compute the generation of an Output Set, while write actions are events resulting from the generation of an Output Set. Section \ref{sec:generate-general} describes the general steps for the Generation phase, while Sections \ref{sec:generate-reader} and \ref{sec:generate-writer} respectively describe the processing of read and write actions. 

\subsubsection{General case}
\label{sec:generate-general}

In a stateful operator OP, when a triggering condition evaluates to true for an Input Set, the generation of an Output Set using a function $F$ must follow the processing steps of Algorithm \ref{alg:generate-general}, in which we assume that data lineage capture is enabled for a set of input ports IN and a set of output ports OUT, as defined in Section \ref{sec:data-lineage-config}. Note that the generation phase can be triggered for more than one Input Set.

\begin{algorithm}[tb]
\caption{Generation of an Output Set by operator OP}
\label{alg:generate-general}
\begin{tabular*}{\linewidth}{rp{0.75\linewidth}}
\textbf{input:} & event state and global state; \\
& function $F$ used to generate an Output Set \\
& data lineage enabled for ports IN and OUT \\
\textbf{output:} & Updates to OP global and event state; \\
& Updates to Input Set for $F$ and LOG.io logs; \\
& output events sent on output ports \\
\end{tabular*}
\begin{enumerate}

    \item \textit{Read actions:} If $F$ requires a read action $A$, OP must implement the processing steps of Algorithm \ref{alg:generate-reader}

    \item \textit{Global state:} If $F$ uses a global state then create a new State\_ID $s$, assign to $s$ as InSet\_ID the Input Set for $F$, serialize the global state, and add an event in EVENT\_LOG with event\_ID=$s$ and status "undone";

    \item \textit{Output Set:} $F$ computes the Output Set; assign a new event\_ID (SSN) to each output event 

    \item \textit{Update Logs:} use a transaction to: (1) log all output events in EVENT\_LOG with status "undone"; (2) store the data of output events in EVENT\_DATA, (3) log state\_ID $s$ in STATE with its state, (4) set the status to "done" in EVENT\_LOG for all events of the Input Set, and (5) If data lineage is enabled: (a) for each read action\_ID, add an entry in EVENT\_LOG  with status "done" and an entry in EVENT\_DATA with a reference to the store of the effect of the read action, (b) for each output event $o$ on a port in OUT, create an entry in EVENT\_LINEAGE for $o$ and the InSet\_ID of the Input Set. After that, Input Sets with "done" events are emptied. 

    \item \textit{Send output events:} Output events are sent on their respective output port.     

    \item \textit{Write actions:} For output events that are an atomic write action, OP must implement the steps of Algorithm \ref{alg:generate-writer}. 
    
\end{enumerate}
\end{algorithm}

In Step 2, a new state\_ID is assigned to the global state, and an event corresponding to the global state is created in EVENT\_LOG with null values for the two port IDs. The event is also assigned the InSet\_ID for $F$ to keep track of the global state used for the generation of the Output Set. 
In Step 3, the operator's data batching strategy determines the number of output events generated from an Output Set (dynamic batching). The event\_ID assigned to a write action $A$ on a connection $Cx$ must be unique for $A$ and $Cx$ within OP. 

Step 4 is the core requirement of the LOG.io protocol. It requires logging the events of an Output Set $O$ as "undone" and marking the events of the Input Set for $O$ as "done" within an atomic transaction. The output event created in EVENT\_LOG for a write action in an operator OP uses "OP.Cx" for the receiver port\_ID and a \textit{null} value for the sender port\_ID. This is how a write action can be distinguished in the log during recovery in Algorithm \ref{alg:recover-writer}. The event data for a write action in EVENT\_DATA consists of the description of the write action that must be performed. At the end of Step 4, Input Sets that contain "done" events are emptied because they cannot be used anymore to generate output events. 
One case of failure of the transaction is when no event with a given InSet\_ID can be found in EVENT\_LOG. This case will be explained later in Section \ref{sec:dynamic-scaling}. 

When data lineage capture is enabled, action (5) is added to the atomic transaction of Step 4. Action (5.a) creates an event in EVENT\_LOG and EVENT\_DATA for each read action identified by its action\_ID $r$. The entry in EVENT\_LOG has an event\_ID $r$, an InSet\_ID defined by the Input Set for $F$, "OP.Cx.r" as sender port, and a \emph{null} value for the receiver port. A reference to the store of $r(A,S)$ is kept in the entry created in EVENT\_DATA. 
Action (5.b) records the data lineage relationship between the events of the Output Set sent on a port of OUT and the Input Set. 

The sending of output events follows their successful logging (pessimistic logging). The conditions under which optimistic logging can be used are discussed in Section \ref{sec:recovery-with-replay}. Finally, in Step 6, write actions are processed after the sending of events to avoid an additional latency in the processing of output events by downstream operators. However, if there exists a causal dependency between an operator OP1 that writes data to an external system and a downstream operator OP2 that reads the same data, Steps 5 and 6 can be reordered.

\begin{algorithm}[tb]
\caption{Side-effect read actions in an operator OP}
\label{alg:generate-reader}
\begin{tabular*}{\linewidth}{rp{0.75\linewidth}}
\textbf{input:} & read action $A$ on connection $Cx$;  \\
& function $F$ used to generate an Output Set; \\
& Input Set for $F$ \\
\textbf{output:} & update to Input Set for $F$; \\
& Updates to LOG.io logs \\
\end{tabular*}
\begin{enumerate}

    \item \textit{Initialization:} If data lineage is enabled then assign a new action\_ID $r$ to $A$ and assign to $r$ the InSet\_ID of the Input Set for $F$. Then execute action $A$ on connection $Cx$ and compute the effect of $A$

    \item \textit{If $A$ is not replayable:} If data lineage is enabled, store the effect of $A$. If $A$ fails, discard the store for $A$ (if any), replay $A$ and proceed with Step 2. 

    \item \textit{If $A$ is  replayable:} If data lineage is enabled, store the effect of $A$. If $A$ fails, then replay $A$ and resume the computation of the effect of $A$ from its last event.   
  
\end{enumerate}
\end{algorithm}

\subsubsection{Side-effect read actions}
\label{sec:generate-reader}

When the generation of an Output Set requires one or more read actions, an operator OP must implement the processing steps of Algorithm \ref{alg:generate-reader} for each read action. The idea is that the effect of each read action creates additional (side-effect) input events for the generation of an Output Set, and therefore, the read action must be added to the corresponding Input Set. 

Like a Source operator, the processing steps distinguish the cases when $A$ is replayable or not. In Step 1, an action\_ID is created for $A$ only if data lineage is enabled, because only in this case we are interested to capture the data lineage relationships between events in the Output Set sent to a port in OUT and the read action\_ID. To achieve this, the InSet\_ID of the Input Set for $F$ is assigned to the event corresponding to the read action. 
In the case when data lineage capture is enabled, since $A$ is not replayable in Step 2, OP must store the effect of $r$ because otherwise we will not know the records in the effect of $r$ that contributed to the generation of an output event sent on a port in OUT.  
Note that even when a read action $A$ is replayable in Step 3, we also need to store its effect $r(A,S)$ because if it is replayed later over an observable state $S'$ we may have $r(A,S) \preceq r(A,S')$, which does not accurately describe the data lineage relationships for the output events.  

\subsubsection{Side-effect write actions}
\label{sec:generate-writer}

When an Output Set contains write actions, an operator OP must implement the processing steps of Algorithm \ref{alg:generate-writer} for each pending write action, which is captured as an output event in EVENT\_LOG that has a \textit{null} sender port and status "undone". The description of the write action to execute on a connection is obtained from EVENT\_DATA. 
A "failure" of a write action $w$ occurs when there is either a failure in the connection to the external system or the external system fails before acknowledging the successful execution of $w$. 
Because of our assumption that write actions are durable (see Section \ref{sec:atomic-actions}), we ignore the other cases.  

\begin{algorithm}[tb]
\caption{Side-effect write actions in an operator OP}
\label{alg:generate-writer}
\begin{tabular*}{\linewidth}{rp{0.75\linewidth}}
\textbf{input:} & write action $w$ on connection $Cx$  \\
\textbf{output:} & Updates to LOG.io log and external system \\
\end{tabular*}
\begin{enumerate}

    \item \textit{Execution:} execute action $w$ on connection $Cx$ 

    \item \textit{Failure:} If $w$ execution fails it is replayed
   
    \item \textit{Success:} If $w$ succeeds, use an atomic transaction to update the status of event $w$ in EVENT\_LOG as "done"

\end{enumerate}
\end{algorithm}

In Algorithm \ref{alg:generate-writer}, we assume that write actions are executed individually. If they are executed as a group, we require getting a successful acknowledgment for the group.  
In Step 2, there is an arbitrary number of attempts to connect to the external system before raising a fatal error. 
In Step 3, the status update of $w$ to "done" indicates that $w$ was successfully (and durably) acknowledged by the external system. 


\begin{example}
\label{full-example}
We illustrate Algorithms \ref{alg:input-events} and \ref{alg:generate-general} with the stateful operator OP3 of Example \ref{ex:two-input-sets}.  
When a first event $a_1$ is received from OP1, in Step 2 of Algorithm \ref{alg:input-events}, it is de-serialized and the "date" field of all its records is checked. Suppose that all records are within the same hour H, then a new InSet\_ID $I_1 = H$ is created and assigned to $a_1$, and all records of $a_1$ are inserted into the window for H. Then, next event $a_2$ is processed. Suppose that some records have a date within the hour H and others have a date within the hour H + 1. Then, a new InSet\_ID $I_2 = H+1$ is created, $a_2$ is assigned to both $I_1$ and $I_2$ and the records of $a_2$ are inserted into the two windows accordingly. Now, the triggering condition is true (Step 3) and the Generation phase can start for window H, that is, Input Set $I_1$. 

Steps 1 and 2 of Algorithm \ref{alg:generate-general} are omitted because they are not relevant. Let's assume that in Step 3, the data batching strategy of OP3 consists of generating a single output event $b_1$ from the window for hour H. In Step 4, $b_1$ is logged into the EVENT\_LOG with status "undone", events assigned to $I_1$ are marked as "done", and if data lineage is enabled on OP3's output port, a new entry is created in EVENT\_LINEAGE for $b_1$ with Input Set $I_1$. The window for hour H is emptied in the event state. In Step 5, event $b_1$ is sent to OP3's output port.  

Next, input event $a_3$ is processed. Suppose that its records have a date within hour H + 2. Then a new InSet\_ID $I_3=H+2$ is created and assigned to $a_3$ and the records of $a_3$ are inserted into a new window for hour H+2. Afterwards, the triggering condition is true for window H+1 and the generation of an Output Set can start for Input Set $I_2$. Assuming that a single output event $b_2$ is created and sent, the LOG.io log would be as depicted in Tables \ref{tab:event-log-2} and \ref{tab:event-log-3}, where we assume that OP4 is a successor to OP3. 
\end{example}

\begin{table}[htbp]
\centering
  \caption{Table EVENT\_LOG}
  \label{tab:event-log-2}
    \setlength{\tabcolsep}{4pt}
\resizebox{0.45\textwidth}{!}
{%
  \begin{tabular}{l|l|l|l|l|l|l}
\toprule
\attr{Event\_ID}&	\attr{Status}& \attr{Send\_Op}&	\attr{Send\_p}& \attr{Rec\_Op}&	\attr{Rec\_p}& \attr{InSet\_ID}\\
\midrule
    $a_1$&	done & OP1 & out &	OP3 & in & $I_1$\\
    $a_2$&	done & OP1 & out &	OP3 & in & $I_1$\\
    $a_2$&	undone & OP1 & out & OP3 & in & $I_2$\\
    $b_1$&	undone & OP3 & out & OP4 & in & null \\
    $a_3$&	undone & OP1 & out & OP3 & in & $I_3$\\
    $b_2$&	undone & OP3 & out & OP4 & in & null \\    
 \bottomrule
\end{tabular}
}
\end{table}

\begin{table}[htbp]
\centering
  \caption{Table EVENT\_LINEAGE}
  \label{tab:event-log-3}
    \setlength{\tabcolsep}{4pt}
\resizebox{0.28\textwidth}{!}
{%
\begin{tabular}{l|l|l|l|l|l|l}
\toprule
\attr{Event\_ID}& \attr{Send\_Op}& \attr{Send\_p} & \attr{InSet\_ID}\\
\midrule
    $b_1$&	OP3 & out &	$I_1$ \\
    $b_2$&	OP3 & out &	$I_2$ \\
 \bottomrule
\end{tabular}
}
\end{table}

\subsection{Garbage collect of LOG.io logs}
\label{sec:garbage-collect}

Tables of LOG.io logs can be cleaned up during or after the execution of a data pipeline depending on whether the data pipeline is long-running or not. 

Events in the EVENT\_LOG table can be removed as soon as they are "done". We shall see later an exception to this rule when we present the notion of "replayable operator" in Section \ref{sec:replay-operators}. 

Events in EVENT\_DATA that correspond to "done" events in EVENT\_LOG can also be removed, unless their sender operator has its data lineage capture enabled for the ports on which the events are sent, because access to the event's body is needed to answer data lineage queries. 
However, a data pipeline that consumes unbounded data (aka streaming or infinite data) can run forever, and the event's data can grow arbitrarily. A simple solution is to use a retention period $\tau$ as follows: the  lineage of events along the data lineage paths defined by a data lineage scope $(start, target)$ is kept for the output events produced at port $target$ within a sliding window of time $\tau$. Then, tables EVENT\_DATA and EVENT\_LINEAGE can be asynchronously cleaned up accordingly.

A read action in READ\_ACTION table can be deleted when it is "complete" and its corresponding event in EVENT\_LOG is "done" unless the operator issued the read action to generate output events on a port for which data lineage capture was enabled. In that case, the read action must be kept, using the same rules as for events in EVENT\_DATA when the data pipeline is long-running.

When data lineage capture is disabled, it is sufficient to keep the latest state\_ID record in STATE. Thus, logging a new global state can overwrite the previous stored state and a single state\_ID value is needed. When data lineage capture is enabled, the deletion of STATE records follows the same rule as for READ\_ACTION records when the data pipeline is long-running.

\section{LOG.io rollback recovery protocol }
\label{sec:logio-recovery}

In Section \ref{sec:recovery-overview}, we give an overview of the recovery process. Then in Sections \ref{sec:recovery-source} to \ref{sec:receovery-middle}, we describe the detailed recovery steps for each type of operator. Finally, we provide in Section \ref{sec:correctness-conditions} the conditions required by LOG.io to ensure a correct rollback recovery. 

\subsection{Overview of rollback recovery}
\label{sec:recovery-overview}

As explained in Section \ref{sec:intro-pipelines}, operators can be grouped to execute together in one execution environment assigned to a processing node (for instance, within a Docker image deployed as a Kubernetes pod). 
When an unexpected system failure occurs in an operator OP, the group to which OP belongs also fails and must be restarted during rollback recovery. 
Given an operator OP, we denote $Pred(OP)$ the set of operators having at least one output connection to OP and $Succ(OP)$ the set of operators having at least one input connection to OP.
When an operator OP belongs to a group that failed, OP is restarted as follows:

\begin{enumerate}
     \item OP must \textbf{recover its output events} to each successor operator OP' in $Succ(OP)$. That is, OP must (re-)send to OP' all the events which are still "undone" and have not yet been acknowledged by OP' (i.e., In\_Event\_ID $=$ null in EVENT\_LOG)
     \item if $Pred(OP) \neq \emptyset$, OP must \textbf{recover its processing} for each OP' in $Pred(OP)$. That is, OP must (re-)process all events received from OP' that are "undone" and acknowledged by OP.  
\end{enumerate}

In the subsequent subsections, we describe how the recovery of output events and the recovery of processing are done for each type of operator. In our protocol, we shall distinguish three states for an operator: "running", "dead" and "restarted". 

\subsection{Recovery of a Source operator}
\label{sec:recovery-source}
By definition, a Source operator has no predecessor operator and can only recover its output events. The processing steps of Algorithm \ref{alg:recover-source} must be followed by OP. 

\begin{algorithm}[tb]
\caption{Recover output events for a Source operator OP}
\label{alg:recover-source}
\begin{tabular*}{\linewidth}{rp{0.75\linewidth}}
\textbf{input:} & operator OP in state "restarted" \\
\textbf{output:} & send output events on "OP.out" connected to "OP'.in"
\end{tabular*}
\begin{enumerate}

    \item \textit{Resend:} send to OP' with increasing event\_ID value output events with status "undone", receiver operator OP' and InSet\_ID $=$ null 

    \item \textit{Restore:} Get last read action $r$ from READ\_ACTION; Get last global state from STATE; initialize global state 

    \item \textit{Resume "complete":} if $r$ is "complete" then get event $r$ from EVENT\_LOG
        \begin{itemize} 
        \item[a.] If $r$ is "done" : discard store for the effect of $r$ 
        \item[b.] Otherwise: resume execution from Step 2.c of Algorithm \ref{alg:normal-source} from the last offset of the effect of $r$ stored in STATE
        \end{itemize}   
    \item \textit{Resume "incomplete":} if $r$ is "incomplete"
        \begin{itemize}     
        \item[a.] if $r$ is non-replayable: discard the store for the effect of $r$ and replay $r$; resume execution from Step 2.a of Algorithm \ref{alg:normal-source}
        \item[b.] Otherwise: replay $r$ and resume execution from Step 3 of Algorithm \ref{alg:normal-source} using the last offset stored in STATE
        \end{itemize}

\end{enumerate}
\end{algorithm}

In Step 1, the output events that have not been acknowledged by a successor operator OP' must be resent. This case may happen when a failure occurred after the "undone" output event was logged in EVENT\_LOG but before the event was sent on the output port. 
In Step 2, the last read action $r$ from which OP must resume the generation of output events and the last global state are retrieved from the log.
Step 3 handles the case where $r$ is "complete". The recovery action for sending pending events on "OP.out" is already handled by Step 1 (whether $r$ is replayable or not). No other recovery action is needed if $r$ is replayable because $r$ is marked as "complete" when the last output event has been logged. If there exists an event for $r$ in EVENT\_LOG (i.e., $r$ is non-replayable), then in Step 3.a ($r$ is "done"), the store for the effect of $r$ must be discarded and the event for $r$ in EVENT\_DATA must be deleted, if this was not done already in Step 2.d of Algorithm \ref{alg:normal-source} before the failure occurred. In Step 3.b ($r$ is "undone"), the execution of the generation of output events in OP must be resumed using the restored global state. 
In Step 4.a, if $r$ is "incomplete", then either the effect of $r$ was not completely stored or OP failed before marking $r$ as "complete". In Step 4.b, the generation of output events is resumed. 


\subsection{Recovery of a Middle operator}
\label{sec:receovery-middle}

\paragraph{Recovery of output events} 
When a Middle operator OP is restarted after a failure (i.e., in state "restarted"), it must first recover its output events to all its successor operators and then recover the processing of its pending write actions, as described in Algorithm \ref{alg:recover-middle-output}. 

\begin{algorithm}[tb]
\caption{Recover output events for a Middle operator OP}
\label{alg:recover-middle-output}
\begin{tabular*}{\linewidth}{rp{0.75\linewidth}}
\textbf{input:} & $Succ(OP)$; state of operator OP \\
\textbf{output:} & send output events
\end{tabular*}
\begin{enumerate}


        \item \textit{Resend:} if OP is in state "restarted": for each operator OP' in $Succ(OP)$, resend to OP' (with increasing event\_ID value) all output events from EVENT\_DATA which have in EVENT\_LOG status = 'undone', sender operator ID = OP, receiver operator\_ID = OP' and InSet\_ID $=$ null

       \item \textit{Write actions:} Process pending write actions as specified in Algorithm \ref{alg:recover-writer}
       

 \end{enumerate}
\end{algorithm}

In Step 1, events are retrieved from EVENT\_LOG, their data is obtained from EVENT\_DATA, de-serialized and sent on the required output port. Events are sent to each output port of OP in the same order (increasing event\_ID) as before OP failed. 
Note that OP may send an obsolete event to OP' because after reading an event $e$ as "undone" with InSet\_ID $=$ $null$, OP' (which is still running) may have acknowledged $e$ and assigned it an InSet\_ID value. 
This is why the filtering condition of Step 2 in Algorithm \ref{alg:input-events} is needed. It exploits the assumptions (see Section \ref{sec:premininaries}) that each connection between two operators is reliable, respects a FIFO delivery, and that events are sent on each connection with an increasing event\_ID.

\paragraph{Recovery of write actions} The processing steps for the recovery of a Writer operator OP are described in Algorithm \ref{alg:recover-writer}. By Step 4 of Algorithm \ref{alg:generate-general}, each write action is an output event that is logged with its description in EVENT\_DATA and logged in EVENT\_LOG with a status "undone" and a \emph{null} sender port\_ID value. Afterwards, as described in Algorithm \ref{alg:generate-writer}, each write action must be executed on the external system and its status is tracked in EVENT\_LOG. Therefore, two cases can happen: (1) OP failed before proceeding with the execution of the write action, or (2) OP failed while proceeding with the execution of the write action. We handle both cases by re-executing the processing steps of Algorithm \ref{alg:generate-writer}. 

However, we still have to capture the case where OP failed before receiving the success response from the external system, or OP failed before executing the atomic transaction that marked the write action as "done" in EVENT\_LOG. To handle this case, OP must be able to access the status of the write action in the external system to know if it has been successfully acknowledged. When this is possible, we say that $w$ is \emph{checkable} on the external system. Different techniques can be employed to make a write action checkable, which are not in the scope of this paper. Step 2.a ensures that the pending atomic write actions are done exactly once. Finally, in Step 2.b, if the write action is idempotent, a correct rollback recovery is still possible.

\begin{algorithm}[tb]
\caption{Recover output of a Writer operator OP}
\label{alg:recover-writer}
\begin{tabular*}{\linewidth}{rp{0.75\linewidth}}
\textbf{input:} & LOG.io logs \\
\textbf{output:} & send write actions; update LOG.io logs
\end{tabular*}
\begin{enumerate}

        \item \textit{Get write actions:} Access in increasing event\_ID order write actions from EVENT\_DATA that have status "undone" and sender port\_ID = null in EVENT\_LOG. 
        
        \item \textit{Execute:} For each returned write action $w$ do:
      
        \begin{itemize} 
        \item[a.] if the status of $w$ in the external system is "success" then use an atomic transaction to update the status of event $w$ in EVENT\_LOG as "done"
        
        \item[b.] Else replay $w$ by executing the processing steps of Algorithm \ref{alg:generate-writer}
        \end{itemize}                

\end{enumerate}
\end{algorithm}

\paragraph{Recovery of processing} 
After recovering its output events, an operator OP that failed must follow the processing steps of Algorithm \ref{alg:recover-processing} to recover its processing of input events. 
Step 1 restores the global state and the array of the latest event\_ID from each input port that was used to update the global state. It also removes the stores of the effect of read actions that were never used to generate output events because a failure occurred before Step 5 of Algorithm \ref{alg:generate-general}.


\begin{algorithm}[tb]
\caption{Recover processing of operator OP}
\label{alg:recover-processing}
\begin{tabular*}{\linewidth}{rp{0.75\linewidth}}
\textbf{input:} & LOG.io logs \\
\textbf{output:} & output events sent on output ports; \\
& Updates to OP state and LOG.io logs \\
\end{tabular*}
\begin{enumerate}

        \item \textit{Restore state:} global state and LOG.io context are restored from STATE and the latest event\_ID sent on each output port is retrieved from EVENT\_LOG. Discard the stores of the effect of read actions that are not in EVENT\_LOG.
        
        \item \textit{Recover input events:} get from EVENT\_LOG all events sent to OP with status "undone" and InSet\_ID $\neq$ null. Events from different port IDs are processed in an order that respects the deterministic logic of OP when it exists, and the increasing event\_ID order for each port ID. Each event $e$ with InSet\_ID $I$ is then processed as follows:

        \begin{itemize} 
        \item[a.] Get and de-serialize event $e$ from EVENT\_DATA
        
        \item[b.] Use $e$ to update the global state and LOG.io context, if not already done; Use $e$ to only update the event state corresponding to Input Set $I$. 
        
        \item[c.] If a triggering condition evaluates to true, then the Generation phase is triggered as in Algorithm \ref{alg:generate-general}.
        \end{itemize}                

        \item \textit{Resume processing:} The state of OP is set to "running"; resume normal processing of input events received on input ports, as defined in Algorithm \ref{alg:input-events}
        
\end{enumerate}
\end{algorithm}

Step 2 restores the event state of OP by (re-)processing all the events that it previously received and incompletely processed (otherwise those events would be marked as "done"). 
If OP enforces a deterministic order to process the events received on different input ports, that order is also enforced during recovery. Otherwise, events are processed in a round-robin fashion between port IDs and in an increasing event\_ID order for each port ID. This may differ from the order in which OP processed the events before the failure. However, the rollback recovery is still correct because this processing order could occur in a failure-free execution. During Step 2.b, the restored global state is only updated by acknowledged events that did not update it before the state was logged in STATE during Step 4 of Algorithm \ref{alg:generate-general}. This is checked by Step 2 of Algorithm \ref{alg:input-events} using the array of event\_IDs restored from STATE. In Step 2.c, the State Update phase is adapted: the event state is updated only for the Input Set assigned to the recovered event. 

Step 3 resumes the normal processing of input events. 
An input event that was successfully sent to OP and not acknowledged is in the reliable communication channel and will be processed during Step 3. Alternatively, if the sender operator failed before sending the event (and hence the event could not be acknowledged), it will be sent again when the sender recovers its output events and processed during Step 3. 
Here again, the filtering condition in Step 2 of Algorithm \ref{alg:input-events} guarantees that a given input event will not be processed twice by OP.  

\begin{example}
\label{ex:recovery}
Consider the data pipeline of Figure \ref{fig:Example4} and the log described in Table \ref{tab:event-log-4} below. Suppose that a failure occurs before event $a_3$ is acknowledged by OP3. When OP3 is restarted, it first recovers its output events. By Step 1 of Algorithm \ref{alg:recover-middle-output}, it resends event $b_1$. Step 2 is not applicable. Then, OP3 recovers its processing. By Step 1 of Algorithm \ref{alg:recover-processing}, LOG.io context is restored from STATE. The latest event sent is $b_1$. Then by Step 2.a, event $a_2$ with InSet\_ID $I_2=H+1$ is retrieved from the log and used in Step 2.b to update the window corresponding to hour H+1. Thus, records of $a_2$ that belong to hour H are ignored. No triggering occurs and, by Step 3, OP3 resumes its processing and processes event $a_3$. 

\begin{table}[htbp]
\centering
  \caption{Table EVENT\_LOG}
  \label{tab:event-log-4}
    \setlength{\tabcolsep}{4pt}
\resizebox{0.45\textwidth}{!}
{%
  \begin{tabular}{l|l|l|l|l|l|l}
\toprule
\attr{Event\_ID}&	\attr{Status}& \attr{Send\_Op}&	\attr{Send\_p}& \attr{Rec\_Op}&	\attr{Rec\_p}& \attr{InSet\_ID}\\
\midrule
    $a_1$&	done & OP1 & out &	OP3 & in & $I_1$\\
    $a_2$&	done & OP1 & out &	OP3 & in & $I_1$\\
    $a_2$&	undone & OP1 & out & OP3 & in & $I_2$\\
    $b_1$&	undone & OP3 & out & OP4 & in & null \\
    $a_3$&	undone & OP1 & out & OP3 & in & null \\
 \bottomrule
\end{tabular}
}
\end{table}

\end{example}

\subsection{Correctness of LOG.io}
\label{sec:correctness-conditions}

As explained in Section \ref{sec:intro-rollback}, when a failure occurs, a correct recovery protocol must bring a data pipeline to a \emph{consistent} global state, that is, a state such that if an operator's state reflects that an input event was received, or if the state of a communication channel reflects that an event was sent, then the state of the sender operator reflects sending that event. In addition, the result of the execution of the recovered data pipeline on external systems must be equivalent to a failure-free execution of the failed pipeline, possibly scheduled at a point in time that follows the time at which it was originally scheduled. The latter condition ensures the correctness with respect to the external systems, as in the "end-to-end exactly once processing" property of \cite{AKIDAU:VLDB:15, CARBONE:VLDB:17}. 

During an error-free execution of a data pipeline, at any point in time, the global state is consistent. Indeed, our protocol firstly ensures that when an event $e$ is received (i.e., acknowledged) by an operator, an InSet\_ID has been assigned to $e$ and $e$ is logged with status "undone" by the sender operator. Secondly, if an event $e$ is in the communication channel then no InSet\_ID has been assigned to $e$ by the receiver operator and $e$ is logged with status "undone" by the sender operator. Since this state is persisted in both the logs and the reliable buffer of the communication channel, it can always be restored after a failure. 

In the remaining of this section, we focus on the equivalence to a failure-free execution. 
In Section \ref{sec:def-correctness},  we clearly define the correctness condition about the equivalence to a failure-free execution and then establish in Section \ref{sec:correctness-logio} the conditions that must be satisfied by a data pipeline to prove the correctness of LOG.io. 

\subsubsection{Preliminary definitions}
\label{sec:def-correctness}
In the following, we reason on a data pipeline that processes a fixed arbitrary long amount of input data ingested by Source operators. We introduce the notion of schedule of a Source operator OP to describe the sequence of read actions performed by OP to generate events on its output ports. 


\paragraph{Schedule of a Source operator}
Let $R$ be a Source operator in a data pipeline DP.
A \emph{schedule of R}, noted $\sigma(R) = [(A_1,S_1), ..., (A_n,S_n)]$, is a sequence in which each $(A_i,S_i)$ represents a successful read action $A_i$, executed over the observable state $S_i$ of an external system, whose effect has been used to generate events on the output ports of R during the execution of DP. 

Read actions that have aborted before completion 
do not appear in a schedule $\sigma(R)$. 

\paragraph{Complete execution of a data pipeline} Let DP be a data pipeline with Source operators $R_1, ..., R_n$. 
We denote by $\Gamma(DP, \sigma(R_1), ..., \sigma(R_n))$ a complete failure-free execution of DP, where each $\sigma(R_i)$ is the schedule of operator $R_i$ during the complete execution of DP. 
We denote by $\Gamma^{rec}(DP, \sigma(R_1), ..., \sigma(R_n))$ a complete recovered execution of DP, that is, a complete execution of DP with recovery from one or more failures. 

In a complete execution of a data pipeline, the effect of all the read actions occurring in the schedule of each Source operator has been completely used to ingest events, and the data pipeline has finished processing all the ingested events.   


In the next definition, we use the term "destination" of an operator to designate either an output port or a connection to an external system on which write actions are sent. 


\paragraph{Correct recovered execution} 
Let DP be a data pipeline with Source operators $R_1, ..., R_n$ and other operators $OP_1,...,OP_p$ .
Let $OP_i.out_j$, $1 \leq j \leq q$, denote the output destinations of $OP_i$, and 
$O_{i, j} = \Gamma^{rec}(DP, \sigma(R_1), ..., \sigma(R_n)).OP_i.out_j$ be the sequence of all output events generated on destination $OP_i.out_j$ in $\Gamma^{rec}$.
We shall say that $\Gamma^{rec}(DP, \sigma(R_1), ..., \sigma(R_n))$ is a \emph{correct recovered execution} of DP if, 
for each $i$, $1 \leq i \leq n$, there exists some schedule $\sigma'(R_i)$ in a failure-free execution $\Gamma$ of DP, such that for each $j$, $1 \leq j \leq p$, and each $k$, $1 \leq k \leq q$:
\begin{center}
$\Gamma^{rec}(DP, \sigma(R_1), ..., \sigma(R_n)).O_{j,k} = \Gamma(DP, \sigma'(R_1), ..., \sigma'(R_n)).O_{j,k}$
\end{center}

Our definition of correct rollback recovery establishes the requirement so that the result of a complete recovered execution of data pipeline is identical, at the destinations of any operator, to a failure-free execution of that pipeline scheduled at some point in time that could differ from the time at which the pipeline was initially scheduled \cite{ELNOZAHY:02}. 
In our definition, we pay attention to the states of the external systems observed by Source operators (notion of schedule) and the states modified by Writer operators (notion of destination that includes output events on connections to external systems). 


\subsubsection{Conditions for correctness}
\label{sec:correctness-logio}

We first show that LOG.io provides a correct recovered execution for a Source operator. We assume without loss of generality that a Source operator has a single output port. The case of multiple output ports can be handled by a multiplexer Middle operator.

\begin{proposition}
Let R be a Source operator in a data pipeline DP, and $\Gamma^{rec}(DP, \sigma(R))$ be a complete recovered execution of DP with schedule $\sigma(R)$. 
Let $out$ be the output port of R. Let $O = \Gamma^{rec}(DP, \sigma(R)).out$ be the sequence of output events produced by R on port $out$, then there exists a schedule $\sigma'(R)$ in a complete failure-free execution $\Gamma$ such that: 
\begin{align}
    \label{eq:prop1}
    \Gamma^{rec}(DP, \sigma(R)).out= \Gamma(DP, \sigma'(R)).out
    \end{align}
\end{proposition}

\begin{proof}
Let $I=[A_1, ..., A_k]$ be the sequence of read actions executed by R during a complete failure-free execution $\Gamma$ of DP. Any schedule in $\Gamma$ is a sequence that contains a single element for each action of $I$ and each output event is the result of a function over $r(A_i, S_i)$ (effect of $A_i$ on state $S_i$) for some action $A_i$ of $I$. 
Thus, if Equation \ref{eq:prop1} is violated, there exists a read action $A_i$ such that both $(A_i, S_i)$ and $(A_i, S'_i)$ belong to $\sigma(R)$ with an event $e$ that is in the effect of $(A_i, S_i)$ and not in the effect of $(A_i, S'_i)$, or conversely. We assume that $(A_i, S'_i)$ follows $(A_i, S_i)$ in $\sigma(R)$. 

We shall prove that one of these two properties hold: (P1) there is a single element in $\sigma(R)$ for every action of $I$, or (P2) if two elements $(A_i, S_i)$ and $(A_i, S'_i)$ exist in $\sigma(R)$, the output events produced from $r(A_i, S_i)$ and $r(A_i, S'_i)$ are the same as those produced from $r(A_i, S'_i)$.   

Let $(A_i, S_i)$ be the action started by R before DP failed for the first time. If R did not fail, then R continues its normal processing, $(A_i, S_i)$ is added to $\sigma(R)$ and $A_{i+1}$ is processed. Thus, there will be a single element in $\sigma(R)$ for $A_i$. If R fails, the only following cases may occur:
\begin{itemize}
    \item If R fails in Steps 1, 2.a or 2.b of Algorithm \ref{alg:normal-source}. Then $A_i$ is "incomplete" and by Step 4.a of Algorithm \ref{alg:recover-source}, $A_i$ is aborted and replayed over a state $S'_i$ that is subsequent to $S_i$. Thus, $(A_i, S_i)$ will not be in $\sigma(R)$. This proves property P1.
    \item If R fails in Step 2.c or 2.d of Algorithm \ref{alg:normal-source}. Then $A_i$ is "complete" and by Step 3 of Algorithm \ref{alg:recover-source}, the effect of $(A_i, S_i)$ will eventually be used to send all output events and action $A_{i+1}$ is executed. Thus, $(A_i, S_i)$ is added to $\sigma(R)$ and there will be no other element in $\sigma(R)$ for $A_i$. This proves property P1.
    \item If R fails in Step 3 of Algorithm \ref{alg:normal-source}, by Step 4.b of Algorithm \ref{alg:recover-source}, $A_i$ is replayed over a state $S'_i$ that is subsequent to $S_i$. If output events are produced from $r(A_i, S'_i)$ and $r(A_i, S_i)$, the two actions are in $\sigma(R)$. However, since $A_i$ is replayable, any output event generated from $r(A_i, S_i)$ is also an output event generated from $r(A_i, S'_i)$. Furthermore, no event can be duplicated because the generation of output events is resumed from the last event generated, sent (and logged) from $r(A_i, S_i)$. This proves property P2.   
\end{itemize}

Thus, there always exists a schedule of read actions such that the complete execution of DP without any failure provides the same result on the output port of R as the complete recovered execution of DP.
\end{proof}

We now show that LOG.io provides a correct recovered execution for a Middle operator.

\begin{proposition}
\label{prop2}
Let DP be a data pipeline with Source operators $R_1, ..., R_n$, and  a Middle operator $OP$. Let $\Gamma^{rec}(DP, \sigma(R_1), ..., \sigma(R_n))$ be a complete recovered execution of DP. 
Let $out_j$, $1 \leq j \leq q$, be the destinations of $OP$ and $\Gamma^{rec}(DP, \sigma(R_1), ..., \sigma(R_n)).out_j$ be the sequence of output events produced on destination $out_j$. 
Then for each $i$, $1 \leq i \leq n$, there exists some schedule $\sigma'(R_i)$ such that for each $j$, $1 \leq j \leq q$:
\begin{align}
    \label{eq:prop2}
    \Gamma^{rec}(DP, \sigma(R)).out_j = \Gamma(DP, \sigma'(R)).out_j
    \end{align}
\end{proposition}

\begin{proof}
We use a proof by induction on the length $L$ of the path of operators to OP. 

\paragraph{\textbf{Base case}: $L=1$} OP has at least one input port connected to a Source operator among $R_1, ..., R_n$. Since Source operators have a correct recovery, in case of failure of DP caused by any Source operator, the sequence of events received by OP on each input port after recovery is the same as what OP would receive in a failure-free execution of DP. We show that for any sequence of events sent to OP on each input port, if OP fails, then property \ref{eq:prop2} holds. 

When OP fails for the first time, the following cases may occur.

\noindent
\textbf{Case 1:} OP failed during the State Update phase (Algorithm \ref{alg:input-events}) of an input event $e$. Then OP will recover its processing using Algorithm \ref{alg:recover-processing} as follows.  
 
\begin{itemize}
\item In Step 1 of Algorithm \ref{alg:recover-processing}, the global state is restored from the STATE log, including the array of the latest event\_IDs from each input port used to update the global state. 

\item In Step 2 of Algorithm \ref{alg:recover-processing}, all "undone" and acknowledged events are retrieved from the log and processed using Steps 2.a to 2.c.   
If OP failed \textit{after Step 2} during the processing of event $e$ in the State Update phase, then $e$ will be part of this recovery phase. Next, if OP enforces a deterministic order between events coming from different input ports, it will also be enforced during recovery. Otherwise, LOG.io uses a round-robin order, which is one possible non-deterministic order in a failure-free execution of OP. 
Step 2.b of Algorithm \ref{alg:recover-processing} ensures that the global state is only updated by recovered input 
events acknowledged by OP \emph{after} the last output event was logged with the global state (in Step 4 of Algorithm \ref{alg:generate-general}). 
Hence, after Step 2.b, the global state is rebuilt as in a failure-free processing. 

\item Before the failure of OP, each Input Set contained only "undone" and acknowledged events because during Step 4 of Algorithm \ref{alg:generate-general}, Input Sets having "done" events are emptied. By Step 2.b of Algorithm \ref{alg:recover-processing}, each Input Set is rebuilt by successively processing all pending "undone" and acknowledged events in an order that could occur in a failure-free execution. Hence, after Step 2.b, the Input Sets are rebuilt as in a failure-free execution. 

\item From what precedes, the output events that are produced from the recovered global and event states can also be produced in the failure-free execution, even when the generation function of output events is non-deterministic. 

\item If OP failed in Steps 1-2 during the processing of an event $e$ in the State Update phase, $e$ was not acknowledged and it will be processed again when OP resumes its processing in Step 3 of the recovery Algorithm \ref{alg:recover-processing} after all previously received events have been processed, from an operator state that could occur in a failure-free execution. 
\end{itemize}

\noindent
\textbf{Case 2:} OP failed during the Generation Phase of an Output Set. 
\begin{itemize}
\item OP failed during the processing of some Input Set $I$ using the function $F$ in Steps 1 to 4 of Algorithm \ref{alg:generate-general}, before committing the atomic transaction. When OP recovers its processing using Algorithm \ref{alg:recover-processing}, the global and event states are rebuilt as in a failure-free execution (as explained in Case 1 before). If the Generation phase is triggered for Input Set $I$ using $F$, an Output Set is calculated, which may differ from the Output Set that was calculated before the failure occurred. This has no impact on Equation \ref{eq:prop2} since no output event for Input Set $I$ was previously logged and therefore sent. 


\item OP failed after committing the atomic transaction in Step 4 and before completing Step 6 of Algorithm \ref{alg:generate-general}. Since the output events were logged before the failure, they can be recovered using the recovery Algorithm \ref{alg:recover-middle-output} in which Step 1 ensures that events that have not been acknowledged are sent on each output port in the order in which they were produced. If OP failed during Step 4 before emptying the Input Set containing "done" events, that Input Set will not be rebuilt during the recovery of processing (as explained in Case 1). If the failure occurred after sending output events in Step 5 and some sent event $e$ is acknowledged, while $e$ is sent again through Step 1 of Algorithm \ref{alg:recover-middle-output}, then $e$ will be discarded by the receiving operator using Step 1 of Algorithm \ref{alg:input-events}, because it is obsolete. Thus, the sequence of output events that will be sent to each downstream operator will exactly be the sequence of events that would be sent if the failure had not occurred. Hence, Equation \ref{eq:prop2} is satisfied. 

\item OP failed during Step 6 of Algorithm \ref{alg:generate-general}. Since the write actions were logged in Step 4, they can be recovered using the recovery Algorithm \ref{alg:recover-writer}. However, during Step 3 of Algorithm \ref{alg:generate-writer}, a write action can be executed twice in the case where either OP failed before receiving a "success" response from the external system, or OP failed before executing the transaction that marked the write action as "done" in EVENT\_LOG. Recovery Step 2.a of Algorithm \ref{alg:recover-writer} requires that a write action is checkable on the external system. When this is not possible, the only way to ensure that a write action is not executed twice, which would differ from a failure-free execution, is to require that the write action be idempotent. Under these conditions, Equation \ref{eq:prop2} is satisfied.
\end{itemize}

\noindent
\textbf{Case 3:} OP failed at the same time as other operators in DP. Each failed operator is restarted and recovers independently from the others. Since recovery is driven by the contents of the LOG.io logs, the only conflicting cases can occur when two consecutive operators concurrently access the logs. Suppose that operator OP failed as well as its successor operator OP':
\begin{itemize}
    \item During recovery, OP recovers its "undone" unacknowledged output events from the log and sends them to OP' (Step 1 of Algorithm \ref{alg:recover-middle-output}). After recovery, OP' resumes its processing (Step 3 of Algorithm \ref{alg:recover-processing}) and acknowledges its "undone" input events. If an event $e$ is acknowledged by OP' after recovery and before OP recovers its output event, OP will not send $e$e during its recovery of output events. If event $e$ is being processed by OP' after recovery while OP recovers its output events, OP' will filter out the duplicate event $e$ using the filtering condition in Step 1 of Algorithm \ref{alg:input-events}. Thus $e$ will never be processed twice by OP'.  

    \item During recovery, OP' recovers its "undone" input events and marks some of them events as "done" during the Generation phase (Step 2.c of Algorithm \ref{alg:recover-processing}). During recovery, OP recovers its "undone" unacknowledged output events from the log and sends them again to OP'. If an event $e$ is marked as "done" by OP' during recovery, it was necessarily acknowledged before and will not be sent again by OP.  Thus $e$ will never be processed twice by OP'.
\end{itemize}

\paragraph{\textbf{Induction step: }$L=N$} Suppose that Proposition \ref{prop2} holds in a data pipeline DP for any Middle operator OP on a path of length $N$. We prove that this is still the case for an operator OP whose input ports are connected to one or more operators of DP which are at a distance $N$ from Source operators. 

Since Source and Middle operators of DP have a correct recovery, in case of failure of DP caused by any operator, the sequence of events received by OP on each input port after recovery is the same as what OP would receive in a failure-free execution of DP. Using the same case analysis as before, we then show that for any sequence of events sent to OP on each input port, Equation \ref{eq:prop2} holds. 

\end{proof}

\subsubsection{Summary of correctness conditions}

LOG.io provides limited conditions to ensure a correct recovered execution. Firstly, it does not require that the read actions of Source operators be replayable because our protocol takes care of non-replayable read actions. Secondly, LOG.io accepts Middle operators that perform stateful non-deterministic computations with read and write side-effect actions on external systems. 
However, for write side-effects, LOG.io requires that a write action be either checkable on an external system or idempotent. 


\section{LOG.io recovery with operator replay}
\label{sec:replay-operators}

Our protocol requires that an operator logs the data of the output events before actually sending the output events to their destination operator. This is called \emph{pessimistic logging}. As observed in previous work \cite{lineage-stashing:2019}, this has two drawbacks: (1) it creates a latency that grows with the number of output events and the size of events, and (2) it generates concurrent database transactions over the log database. 
In this section, 
we present a version of LOG.io which assumes that the output event data is not necessarily stored during normal processing.

\subsection{Principle of replaying an operator}
\label{sec:operator-replay}

\begin{figure}
    \centering
    \includegraphics[width=0.90\linewidth]{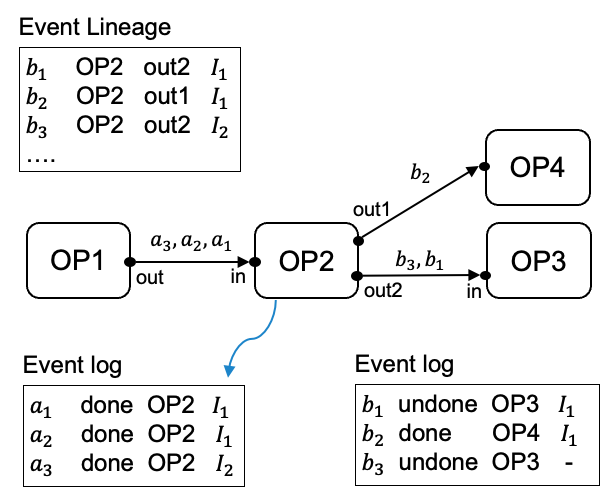}
    \caption{Data pipeline with data lineage enabled}
    \label{fig:replay-1}
\end{figure}

We start with a motivating example.
\begin{example}
\label{ex:replay-1}
We consider the abstract example in Figure \ref{fig:replay-1}.
Events $a_i$ (resp. $b_i$) are output events of OP1 (resp. OP2). The execution state of the pipeline is partially represented by the LOG.io log tables. The EVENT\_LOG table displays an output event\_ID, its status, the receiving operator, and the assigned InSet\_ID, and the EVENT\_LINEAGE table displays an output event\_ID, its sending operator and port, and the InSet\_ID in the sending operator from which it is generated. 
We suppose that operator OP2 does not store the data of its output events. At this point, suppose that OP2 fails, then during the recovery of its output events, OP2 cannot access event $b_3$ (unacknowledged and "undone") from EVENT\_DATA, de-serialize it, and send it to OP3. However, $b_3$ can be regenerated by OP2 from its Input Set $I_2=\{a_3\}$, as shown by the EVENT\_LINEAGE table. If OP1 does not store its data either, then $a_3$ must be re-generated by OP1 from its Input Sent, sent to OP2 which can now consume it to regenerate $b_3$. Alternatively, suppose that OP3 fails, then it must recover the processing of event $b_1$ whose data is not logged. Thus, $b_1$ must be regenerated by OP2 from its Input Set $I_1=\{a_1,a_2\}$ and sent to OP3. However, since event $b_3$ is already in the communication channel to OP3, it must be discarded to process $b_1$ first. Afterwards, $b_3$ must be sent again to OP3 by OP2. 
\end{example}

\paragraph{Replay operator} An operator is said to be a \emph{replay operator} if it does not store the data of its output events (in EVENT\_DATA) and it satisfies the following conditions: (1) its data lineage is enabled on all its ports and (2) it is deterministic.  

We illustrate the two conditions imposed on a replay operator using Example \ref{ex:replay-1} in which OP2 failed. Condition (1) is obvious since we need to know from which Input Set event $b_1$ must be regenerated. Condition (2) actually includes two requirements. First, the generation of output events must use a deterministic function. In fact, event $b_2$ has already been processed by OP4 (marked as "done"). So, suppose that OP2 replays the generation of an Output Set from $I_1$ and we obtain a result that does not include the event $b_2$ previously sent to OP4. This could occur, for instance, if OP2 uses a read action that is not replayable. Then, this would compromise the correctness of the recovery execution because it cannot be equivalent to a failure-free execution. As a second requirement of determinism, if an operator is stateful, then the order in which the events received on its input ports are processed must be deterministic, or the processing logic of input events must be insensitive to that order. If this is not the case, a different Output Set can be generated from before the failure, again resulting in an incorrectly recovered execution. If an operator is stateless, this second requirement is useless since, by definition, a stateless operator generates an Output Set from a single input event.

\subsection{Recovery protocol with operator replay}
\label{sec:recovery-with-replay}

Two lessons can be drawn from Example \ref{ex:replay-1}. Firstly, if a replay operator fails, the output events to recover must be regenerated from their respective Input Sets, as indicated by the EVENT\_LINEAGE table. Secondly, if a failed operator OP has a predecessor replay operator OP', then OP' must resend to OP the input events whose processing must be recovered by OP as well as the events it has sent afterwards. This can happen recursively along the chain of replay operators that precede OP. 

Therefore, with the replay-based version of LOG.io, when an operator OP fails, the Pipeline Engine performs the following actions: (1) it restarts OP in state "restarted" (as with normal recovery), (2) if OP is preceded by a replay operator, then it is stopped and restarted in state "replay", and (3) every replay operator that precedes an operator in state "replay" is stopped and restarted in state "replay". Each operator in state "restarted" or "replay" is then asked to recover its output events, and if applicable, recover its processing, as described below. 



\vspace{2mm}
\noindent
During the \textbf{recovery of output events}:
\begin{itemize}
    \item If OP is a replay operator in state "restarted": OP must replay the generation of output events that are "undone" and not acknowledged. So, OP marks the corresponding input events as "replay" and recovers their processing. 
    
    \item If OP is a replay operator in state "replay": OP must replay the generation of output events marked as "replay" and the output events sent after them. So, OP marks the corresponding input events as "replay" and recovers their processing.
    
    \item If OP is not a replay operator: OP resends its output events that are "undone" and not acknowledged from the EVENT\_DATA log (regular LOG.io recovery). 
\end{itemize}

\noindent
During the \textbf{recovery of processing}:
\begin{itemize}
    \item If OP is a non-replay operator that recovers events from a replay operator: OP marks these input events as "replay" (this is already done if OP is in state "replay"). 
    
    \item If OP recovers events from a non-replay operator: OP processes the input events ("replay" or "undone") to recover from the EVENT\_DATA log. This subsumes the regular LOG.io recovery. 

    \item If OP receives an input event marked as "replay": it processes it without applying the obsolete filtering condition of the regular State Update phase. 
\end{itemize}


        
 

\begin{algorithm}[tb]
\caption{Recover output events in replay mode}
\label{alg:recover-and-replay}
\begin{tabular*}{\linewidth}{rp{0.75\linewidth}}
\textbf{input:} & OP in state "restarted" or "replay" \\
\textbf{output:} & LOG.io log updates
\end{tabular*}
\begin{enumerate}

        \item \textit{Regular case:} if OP is not a replay operator, follow the processing steps of Algorithm \ref{alg:recover-middle-output}. Otherwise, continue. 
        
        \item \textit{Input Sets to replay:} if OP is a replay operator in state "restarted" (resp. in state "replay"), use EVENT\_LINEAGE to get the Input Sets, noted In\_Rec, of all output events which have status = "undone" and InSet\_ID $=$ \textit{null} (resp. status = "replay"). If OP is in state "replay", add to In\_Rec the Input Sets of output events sent after those marked as "replay". Exit if no Input Set is returned. 

        \item \textit{Set LOG.io context:} restore LOG.io context from STATE and update it with the smallest event\_ID of the output events generated from In\_Rec on each output port of OP and the smallest InSet\_ID
        
        \item \textit{Prepare replay:} use a transaction to: (1) set the status of all events of Input Sets in In\_Rec to "replay" and set InSet\_ID to \textit{null}, (2) set all output events to recover to status "replay", and (3) store in STATE the LOG.io context

        \item \textit{Write actions:} Process pending write actions as specified in Algorithm \ref{alg:recover-writer}
 
\end{enumerate}
\end{algorithm}

More specifically, with the replay-based version of LOG.io, when an operator recovers its output events, it must execute the processing steps of Algorithm \ref{alg:recover-and-replay}. 
In Step 1, if OP is not a replay operator, it can only be in state "restarted" during recovery. 
If OP fails during recovery before Step 4 then Steps 1 to 3 are reprocessed. 
After Step 4, all input events that must be reprocessed are marked as "replay" consistently with the LOG.io context persisted in STATE. If OP fails at this point, during recovery, Step 2 will return an empty set which stops the recovery of output events. Finally, in Step 5, after being restarted, it must process its pending write actions. 

\begin{algorithm}[tb]
\caption{Recover processing in replay mode}
\label{alg:recover-processing-replay}
\begin{tabular*}{\linewidth}{rp{0.75\linewidth}}
\textbf{input:} & OP in state "restarted" or "replay" \\ 
& replay operators connected to input ports of OP \\
\textbf{output:} & output events sent on output ports; \\
& LOG.io log updates \\
\end{tabular*}
\begin{enumerate}

        \item \textit{Regular case:} if OP is not a replay operator and has no preceding replay operator, follow the processing steps of Algorithm \ref{alg:recover-processing}. Otherwise, continue. 
        
        \item \textit{Restore state:} restore the global state from STATE including LOG.io context; If OP is not a replay operator then discard the stores of the effect of read actions which are not in EVENT\_LOG.
        
        \item \textit{Inputs from replay operator:} if OP has a preceding replay operator, mark input events coming from a replay operator that are "undone" and acknowledged as "replay"

        \item \textit{Process input events:} input events coming are processed as follows:

        \begin{itemize} 
        \item[a.] non "replay" events received on a port connected to a replay operator OP' are discarded while "replay" events are processed through the State Update phase as in Algorithm \ref{alg:input-events}
        
        \item[b.] "replay" input events coming from non-replay operators, and "undone" and acknowledged input events coming from non-replay operators, are obtained from the EVENT\_DATA log and processed through the State Update phase     
        \end{itemize}

        \item \textit{Resume processing:} OP is set to state "running" and resumes its normal processing of input events 
\end{enumerate}
\end{algorithm}

When an operator recovers its reprocessing, it must execute the processing steps described in Algorithm \ref{alg:recover-processing-replay}. 
In Step 1, if OP is not a replay operator, then it is in state "restarted" and it could have non-replayable read actions whose effect is logged. 
In Step 2, if OP is a replay operator, then the "done" input events to recover have been marked as "replay" during the recovery of output events in Algorithm \ref{alg:recover-and-replay}.  However, since OP has been stopped and restarted, it must also recover the processing of its "undone" acknowledged input events. 
Step 3 processes events in an order that respects the deterministic logic of OP when it exists, and the increasing event\_ID order for each port. Step 3.a handles the case when input events must be replayed by a non-replay preceding operator. Step 3.b includes the original recovery of input events in Step 2 of Algorithm \ref{alg:recover-processing} when their data is stored. 

To support the LOG.io recovery protocol in replay mode, the normal processing of an operator must be changed as follows. 

\vspace{2mm}
\noindent
During the \textbf{State Update phase}: If an operator OP in state "replay" or "restarted" is preceded by a replay operator, then in Step 1 of Algorithm \ref{alg:input-events}, it discards non-"replay" events, and it accepts "replay" events without applying the filtering condition for obsolete events. An operator in state "running" discards "replay" input events. 

\vspace{2mm}
\noindent
During the \textbf{Generation phase}: if a replay operator OP is in state "replay" or "restarted", then the steps of Algorithm \ref{alg:generate-general} are adapted:
\begin{itemize}
\item In Step 3, OP assigns event\_IDs starting from the values of LOG.io context restored in Step 1 of   Algorithm \ref{alg:recover-processing}, 
\item In Step 4, output events with status "replay" are marked as "undone" and no new output event is appended to the EVENT\_LOG since it already exists, 
\item In Step 5, events with status "replay" are sent in a message with an attribute "replay" in their headers to distinguish them from other events during the State Update phase.
\end{itemize}

\begin{example}
\label{ex:replay-2}
Consider Example \ref{ex:replay-1} in which OP1 is not a replay operator. If replay operator OP2 fails it is restarted in state "restarted" and must recover its output event $b_3$ using Algorithm \ref{alg:recover-and-replay}. Step 1 is not applicable, and Step 2  returns In\_Rec = $\{I_2\}$. Step 3 updates LOG.io context with event\_ID = $b_3$ for port OP2.out2 and InSet\_ID = $I_2$. In Step 4, events $a_3$ and $b_3$ are marked as "replay" and LOG.io context is stored. Then OP2 must recover its processing using Algorithm \ref{alg:recover-processing-replay}. Since OP1 is not a replay operator, then OP2 accesses "replay" event $a_3$ from the log and acknowledges it with InSet\_ID $I_2$. The generation of $b_3$ is then triggered, and OP2 marks $b_3$ as "undone", marks $a_3$ as "done", and sends $b_3$ as a replay event to OP3, which is in the state "running". If in between, OP3 already processed $b_3$, then the replay event will be filtered as obsolete. 

If (non-replay) OP3 fails, it recovers its output events from the log as in Step 1 of Algorithm \ref{alg:recover-and-replay}. Then, OP3 recovers the processing of $b_1$ using Algorithm \ref{alg:recover-processing-replay}. By Step 2, $b_1$ is marked as "replay" and by Step 3, OP3 waits to receive replay events on its input port. Next, OP2 is stopped, set to state "replay", and must recover its output events using Algorithm \ref{alg:recover-and-replay}. By Step 2, In\_Rec = $\{I_1, I_2\}$. Step 3 updates LOG.io context with event\_ID = $b_1$ for port OP2.out2, event\_ID = $b_2$ for port OP2.out1, and InSet\_ID = $I_1$. After Step 4, input events $a_1, a_2, a_3$ are marked as "replay". Then, OP2 recovers its processing using Algorithm \ref{alg:recover-processing-replay}. Steps 1 and 3 are not applicable. By Step 3.b, OP2 accesses and processes the "replay" input events from the EVENT\_DATA log. OP2 regenerates the output events and sends $b_1, b_2$ as replay events to OP3 and OP4 respectively, and sends $b_3$ to OP3 as a regular event. Then, OP2 returns to state "running". OP4 discards event $b_2$ because it is obsolete. OP3 first processes $b_3$ (already in the FIFO channel), which is not a replay event and is discarded. Then, it processes replay event $b_1$, marks it as "undone" and assigns it to InSet\_ID $I_1$. OP3 returns to state "running", and finally processes $b_3$ as a regular event. 
\end{example}

The principle of recovery in replay mode can be applied to support an optimistic logging of output events when an operator satisfies the conditions of a replay operator. In that case, output events are sent before logging their data. At recovery time, if the event data is not logged, recovery is done in a replay mode, otherwise it is done using the regular recovery mode. 



\section{LOG.io Implementation} 
\label{sec:logioImplementation}

In this section, we describe the implementation of the LOG.io protocol made in the SAP Data Intelligence Cloud (SAP DI Cloud) system, which is a comprehensive data management solution \cite{SAP:DI:24}. Section \ref{sec:4.SystemArchitecture} gives an overview of the architecture of the system. Section \ref{sec:4.Protocol-api} describes the API used to implement the LOG.io protocol within operators and illustrates its usage for different types of operators in Sections \ref{sec:4.OperatorImpl} and \ref{sec:4.RecoveryImplementation}.

\subsection{Architecture overview} \label{sec:4.SystemArchitecture}




A main feature of SAP DI Cloud is the ability to create distributed stream processing applications in the form of data pipelines using a high-level graphical interface \cite{SAPDImodeler}. 

An operator can either be predefined in SAP DI or custom defined (e.g., in Python, node.js or Golang). An operator is implemented based on a runtime environment and respective libraries (e.g., an operator executing Python code requires a Python interpreter and libraries). The runtime environment requirements are defined by a Docker file. Operator definitions, libraries and Docker files are stored in a repository. Tags are associated with both operators and Docker files. 

\paragraph{Data pipeline deployment} When a data pipeline is deployed, The \emph{Deployment Engine} translates 
each operator into a process that either runs individually, or within a generic process called a \emph{sub-engine} (e.g., a sub-engine to execute Python, node.js or C++ code). The default engine executes Golang code. All operators running in the same sub-engine run in the same operating system process. A sub-engine can interpret and execute a subgraph of data pipeline with its sub-engine-specific operators. Sub-engines have associated predefined tags.

An \emph{Image Composer} automatically groups operators in such a way that each group of connected operators have their required tags matched by a single Docker file or a sub-engine.
The resulting Docker files are then built and deployed on a Kubernetes cluster, each group of operators being assigned to a different container and pod. 

Inside each group pod, a \emph{Group Manager} process is responsible for managing the life cycle of the operators in its subgraph. 
During the start of a graph, it establishes the connections between operators, runs the initialization methods of operators, and finally starts them. The Group Manager is also responsible for serializing and de-serializing the messages (events) exchanged between different groups.

\paragraph{Data pipeline execution}
Data is transported from operator to operator in a generic message format, which can be refined by structured metadata descriptions (that is, table-type descriptions). The transport medium can be in-process queues, or other low-level communication primitives depending on whether the message crosses sub-engine or group boundaries. In the latter case, they are serialized and delivered via inter-process communication when they cross sub-engine's boundaries, or via an in-house messaging system, built on top of TCP, when they cross group's boundaries.

The \emph{Pipeline Engine} (API server) component keeps track of the running data pipelines, stores their metadata and manages the storage of a data pipeline state for rollback recovery purpose in a database instance. The Pipeline Engine is a user application, i.e., each user runs its own instance of the engine. Therefore, modification to artifacts in the repository can be executed in the scope of the user  (without exposing the modification to other users in SAP DI).
The API server and the group pods communicate with control events through a message-passing framework.

\paragraph{ABS versus LOG.io} 
SAP DI Cloud currently implements a version of the Asynchronous Barrier Snapshotting (ABS) protocol that is quite close to the version implemented in Flink \cite{FLINK:24}. One difference is that SAP DI Cloud does not provide a two-step commit protocol to synchronize the write actions of multiple Writer operators. The consequence is that it imposes stricter constraints than Flink to ensure a correct rollback recovery. 
To support the LOG.io protocol, we made changes to the implementation of the Pipeline Engine that are incompatible with the implementation done to support ABS. A different version of the Pipeline Engine was created in which we disabled the parts of the code that are irrelevant for LOG.io. Therefore, in our experiments, we use two different clusters, each of which runs a different version of rollback recovery. 

The ABS protocol within SAP DI Cloud uses an instance of the in-memory SAP HANA database \cite{SAP-HANA:25} to store the periodic snapshots created by operators. For convenience and simplification of the SAP DI cluster deployment scripts, we use the same database instance to store the logs of LOG.io. However, although the HANA database provides elasticity and distributed processing capabilities, we did not change its installation settings to accommodate the needs of LOG.io. We shall take this into account in our performance evaluation in Section \ref{sec:experiments}

\subsection{LOG.io Protocol API} \label{sec:4.Protocol-api}

A set of methods, called LOG.io API, is provided to support the implementation of the LOG.io protocol in data pipeline operators, as defined in Sections \ref{sec:logio-normal-protocol} and \ref{sec:logio-recovery}. 


\paragraph{LOG.io Interface Methods} Table \ref{tab:logio-api} gives a summary of the LOG.io API exposed to the developers of operators. 

\begin{table}[ht]
\setlength\tabcolsep{1pt}
\caption{LOG.io API interface methods}
\label{tab:logio-api}
\begin{tabular}{c|c|l}
\hline
\rowcolor[HTML]{C0C0C0} 
\textbf{Method} & \textbf{Arguments} & \multicolumn{1}{c}{\cellcolor[HTML]{C0C0C0}\textbf{Description}} \\ \hline

GetActionID & actionInit & \begin{tabular}[c]{@{}l@{}}Returns a new identifier\\ for an atomic action\end{tabular} \\ \hline

\rowcolor[HTML]{EFEFEF} 
GetStateID & procInfo & \begin{tabular}[c]{@{}l@{}}Returns a new identifier\\ for a global state\end{tabular} \\ \hline

\rowcolor[HTML]{FFFFFF} 
BeginTransaction & None & \begin{tabular}[c]{@{}l@{}}Expose the transaction\\ interface\end{tabular} \\ \hline

\rowcolor[HTML]{EFEFEF} 
InitializeReadAction & \begin{tabular}[c]{@{}c@{}}actionInfo,\\ stateInfo, \\ state \end{tabular} & \begin{tabular}[c]{@{}l@{}}add "incomplete" entry in \\ READ\_ACTION and \\  log state in STATE \end{tabular} \\ \hline


CompleteReadAction & \begin{tabular}[c]{@{}c@{}}actionInfo,\\ actionData\end{tabular} & \begin{tabular}[c]{@{}l@{}}Mark a read action as \\ "complete" and log \\ a read action event \end{tabular} \\ \hline

\rowcolor[HTML]{EFEFEF} 
DropReadAction & \begin{tabular}[c]{@{}c@{}}actionInfo \end{tabular} & \begin{tabular}[c]{@{}l@{}}Discard store effect for \\ a read action and delete\\  event data\end{tabular} \\ \hline

LogStateEvent & \begin{tabular}[c]{@{}c@{}} stateInfo, \\ inSetID \end{tabular} & 
\begin{tabular}[c]{@{}l@{}}Log an "undone" event \\ for a state \end{tabular}\\ \hline

\rowcolor[HTML]{EFEFEF} 
UpdateContext & \begin{tabular}[c]{@{}c@{}} eventInfo \end{tabular} & 
\begin{tabular}[c]{@{}l@{}}Update the LOG.io context \\
for an event 
\end{tabular}\\ \hline

GetWriteActions & procInfo & \begin{tabular}[c]{@{}l@{}}Get "undone" write \\ actions \end{tabular} \\ \hline

\rowcolor[HTML]{EFEFEF} 
CheckEvent & eventInfo & \begin{tabular}[c]{@{}l@{}}Check that input event \\ is not obsolete \end{tabular} \\ \hline

\rowcolor[HTML]{FFFFFF}
AssignInSets & \begin{tabular}[c]{@{}c@{}}inSetIDs,\\ eventInfo\end{tabular} & \begin{tabular}[c]{@{}l@{}}Assign InSet\_IDs to input\\  events in EVENT\_LOG\end{tabular} \\ \hline

\end{tabular}
\end{table}

The method \texttt{GetActionID} takes an \texttt{actionInit} argument that contains an action name, a connection\_ID and an operator\_ID and returns a new action\_ID. 
The method \texttt{GetStateID} takes a \texttt{procInfo} argument that identifies an operator\_ID and returns a new state\_ID. 

Method \texttt{BeginTransaction} exposes a transaction interface that will be explained later. 

Method \texttt{InitializeReadAction} creates a new "incomplete" entry in READ\_ACTION for a read action identified by argument \texttt{actionInfo} that contains an action\_ID, a connection\_ID to the external system, an operator\_ID, and a description of the action, i.e., what the action does. It also logs in STATE the global state used to create the read action. 

The method \texttt{CompleteReadAction} marks a read action as "complete" in READ\_ACTION and logs an "undone" event for the read action  in EVENT\_LOG and EVENT\_DATA. It takes as arguments the previously defined \texttt{actionInfo} and an \texttt{actionData} which references the effect of the read action.

The \texttt{LogStateEvent} method logs an "undone" event associated with a state in EVENT\_LOG. It takes as argument a \texttt{stateInfo} to identify the state and an \texttt{inSetID} argument that represents the Input Set ID to which the event is assigned.

The \texttt{UpdateContext} method updates the LOG.io context that is logged in STATE and contains the array of latest event IDs used to update the global state. 


Method \texttt{DropReadAction} drops a read action $r$ by discarding the store for the effect of $r$ and deleting the read action event from EVENT\_DATA (as in Step 2.d of Algorithm \ref{alg:normal-source}). The implementation of this method is specific to the type of (non-replayable) read action.  

Method \texttt{GetWriteActions} returns all output events in EVENT\_LOG for an operator \texttt{procInfo} that are "undone" and have a null value for the sender port\_ID. 

Method \texttt{CheckEvent} checks that an input event is not obsolete as required by Step 1 of the State Update phase. 

Method \texttt{AssignInSets} takes an \texttt{eventInfo} and a \texttt{InSets} as arguments and updates EVENT\_LOG accordingly. The method uses several update statements when several InSet\_IDs are assigned to the same event, as explained in Section \ref{sec:normal-state-update}.

\paragraph{LOG.io Transaction Interface}
Table \ref{tab:logio-transaction} summarizes the methods associated with the LOG.io transaction interface. 

\begin{table}[ht]
\setlength\tabcolsep{1pt}
\caption{LOG.io transaction interface methods and a summary of their functionality.}
\label{tab:logio-transaction}
\begin{tabular}{c|c|l}
\hline
\rowcolor[HTML]{C0C0C0} 
\textbf{Method} & \textbf{Arguments} & \multicolumn{1}{c}{\cellcolor[HTML]{C0C0C0}\textbf{Description}} \\ \hline

Commit & None & \begin{tabular}[c]{@{}l@{}}Commit a transaction \end{tabular} \\ \hline

\rowcolor[HTML]{EFEFEF} 
LogSourceEvent & \begin{tabular}[c]{@{}c@{}}eventInfo,\\ eventData \end{tabular} & 
\begin{tabular}[c]{@{}l@{}}Log "undone" output event \end{tabular}\\ \hline

LogOutputEvents & \begin{tabular}[c]{@{}c@{}}eventInfo, \\ eventData, \\ inEvent\end{tabular} & \begin{tabular}[c]{@{}l@{}}Log "undone" output events \\  and mark input events \\ as "done" \end{tabular} \\ \hline

\rowcolor[HTML]{EFEFEF} 
DoneEvent & \begin{tabular}[c]{@{}c@{}} eventInfo \end{tabular} & Mark an event as "done" \\ \hline

StoreState & \begin{tabular}[c]{@{}c@{}} stateInfo, \\ state \end{tabular} & 
\begin{tabular}[c]{@{}l@{}} Store a serialized state  \end{tabular}\\ \hline

\end{tabular}
\end{table}

Method \texttt{Commit} commits the transaction created by the \texttt{BeginTransaction} method.

The \texttt{LogSourceEvent} method is used to log "undone" output events in EVENT\_LOG and EVENT\_DATA. It takes as input an \texttt{eventInfo} and \texttt{eventData} argument that respectively represent the events to be logged and their data. 

The \texttt{DoneEvent} method marks events in EVENT\_LOG identified by an 
\texttt{eventInfo} argument as "done". 

The \texttt{LogOutputEvents} method logs the output events of an Output Set and marks the events of its associated Input Set as "done". It takes as arguments 
an \texttt{eventInfo}, an \texttt{eventData} representing the events to be logged and their data, and an \texttt{inEvents} representing the events of the Input Set that must be marked as "done".

Method \texttt{DoneEvent} stores the serialized state for a state ID. It performs an upsert into the STATE table. 


\paragraph{LOG.io interface methods for recovery}
Table \ref{tab:logio-api-recovery} lists some methods used for LOG.io recovery.

Method \texttt{FetchAckEvents} returns the "undone" events with an assigned InSet\_ID (non null) from the EVENT\_LOG table for the given \texttt{procInfo}.

Method \texttt{FetchResendEvents} returns the undone events not assigned to an Input Set from the EVENT\_LOG table for the given \texttt{procInfo}. Returned events have a port name, a header and a body.

Finally, method \texttt{GetProcState} retrieves the latest (serialized) global state for a \texttt{procInfo} argument.  

\begin{table}[ht]
\setlength\tabcolsep{1pt}
\caption{LOG.io API interface methods for recovery}
\label{tab:logio-api-recovery}
\begin{tabular}{c|c|l}
\hline
\rowcolor[HTML]{C0C0C0} 
\textbf{Method} & \textbf{Arguments} & \multicolumn{1}{c}{\cellcolor[HTML]{C0C0C0}\textbf{Description}} \\ \hline

FetchAckEvents & procInfo & \begin{tabular}[c]{@{}l@{}}Fetch undone events assigned\\ to an InSet\_ID\end{tabular} \\ \hline

\rowcolor[HTML]{EFEFEF}
FetchResendEvent & procInfo & \begin{tabular}[c]{@{}l@{}}Fetch undone events without \\any assigned InSet\_ID\end{tabular} \\ \hline

GetProcState & procInfo & \begin{tabular}[c]{@{}l@{}}Returns latest serialized \\ global state\end{tabular} \\ \hline
\end{tabular}
\end{table}

\subsection{Implementation of operators using LOG.io API} 
\label{sec:4.OperatorImpl}

In this section, we illustrate the implementation of generic operators using LOG.io API. 

\subsubsection{Source Operator}

We give in Listing \ref{lst:sourceOPEx3} an example of a generic Source operator that performs a single non-replayable read action and uses a global state. We assume that there is a single output port. For better readability, comments refer to the numbered steps of the Algorithm \ref{alg:normal-source}. 

\begin{lstlisting}[style=customGO,language=go,caption={Source operator with a a non-replayable read action},label={lst:sourceOPEx3}]
sourceOperator(){
    // Steps 1 and 2.a: execute read action with global 
    // state, store the effect if not replayable read
    readActionID = logio.GetActionID(procInfo)
    stateID = logio.GetStateID(procInfo)
    state = serializeState()
    logio.InitializeReadAction(readActionInfo, stateID, 
        state)
    readActionResult = storeReadAction()
    // Step 2.b: not replayable read
    logio.CompleteReadAction(readActionInfo, 
        readActionResult)
    // Step 2.c and 2.d: Iterate over readActionResult  
    loop over readActionResult{
        // Step 2.c Generate an output event
        outputEvent = generateOutputEvent()
        // Log output event and global state
        state = serializeState()
        transaction = logio.BeginTranaction
        transaction.LogSourceEvent(eventInfo, eventData)
        transaction.StoreState(stateInfo, state)
        // if last event, mark read action event as "done"
        if lastEvent then transaction.DoneEvent(eventInfo)
        err = transaction.Commit()
        // send output event
        sendOutputEvents(port, outputEvent)
        // Step 2.d Discard store
        logio.DropReadAction(readActionID)
        }
}
\end{lstlisting}

In lines 4-5, LOG.io creates new action\_ID and state\_ID values from the last values obtained from the READ\_ACTION and STATE tables during the initialization of the operator. Lines 6 and 9 use custom functions. In line 9, method \texttt{storeReadAction} executes the read action and returns a reference to the stored effect of the read action. In line 11, an atomic transaction is started using method \texttt{CompleteReadAction}: the read action is marked as "complete" and is logged as an "undone" event in EVENT\_LOG with the data of the event logged into EVENT\_DATA being the reference to the stored effect of the read action. Line 14 starts the iteration over the effect of the read action (Step 2.c and 2.d in Algorithm \ref{alg:normal-source}). In line 16, the custom function \texttt{generateOutputEvent} is used to generate some output event (a batch of records). Then in lines 19-24, an atomic transaction is started, in which method \texttt{LogSourceEvent} logs the output event in EVENT\_LOG and EVENT\_DATA, method \texttt{StoreState} stores in STATE the global state used to generate the output events, and if the last output event has been generated, \texttt{DoneEvent} marks as "done" the event for the read action. Output events are sent to the output port using the custom function  \texttt{sendOutputEvents}.

\subsubsection{Middle Operator}

We give in Listing \ref{lst:middleOPEx} an example of a generic Middle Writer operator that receives input events on an input port, sends output events to an output port, and generates write actions that are executed on a single external system. We assume that a single function is used to generate output events and that the data lineage is disabled. We also assume that a global state is used. As before, we highlight only the usage of the LOG.io API. 

\begin{lstlisting}[style=customGO,language=go,caption={Generic Middle Writer operator},label={lst:middleOPEx}]
MiddleWriterOperator(){
onInput(event)
// State Update phase 
  // Step 1. Read and filter event from input port
  eventInfo = getEventHeader(event)
  if logio.CheckEvent(eventInfo) 
     {eventData = getStreamEvent(event)}
  // Step 2. State update
  // next function calls logio.UpdateContext(eventInfo)
  updateGlobalState(eventData)
  // next function calls logio.GetInSetID()
  inSetID[] = updateEventState(eventData)
  logio.AssignInSets(inSetID[], eventInfo)
  // Step 3. Triggering
  trigInSetID = triggeredInSet()
// Generate Phase 
  // Step 2. Save global state
  stateID = logio.GetStateID()
  state = serializeState()
  logio.LogStateEvent(stateInfo, trigInSetID)
  // Step 3. Generate output set
  // next function calls logio.GetEventID()
  outputSet = generateOutputSet()  
  outEventInfo = getEvents(outputSet)
  outEventData = getData(outputSet)
  // Step 4. Update logs
  transaction = logio.BeginTransaction
  transaction.LogOutputEvents(outEventInfo, outEventData, 
      trigInSetID)
  transaction.StoreState(stateID, state)
  transaction.Commit
  // Step 5. Send events to the output port
  sendOutputEvents(portName, outputSet)
  // Step 6. Process write actions
  writeActions = logio.GetWriteActions(procInfo)
  // Iterate over "undone" write actions
  loop over writeActions{
      // get and execute next "undone" write action
      response = executeWriteAction(writeAction)
      if response {logio.DoneEvent(eventInfo)}
  }
}
\end{lstlisting}

In lines 4-5, an input event is read on an input port and checked before extracting its body. Then in lines 9 and 12, two custom functions are used to, respectively, update the global and event state. The InSet\_IDs assigned to the input event within method \texttt{updateEventState} are registered using the LOG.io \texttt{AssignInSets} method. In line 15, the (single) Input Set that is used to trigger the generation of an Output Set is assigned to \texttt{trigInSetID}. If more than one Input Set matches the triggering conditions, the block of code that follows, at least until line 34, is repeated. In lines 21-23, output events are generated using a custom function \texttt{generateOutputEvents} and the event info and event data are prepared.  Next, in lines 27-31, an atomic transaction is started to log the output events in EVENT\_LOG and EVENT\_DATA and mark the events of their Input Set as "done" (through \texttt{}{LogOutputEvents}) and update the global state identified by its \texttt{stateInfo} through \texttt{StoreState}. In line 33, output events are sent on an output port using a custom function. In line 35-40, "undone" write actions are obtained and executed. Custom function \texttt{executeWriteAction} is specific to the external system and must manage the cases of failure of the write action as described in Step 2 of Algorithm \ref{alg:generate-writer}.   
Note that a generic Writer operator would execute all steps in Listing \ref{lst:middleOPEx} except line 33 since the only output events generated by a Writer operator are write actions.


\subsection{Implementation of recovery} \label{sec:4.RecoveryImplementation}

When an operator in a group (pod) fails,  
the Deployment Engine recreates the failed pod, restores connections to the messaging system, and operators inside the pod reconnect with the other pods and the HANA database containing LOG.io logs. 
Immediately after, the engine calls the \texttt{onRecovery} function that must be provided by each operator to execute the specific recovery actions of our protocol described in Section \ref{sec:logio-recovery}. 
A generic example of the \texttt{onRecovery} function for a Middle Writer operator is given in Listing \ref{lst:onRecovery}. Comments indicate the processing steps of Algorithm \ref{alg:recover-middle-output}, \ref{alg:recover-writer}, and \ref{alg:recover-processing}. 

\begin{lstlisting}[style=customGO,language=go,caption={Startup recovery implementation},label={lst:onRecovery}]
onRecovery() {
// Recover output events
 // Step 1. Resend events
 events = logio.FetchResendEvents(procInfo)
 loop over events {
    // send next event
    sendOutputEvents(portName, event)
 }
 // Step 2. Write actions
 if (proc.isWriter()) {
    writeActions = logio.GetWriteActions()
    loop over writeActions{
      // get and execute next "undone" write action
      response = executeWriteAction(writeAction)
      if response then logio.DoneEvent(eventInfo)
    }}
// Recover processing
 // Step 1. Recover latest global state from STATE
 rawState = logio.GetProcState(procInfo)
 deserializeState(rawState)
 // get LOG.io context from STATE
 logio.InitializeContext(procInfo)
 // Step 2. Recover input events
 events = logio.FetchAckEvents(procInfo)
 loop over events {
   proc.onInput(event)
 }
}
\end{lstlisting}

The only new custom function in the Listing \ref{lst:onRecovery} is \texttt{deserializeState}. The other custom functions are also used during normal processing. 

\section{Qualitative analysis of LOG.io}
\label{sec:logio-advantages}



In this section, we discuss the benefits of the LOG.io protocol along three main dimensions: non-blocking recovery (Section \ref{sec:non-blocking}), dynamic scalability (Section \ref{sec:dynamic-scaling}) and data lineage capture (Section \ref{sec:data-lineage}).

\subsection{Non-blocking recovery}
\label{sec:non-blocking}

When an operator fails, its embedding group of operators is automatically restarted,  while all other groups of the data pipeline keep running. In this sense, LOG.io is said to be \emph{non-blocking}, unlike other protocols like ABS that require restarting the entire data pipeline from its last epoch. We discuss the impact of this property. 

\paragraph{Latency of restart} Restarting a group still introduces some latency in the execution of a data pipeline. 
Firstly, the operators that precede a restarted operator OP may be blocked due to the backpressure mechanism if they run faster than OP. 
Secondly, operators that follow a restarted operator OP may become idle because they are waiting for new input events originating from OP if they run faster than OP.
Thus, there is a higher latency when a comparatively slow operator fails and it is surrounded by fast operators. However, only the operators located on the paths that contain a restarted operator are affected by the latency caused by recovery. In Section \ref{sec:experiment-results}, we report experiments (Use Cases 1 and 2) that evaluate the overhead of recovering an operator in the situations described above. 

Anyway, it is essential to keep the restart time of a failed group (i.e., a failed pod) as small as possible. To this aim, we implement a warm restart of a pod by keeping a cache of the image that must be deployed. Once the group is restarted, the recovery actions must be performed by the embedded operators, which entails database read operations from LOG.io's log tables and the de-serialization of the global state and the pending events to recover. The size of the global state is usually quite small (it does not contain event data). The number of output events to recover is also small, since it concerns events that could not be sent successfully, or events that were not acknowledged while the group restarted. Finally, the number of input events to recover is bound by the size of the connection buffer.  Thus, the dominant cost factor is the restart of the failed pod. 

\paragraph{Leveraging data parallelization} A data parallelization strategy is used to improve the total processing time of a data pipeline \cite{Roger-Mayer:2019} when a portion of the data pipeline creates a bottleneck during event processing. To achieve this, an overloaded operator OP (or a group of operators) can be replicated, and each replica is used to process a different part of the events sent to OP. With this approach, a "Dispatcher" operator dispatches the events sent to OP to each replica of OP, while a "Merger" operator receives results
from each replica and bundles them into one output stream of events. In our implementation, each replica is deployed as a separate Kubernetes pod. 
The processing logic of operators must fulfill specific conditions to be parallelized \cite{Roger-Mayer:2019}. 
When a replica of an operator OP fails, it is restarted while the other replicas keep processing their events. However, the analysis of the latency created by the recovery of a replica distinguishes two cases. 

When OP is \emph{stateless} and the processing logic of OP is insensitive to the order in which events are sent to its input ports, the Dispatcher can implement a simple round-robin with load-balancing strategy. When a replica fails, the Dispatcher continues to send events to the failed replica until its connection is blocked due to back pressure. Afterwards, all events are sent to the alive replicas until the connection to the failed replica is unblocked. Hence, the failure of one replica does not block the overall processing, and LOG.io takes advantage of data parallelization to limit the latency caused by recovery.
This is not the case with checkpoint-based rollback recovery protocols, such as \cite{CARBONE:ARXIV:15, CARBONE:VLDB:17}, which require stopping and restarting the processing of a data pipeline in case of failure. 

When OP is \emph{stateful}, the Dispatcher must ensure that the events sent to a replica respect the constraints imposed by the processing logic of OP (e.g., key-based, window-based). For instance, if OP builds a key-based event state, all events with the same key value must be sent to the same replica. In such a case, when a replica fails, only the events whose key value is not assigned to the failed replica can be processed by the remaining replica. This entails a higher latency during recovery than with stateless operators. However, LOG.io still takes advantage of data parallelization to keep the data pipeline running.  

\subsection{Dynamic scalability}
\label{sec:dynamic-scaling}

When data pipelines are long-running, the context in which they execute can change over time. First, the rate at which source data is ingested in a data pipeline can fluctuate over time, including periods of idleness. Secondly, the distribution of key values within the data can evolve and become skewed, with a high proportion of data having a few distinct key values during a period of time. 
Therefore, the use of data parallelization should be dynamically adjusted to the execution context because underloaded replicas mobilize useless and costly computing resources, while overloaded operators create execution bottlenecks and increase the total processing time. This is particularly true in cloud computing platforms that use a pay-as-you-go business model in which, for instance, unused deployed pods are charged because they reserve computing resources. With such a pricing model, underused pods should be deleted.   

What is needed is a vertical elastic data parallelization strategy, whereby the number of replicas for an operator, or a group of operators, can dynamically scale up (by adding a new replica) or scale down (by removing a replica), depending on the load of replicas. 
The main problem of dynamic scaling with respect to rollback recovery occurs with stateful operators because the Dispatcher operator that is responsible for sending input events to replicas is also stateful. When a decision to scale up is made, a new replica is added (hence, the topology of the pipeline is modified), and the internal state of the dispatcher is updated to start sending events to this replica. For instance, a key-based Dispatcher will start sending data with a key value to the new replica. If a system failure occurs,  we must recover the data pipeline in such a way that the state of the Dispatcher is \textit{consistent} with the topology of the pipeline. 

Similarly, when a replica is removed, the state of the Dispatcher must be updated to stop sending events to this replica and redirect them to other replicas. In addition, the replica to be removed must be physically deleted only when all the events that it received have been processed or safely recovered and sent to another replica. Otherwise, some events will never be processed, which violates the correctness condition of rollback recovery. 

In the following, we show how LOG.io enables the dynamic scaling of operators to scale-down operators to zero during a period when a long-running data pipeline is idle. We assume that the monitoring of the load of operators (e.g., traffic on connections between groups) is done by a central process, henceforth called a Controller, which is also responsible for the decision to add or remove a replica. We do not discuss the strategies used by the Controller, see for instance \cite{Kalavri:2018}, but rather focus on rollback recovery. 


\begin{algorithm}[tb]
\caption{Scaling up the replicas of OP}
\label{alg:scale-up}
\begin{tabular*}{\linewidth}{rp{0.75\linewidth}}
\textbf{input:} & Image for replica OP;  \\
& Merger MP, Dispatcher DP \\
\textbf{output:} & update pipeline topology;  \\ 
& update LOG.io STATE  \\
\end{tabular*}
\begin{enumerate}

        \item \textit{Update topology:} Deploy a new image for OP using warm start; create the two connections between the Dispatcher (resp. Merger) and the new replica
        
        \item \textit{Merger update:} request MP to update its state

        \item \textit{Dispatcher update:} request DP to update its state
\end{enumerate}
\end{algorithm}

\paragraph{LOG.io dynamic scale up} When the Controller decides to scale up an operator OP, it must follow the processing steps described in Algorithm \ref{alg:scale-up}. A Dispatcher (or Merger) acknowledges a state update request from the Controller only after successfully storing its new state in the STATE log. If a failure occurs after Step 1 or Step 2, the Controller restarts at that step and there is no impact on the execution of the pipeline. Once Step 3 is completed, the scale-up decision is effective. After that, if a failure occurs in the Dispatcher or Merger, the LOG.io recovery protocol presented in Section \ref{sec:logio-recovery} can safely recover a consistent state. Indeed, upon recovery, the global states of the Dispatcher and the Merger are restored from the STATE log. 

\paragraph{LOG.io dynamic scale down} When the Controller decides to scale down an operator OP it must follow the processing steps described in Algorithm \ref{alg:scale-down}. In Step 1.a, the state of the Dispatcher DP is updated and used in Step 1.b to re-assign all events that were sent to OP but are still "undone", which results in a set $O$. In Step 1.c, the atomic transaction may run concurrently with the transaction in OP that executes Step 4 of the Generation phase described earlier in Algorithm \ref{alg:generate-general}. There, OP's transaction must log output events and mark as "done" the events of the Input Set that are used to produce these output events. If the Dispatcher's transaction takes place first, then OP's transaction will fail because the events of the Input Set do not exist anymore and cannot be marked as "done". Hence, output events cannot be created and sent by OP. On the opposite, if OP's transaction executes first, some events of $O$ will be outdated because they have been marked as "done" by OP's transaction. In Step 1.d, only the events in $O$ that remain "undone" are sent to their new destination port. The mutual exclusion between the two atomic transactions of DP and OP enables LOG.io to precisely synchronize the removal of a replica and the reallocation of its events to other replicas without losing any event. 

After that, in Step 2, the Controller can request Merger MP to update its state with the deletion of OP. Finally, the topology of the data pipeline is updated in Step 3.   

If the Dispatcher fails before completing Step 1, the Controller will send the request again after the Dispatcher recovers. If the failure occurs before Step 1.c, Steps a and b are processed again. If the failure occurs after Step 1.c, the recovered state of the Dispatcher already includes the intention to remove replica OP, so Step 1.a is done. Step 1.b will return an empty set and Steps 1.c and 1.d will be ineffective. The output events that were updated during Step 1.c, before the failure occurred, will be sent during the normal recovery of the output events of the Dispatcher (using Algorithm \ref{alg:recover-middle-output}). 

\begin{algorithm}[tb]
\caption{Scaling down the replicas of OP}
\label{alg:scale-down}
\begin{tabular*}{\linewidth}{rp{0.75\linewidth}}
\textbf{input:} & replica OP to delete;  \\
& Merger MP, Dispatcher DP \\
\textbf{output:} & update pipeline topology; \\ 
& update LOG.io logs \\
\end{tabular*}
\begin{enumerate}

        \item \textit{Dispatcher preparation:} Request DP to execute:
            \begin{itemize} 
             \item[a.] update state of DP with the deletion of OP; 
             \item[b.] generate a set $O$ containing all "undone" events in EVENT\_LOG that were sent to OP with their new assignment (i.e., destination port ID, event\_ID) 
             \item[c.] use a transaction to (1) update all events of $O$ in EVENT\_DATA and EVENT\_LOG and (2) store the state of DP in STATE
            \item[d.] send to their destination port all events of $O$ that are still "undone" in EVENT\_LOG
        \end{itemize}                
    
        \item \textit{Merger update:} request MP to update its state

        \item \textit{Update topology:} Delete the two connections between the dispatcher (resp. merger) and OP; Delete replica OP 

\end{enumerate}
\end{algorithm}

\subsection{Fine-grain data lineage capture}
\label{sec:data-lineage}

As explained before, given a set of input ports IN and a set of output ports OUT of an operator on which data lineage capture is enabled, LOG.io stores in the EVENT\_LINEAGE table the relationships between any output event sent on a port in OUT and the Input Set (InSet\_ID) used to produce it. Data lineage relationships between events are then obtained by joining the EVENT\_LINEAGE and EVENT\_LOG tables on (receiving) operator ID, port ID and InSet\_ID, and filtering on the port IDs listed in the set IN. LOG.io guarantees that there is no data lineage relationship $(a, b)$ if $a$ does not provide any contribution to some record of $b$. 

As explained in Section \ref{sec:intro-data-lineage}, this is not the case with the data lineage information captured by streaming systems such as Spark Streaming, StreamS, ChronoStream, and lineage stashing. To illustrate this point, consider the data pipeline in Figure \ref{fig:Example4} and the logs portrayed in Tables \ref{tab:event-log-2} and \ref{tab:event-log-3}. The streaming systems we mentioned would maintain the data lineage between input events $a_2$, $a_3$ and output events $b_2$ because $a_2$, $a_3$ can be used to reconstruct the event state of OP that is used to generate $b_2$. However, no record of $a_3$ is used to produce $b_2$, which is correctly captured by LOG.io. 
Furthermore, unlike the previous dedicated data lineage methods of \cite{glavic:2014, genealog:2018, ananke:2020} designed for dataflow processing systems, LOG.io addresses the general case of custom, and possibly non-deterministic, operators.

The granularity of data lineage captured by LOG.io is defined by the size of the events that flow through a data pipeline. For Source operators, the size of output events (data batches) is a configuration parameter. For Middle operators, the size of an Output Set is determined by the function that generates an Output Set, the size of the event state, and the triggering condition. As explained in Section \ref{sec:generate-general}, each operator decides how data batching is done, that is, how many output events (data batches) are created from the Output Set. Thus, unlike previous dedicated data lineage capture methods defined for dataflow of relational database queries \cite{Titian:2018, Glavic-ICDE:2009, Glavic-ICDE:2017}, the granularity of data lineage in LOG.io is not at the record level. 

A finer grain data lineage, at the record level, can however be obtained by exploiting schema-based data lineage properties, such as dependencies between the output and input event schemas of an operator, as defined, for example, in \cite{Ikeda:2013}. Schema-based data lineage properties can either be automatically extracted from built-in operators (this is often called column-level data lineage) or explicitly provided by the developer of custom operators. 

\begin{example}
Consider again the data pipeline in Figure \ref{fig:Example4} and
the logs portrayed in Tables \ref{tab:event-log-2} and \ref{tab:event-log-3}. 
For a given record $r$ = [2025-02-23T10:00:00+03:00, BZ234, Ohio, 1,429] of $b_1$, we do not know which records of $a_1$ or $a_2$ were used to produce it. Now, suppose that some attribute dependencies \cite{Ikeda:2013} are defined between the input schema $S_{in}=[store\_ID, item\_ID, city, state, date, amount]$ of input events and the schema $S_{out}=[hour, item\_ID, state, total]$ of output events of OP3. An attribute dependency has the general form $S_{in}.A \leftrightarrow S_{out}.B$, where $S_{in}$ and $S_{out}$ are the schemas of the input and output events of an operator OP, and $A$ and $B$ are attributes. It means that whenever $I$ is an instance of $S_{in}$ that resulted through OP into an instance $O$ of $S_{out}$, then the subset of $O$ where $B=x$ is unaffected by all elements in $I$ except those where $A=x$. Using our example, suppose to have the following dependencies: $S_{in}.item\_ID \leftrightarrow S_{out}.item\_ID$ and $S_{in}.state \leftrightarrow S_{out}.state$. Then, they indicate that the records $p$ in $a_1$ and $a_2$ that contributed to the generation of the record $r$ are such that: $p.item\_ID = 'BZ234'$ and $p.state='Ohio'$. 
This refines the data lineage although there could still be a record in $a_2$ that has a date such as "2025-02-23T11:05:00+03:00" (i.e., 11:05 am) and did not contribute to the window associated with hour 10.00 am from which event $b_1$ was generated.     
\end{example}

\section{Related Work} \label{sec:related_work}

In this Section, we review previous work done on rollback recovery for distributed data pipelines. 
Because they can be logically considered as an instance of message-passing systems, we follow the classification of rollback recovery methods used in \cite{ELNOZAHY:02} that distinguishes between checkpoint-based (Sections \ref{sec:coordinated-checkpoint-based} and \ref{sec:log-based}) and log-based (Section \ref{sec:log-based}) rollback recovery protocols. 



\subsection{Checkpoint-based rollback recovery}
\label{sec:coordinated-checkpoint-based}

Checkpoint-based methods save the state of the processes (also called checkpoints) during the normal execution of a data pipeline, and when a failure occurs, restore the most recent consistent set of checkpoints to resume the execution of the data pipeline. 

\subsubsection{Apache Flink}
The most popular checkpoint-based method is the coordinated checkpoint-based method, called Asynchronous Barrier Snapshot (ABS) \cite{CARBONE:ARXIV:15, CARBONE:VLDB:17}, implemented in Flink \cite{FLINK:24}, which is an evolution of the previous work of \cite{LAMPORT:ACM:85} on “distributed snapshots”.  Variants of the ABS protocol have been implemented in systems such as Spark Structured Streaming \cite{SPARK:structured-streaming:23}, Storm \cite{STORM:Checkpointing:2024}, and SAP Data Intelligence \cite{SAP:DI:24}.

The programming model of Flink is conceptually similar to LOG.io, with the added possibility of having cycles in a data pipeline. At the execution level, each operator is mapped to a number of physical tasks, each of which being deployed to available containers throughout a cluster up to the desired degree of parallelism decided at deployment time. The data streams are then partitioned according to the configuration of the physical tasks. 

During normal execution \cite{CARBONE:VLDB:17}, 
special events, called "markers", are injected into each data source operator to divide the data stream into logical periods of finite records denominated \textit{epochs}. Stream operators with multiple inputs execute an alignment phase, where once the operator receives a new snapshot marker, it blocks the channel from where the new marker has been received and waits for the same marker to be received from all its input channels before broadcasting the marker to its output channels and proceeding with the writing of a snapshot of its state on a reliable storage. Once the snapshot is initiated, the input channels are unblocked, and the operator proceeds with its regular execution to the next epoch. An epoch is said to be \emph{complete} once all the sink operators have received the snapshot marker and executed their relative snapshots. The global state of a data pipeline consisting of all state snapshots for a complete epoch is always consistent. 

To handle the consistency of the state of external systems which are not MVCC-enabled (Multi-Version Concurrency Control) databases, Apache Flink uses a two-phase commit protocol. Every time a Writer operator makes a snapshot, the changes to the external state (i.e., the write actions) are inserted into a transaction log (Write-Ahead Log) and are pre-committed. Once a global epoch is reached, all the pre-committed changes are fully committed using an atomic transaction. 

When a failure occurs, the data pipeline is simply restarted from the last complete epoch stored in the snapshot storage. To achieve end-to-end exactly-once correctness, ABS requires that the data sources be replayable and that all operators that update an external state are idempotent or use atomic transactions in the case of multiple write actions. In addition, as in LOG.io, write actions are assumed to be durable and checkable. However, LOG.io does not require a Writer operator to be transactional since it manages the correct recovery of write actions. 

The main advantage of ABS is its low overhead during normal execution because it uses an asynchronous snapshot mechanism, allowing the system to save snapshots while continuing the regular processing of input events. Several factors still influence the overhead. One is the size of the epoch, that is, the period of time that separates two consecutive markers and the size of the local state of stateful operators that must be checkpointed. 
When an operator has multiple inputs, another factor is the back pressure of connections due to the alignment phase explained earlier. Finally, there can be a latency on the external state due to the two-phase commit protocol: the update of any external state is delayed until a complete epoch is reached. 

The overhead of rollback recovery in case of failure is mainly influenced by the processing time of operators, i.e., the time it takes an operator to generate a set of output events from its local state and the length of a data pipeline. This is because the time taken by a marker to cross the pipeline can be significant, which directly impacts the time overhead when a failure occurs. Our measurements in Section \ref{sec:experiments} show this impact. Note that very recently, \cite{TAKDIR:abs-optimization:25} has proposed to perform specialized partial snapshots within global snapshots to recover faster during failures, which shows that this is still an active area of research.

The recovery overhead is aggravated when the frequency with which data are ingested varies greatly over time for two reasons. 
First, with an average size of an epoch, when the frequency of events increases, the time needed to reach a complete epoch also increases because more events occur between two consecutive markers. 
Second, Apache Flink supports limited dynamic scaling (up or down) of operators during the execution of a data pipeline and cannot timely react to avoid the negative effect of a back pressure. Indeed, Flink's Adaptive Scheduler tries to use the maximum parallelism value specified for each operator or an entire job. However, to scale up the number of processes used to execute an operator, it is necessary to take a savepoint (i.e., a consistent snapshot of the pipeline), stop the computation, increase the maximum parallelism value for that operator, and redeploy the data pipeline. This rescaling operation can be done automatically with the experimental "reactive mode", but as explained in \cite{FLINK:elastic:25}, its usage comes with some limitations.

\subsubsection{Other variants of ABS}
Spark Structured Streaming \cite{SPARK:structured-streaming:23} is built on top of the Spark SQL Engine.
Only the default mode, called \emph{micro-batching}, provides an end-to-end exactly-once guarantee. The differences with our (and Flink's) programming model are that the batching of source data is statically defined, operators run Spark SQL queries possibly including UDFs that are deterministic, and each operator fully processes its input batches before the next operator proceeds (aka synchronous stages).  

During the regular execution of a Spark job \cite{ARMBRUST:ACM:18}, as a source operator reads the data, the master node of the Spark application assigns epochs based on the input source offsets, which represent the beginning and end positions of each epoch. The master node stores these offset positions on a durable log. Any stateful operator executes periodic asynchronous checkpoints associated with its current epoch into a state store, using incremental checkpoints when possible. Stateful operations can occur across batches. The master node waits for all operators to report a commit for the current epoch before allowing commits for the next epoch. 

Upon recovery, when a new instance of the job is started, Spark searches the log for information regarding the last not-committed epoch. Spark reconstructs the application context by loading the old state information and reprocessing the epochs while turning off the outputs. Finally, it reprocesses the last epoch and relies on the sink's idempotency to write its results. 
The requirements to achieve correct rollback recovery are the same as Flink: data sources must be replayable, and sink operators must support idempotent writes.  



Apache Storm \cite{STORM:Checkpointing:2024} provides a checkpoint mechanism through which a special message flows through an internal stream throughout the data pipeline. In this mechanism, a special checkpoint source regularly emits a new checkpoint message. Upon receiving a checkpoint message, the operators prepare a transaction containing its current state, send an acknowledgment message back to the checkpoint source, and then forward it to the next operators. The checkpoint is completed when the checkpoint source receives an acknowledgment from all operators, allowing them to commit their transactions.

The recovery phase is triggered the first time the topology is started or when a failure is detected. During recovery, if all operators have not successfully prepared the previous transaction, the checkpoint source sends a rollback message, forcing the operators to abort their transactions and re-send any unprocessed batch of data. However, if all operators have successfully prepared the previous transaction but have not committed, the checkpoint source sends a commit message, allowing any prepared transaction to be committed. 




\subsubsection{Naiad}

The programming model of the Naiad system \cite{MURRAY:ACM:13} is a directed graph in which stateful operators send and receive logically time-stamped messages through connectors, which optionally have a partitioning function to control the exchange of messages between operators. 
At the execution level, a logical data pipeline is decomposed into a set of partitions depending on the degree of parallelism. Each partition contains an instance of the data pipeline that is managed by a worker running in a cluster process. Connectors with partitioning functions are translated in methods that route messages between workers. 

During a normal execution, each message produced by an external system for a source (input) operator is labeled with an integer epoch, which is part of the logical timestamp associated with every message flowing in the graph (another part is related to the possible loop contexts embedding a message). 
Each operator can asynchronously send and receive timestamped messages and may request and receive notifications that they have received all messages bearing a specific timestamp. An operator sends a notification for a timestamp $t$ only when it cannot receive any new message with a timestamp less than or equal to $t$ (a notification plays the role of a low watermark). The notification mechanism allows an operator to trigger work that depends on all the data of a given time, similar to a stateful operator.
The processes hosting the workers participate in the distributed progress tracking protocol, to coordinate the delivery of notifications. Before delivering a notification, a worker must know that there is no outstanding event at any worker in the system with a timestamped message that could result in the timestamp of the notification. 

Naiad uses a coordinated checkpointing, whereby each operator implements a checkpoint and restore interface. When the system performs a periodic checkpoint, all processes and message channels are paused. Then, the system flushes the message channels and invokes a checkpoint on each operator. Once the checkpoint is completed by all operators and the files are flushed to disk or replicated to other computers, the system resumes the worker and message delivery threads. Unlike Flink, both in-transit messages and operator states are checkpointed, and the triggering of a periodic checkpoint pauses the processing. The progress tracking protocol of Naiad makes any complete checkpoint a consistent checkpoint. 
When a process fails, all live processes revert to the last completed checkpoint and the operators of the failed process are reassigned to the remaining processes. The restore method reconstructs the state of each stateful operator using its respective checkpoint file. 




\subsection{Log-based rollback recovery}
\label{sec:log-based}

Log-based rollback recovery methods use both state checkpointing and event logging to enable processes to replay their execution after a failure. 
They rely on the assumption that they can identify all the non-deterministic events executed by each process, and for each such event, log all information necessary to replay the event in case of recovery \cite{ELNOZAHY:02}. Using this assumption, it is possible to recover a failed process and replay its execution as it occurred before the failure. Examples of such events include the order in which events from multiple streams are received by a process, interactions with external systems, and the generation of events using a non-deterministic function.

\subsubsection{MillWheel and Google Dataflow}

The programming model of MillWheel \cite{AKIDAU:VLDB:13} enables the definition of a directed graph of user-defined operators (called \emph{computations}) that can independently consume and produce data records. An operator subscribes to zero or more input streams and publishes one or more output streams. Each operator encapsulates an arbitrary user-defined code, invoked upon receipt of input data, that can update a local state, interact with external systems, and produce output data (called \emph{productions}). The management of low watermarks is not discussed here since it is orthogonal to the rollback-recovery strategy.
At the execution level, each operator runs on one or more machines, and data streams are delivered via RPC. Each output record in a stream has a unique system-assigned ID. Key extraction functions, which assign a key to an input record, must be specified by each operator for each input stream, and the computation code is run in the context of a single key and is only granted access to the operator state for that specific key. The system divides each operator into a set of key-based intervals and assigns these intervals to a set of machines to run them in parallel. All operators have access to a common distributed data store managed on a per-key basis. 

During normal processing, when an operator receives an input record, it checks, based on its record ID, that it is not a duplicate of a previously received record. Then, user code is run, possibly resulting in changes to timers, internal state, and output records. These changes and the record ID are checkpointed into the data store within a single atomic transaction. Then, the sender of the input record is ACKed (upstream backup), which enables to remove the corresponding logged records from the data store. A sender will send again the output records for which it did not receive an ACK after a timeout period. Finally, pending output records are sent to downstream operators. This pattern is called "strong productions". The "weak productions" pattern allows an optimistic checkpointing, whereby output records can be sent \emph{before} checkpointing and then ACKing senders. However, it requires that the computation done by the operator be deterministic so that if a failure occurs before an ACK was sent, the operator can replay the processing of its input records and send exactly the same output records as before.  

When a failure occurs, the system automatically restarts the failed virtual machine and each operator in it is recovered by scanning the data store for its last checkpointed key-based computation intervals, initializing its internal data structures, and resuming its processing by sending ACKs to upstream operators and sending output records to downstream operators. Duplicate elimination of input records in the receiver handles the case of records sent twice. Hence, only the failed operators are restarted without disrupting the execution of the other operators. This strategy allows to balance the load by moving, splitting, or merging computation intervals to different processing nodes. 
To achieve exactly-once semantics, data sources must be replayable. In addition, if the user code within an operator interacts with an external system, it is up to the user to ensure that the effect of the code on this system is idempotent. 


Google Cloud Dataflow \cite{AKIDAU:VLDB:15, DATAFLOW:Overview:2024} uses the same rollback-recovery strategy as MillWheel, except that each output record is tagged with a timestamp, instead of a sequence number, to leverage the processing time watermark capabilities of the system. The programming model of MillWheeel and LOG.io are comparable (non-deterministic operators, event-based triggering semantics, side-effects in intermediate operators). Like LOG.io, output events are logged before sending them (pessimistic logging) and upstream backup is an alternative to the way LOG.io acknowledges the input events received by an operator. However, unlike LOG.io, MillWheel also snapshots operator states. The recovery strategy for weak productions in the case of deterministic operators (with optimistic logging) resembles the replay strategy of LOG.io, although MillWheel relies on ACK timeouts to trigger a replay. Finally, MillWheel and Google Dataflow support dynamic scaling in a cluster but do not provide any built-in data lineage capture facility. 




\subsubsection{Apache Spark Streaming}


Spark Streaming programming model \cite{SPARK:streaming:24} revolves around discretized streams (D-streams) \cite{ZAHARIA:USENIX:12}, which are high-level abstractions representing a continuous stream of data as a sequence of immutable partitioned resilient distributed data sets (RDDs). An extension of the core Spark API is used to define streaming computations (afterwards, called operators) by applying deterministic \emph{transformations} and \emph{output operations} to these D-Streams. 
At the execution level, the system divides live input data streams into a sequence of batches (the D-stream) corresponding to small statically defined time intervals and stores them as RDDs. It then executes a streaming application by generating Spark jobs to process the batches as follows. 
A batch of data is partitioned and processed by executing the DAG of parallelized operators, which produces datasets stored as RDDs. Afterwards, the next batch of data is processed along with the result datasets of the previous batch to produce new datasets, and the same process is repeated. Thus, each time interval's output RDDs reflect all of the input received in that time interval and the previous time intervals. Within a batch job, an operator starts executing when all its upstream operators are finished. 

During normal processing, input data from network sources is replicated to two worker nodes for fault-tolerance by default. Write-ahead logs (WALs) can also reliably save all received data into a fault-tolerant log in the checkpoint directory. While operators are executed in a batch job, the graph of deterministic operations used to build each RDD is tracked in a \emph{lineage graph} at the level of partitions \cite{RDD:2012}. The system can also periodically and asynchronously checkpoint the RDDs produced by an operator, in which case the lineage before the checkpoint can be forgotten. 

When a worker node failure occurs, the lost RDD partitions are re-computed by the operators that built them from their lineage using the original fault-tolerant dataset and the most recent checkpoints. The recomputation is correct because operators are deterministic. Only the failed node and its upstream operators are involved in the recovery which can be executed in parallel within a job using RDD partitions and between jobs for independent operations (e.g., maps). To achieve end-to-end exactly once correctness, Spark Streaming requires that output operators are either idempotent, ensuring that multiple attempts always result in the same output, or perform transactional updates, where all updates are made using atomic transactions.



Unlike LOG.io, Spark streaming is limited to deterministic operators, although the checkpointing of RDDs could be used to relax this limitation by systematically checkpointing the output of any non-deterministic operator. In that case, checkpointing must be synchronous (pessimistic logging) and is similar to the logging of output events in LOG.io. Spark Streaming tracks the lineage of RDD partitions, which is in principle similar to the data lineage capture of LOG.io. However, for stateful operators, the granularity of the lineage is dictated by the statically defined micro-batching of streaming data. In LOG.io, data batching (i.e., the granularity of events) is dynamically defined by each operator. In addition, LOG.io tracks the actual dependency between output and input events (i.e., input events that do not contribute to an output event are not represented in the lineage). As mentioned in \cite{adapted-batching-survey:2022, Das:2014, Zhang:2016}, the use of statically defined batch sizes or intervals raises several challenges due to the dynamic nature of real-time data and workloads. For instance, if the fixed batch interval is too small or the incoming data rate unexpectedly increases, the processing time for a batch can exceed the allocated batch interval. This causes subsequent batches to queue up, leading to an unstable system and increased end-to-end latency. If a failure occurs during such a backlog, the system must recover and process a larger volume of queued data, which inherently leads to longer recovery times as it attempts to catch up. 
Finally, tuning the static configuration of checkpointing can be difficult \cite{SPARK:streaming:24}. Indeed, stateful transformations can cause the dependency chain of RDDs to grow indefinitely. To prevent unbounded recovery times, periodic data checkpointing effectively "cuts off" these dependency chains. However, if checkpointing is set too infrequently (a consequence of static definition), it means a larger portion of the RDD lineage must be re-computed upon a failure, resulting in significantly longer recomputation times for lost RDD partitions. 

\subsubsection{TimeStream}


TimeStream \cite{QIAN:ACM:13} follows the programming model of StreamInsight \cite{ALI:VLDB:09}, in which programmers use a declarative SQL-like language to represent a pipeline of time-aware operators (i.e., windowing or aggregations) over continuous data streams. For execution, TimeStream compiles a continuous query into a streaming DAG, mapping operators to vertexes that execute on physical machines and allow dynamic reconfiguration.

During normal processing, each vertex or operator keeps track of the meta-data information of dependencies and progress, and saves them to reliable storage periodically. Dependencies represent the set of input events that trigger the generation of an output event, and are stored in a compact form using a range set of entry labels. Dependencies are persisted asynchronously, overlapping the tracking logic with computation. Both input and output streams are always persisted reliably but are garbage collected when all of the output streams resulting from an input set have been computed.

When restarting from failure, TimeStream reconstructs the operator state (i.e., window buffers or aggregation tables) by recomputing the logged input events in event-time order based on its dependency path metadata. Furthermore, if a downstream operator fails and requests recovery after a sub-graph substitution to scale operators up or down, TimeStream recursively triggers upstream recovery until it recovers all the necessary input set to reconstruct an output set. Therefore, a correctness guarantee is that every single operator must be deterministic. Another factor is that since TimeStream persists dependencies asynchronously, during failure recovery, it may lead to the centralized query coordinator having a slightly stale view of the graph state. However, it does not hurt correctness because, in the worst case, it will trigger unnecessary recomputation whose duplicate output events will be discarded by downstream operators.

TimeStream programming model is limited with respect to LOG.io because operators must be deterministic, are restricted to SQL statements and do not use a global state (as defined in LOG.io). No side effect is allowed in intermediate operators. Similarly to LOG.io, TimeStream logs output events but does it asynchronously by exploiting the determinism of operators. 
Finally, both protocols support scaling operators up or down during runtime. However, as explained before, rescaling operations may require to recursively recover upstream operators, whereas recovery remains local with LOG.io.

\subsubsection{ChronoStream}

The programming model of ChronoStream \cite{WU:IEEE:15} enables the creation of an acyclic DAG of deterministic logical operators connected by streams of events having user-defined keys. Input streams are consumed by an operator in a deterministic fashion. At the execution level, a logical DAG is compiled into an execution plan where each operator is deployed into one or more operational units, called containers, that run on different processing nodes and form an \emph{operator stage}. The containers of an operator stage collaboratively hold the stage-level internal state of the operators, each maintaining parts of the state and accepting input streams from upstream operator stages in a deterministic manner. Containers of neighboring operator stages are fully connected. The internal state of an operator has two forms. Firstly, the \emph{computation state} is modeled as a key-value store. At deployment time, ChronoStream partitions the computation state maintained in an operator stage into an array of \emph{computation slices}, and distributes them to multiple containers in a balanced fashion. 
Secondly, the \emph{configuration state} maintained in each container includes: an input routing table that directs input events to computation slices, an output routing table which routes output events from an internal slice to the corresponding container in the downstream operator stage. 

During normal processing, each container of an operator stage periodically checkpoints its active computation slices into one or more (defined by a backup factor) of its peer containers located on different processing nodes. This avoids the latency of a persistent storage system. 
For each computation slice, there is a computation \emph{progress vector} that holds the last event number consumed so far from each input stream. To do so, each event output to a stream is assigned a monotonically increasing sequence number. When a computation slice is periodically checkpointed, its progress vector is recorded with the slice snapshot. Checkpointing is done asynchronously, without blocking normal processing, and can therefore be out of sync. This is done by updating the internal state in a shadow buffer and then merging it with the regular key-value store after checkpointing. 
ChronoStream also persists the output events produced from each computation slice periodically. 

Dynamic horizontal scaling is enabled by migrating a computational slice of an operator stage to a new container of the stage as follows. A new container is first created in a processing node. Then, the new container rebuilds the computation slice from one of its checkpoints, using the progress vector of the migrated slice, and notifies the job master when it is done. Then, the job master requests upstream operators to send all corresponding past logged output events (as per the progress vector), as well as new output events, to the new container of the migrated slice. The past events will be processed twice by the former and the new containers, but will be filtered out by the downstream operators due to the deterministic property of operators. Finally, the job master requests the former container to complete and stop the processing of the migrated slice.  

When a container fails, the job master requests the peer containers of the operator stage to reconstruct the computation slices of the failed container from the snapshot checkpoints that they hold. The reconstruction is done as for the horizontal scaling: once each computation slice is recovered, upstream operators reroute their output events to the new destination containers. The reconstructed computation slices can later be migrated to the failed container after recovery. 

ChronoStream programming model is limited with respect to LOG.io because operators must be deterministic and the order in which input events are consumed must also be deterministic. In addition, no side-effect is allowed in intermediate operators. Similarly to LOG.io, ChronoStream logs output events but does it asynchronously by exploiting the determinism of operators. However, it also asynchronously checkpoints the internal state (computation slices) of operators, which LOG.io does not. As in LOG.io, the multiple physical instances of a logical operator are allocated to different processing nodes. When a failure occurs, ChronoStream limits the impact of recovery to the failed container and its upstream operators (for replay) and recovery can proceed in parallel using peer containers. In LOG.io, recovery is local to the failed operator, while rerouting of upstream events can leverage the existence of replicas of the failed operator. Dynamic scaling is different in ChronoStream and LOG.io. Indeed, scaling in ChronoStream is done at the level of computation slices whose granularity is determined at compile time and kept constant throughout the system runtime. As noted in \cite{WU:IEEE:15}, setting the granularity statically to achieve the best elasticity is not easy. In LOG.io, scaling is driven by the dynamic data partitioning strategy of dispatcher operators. Furthermore, no dual computation of output events occurs in LOG.io.

\subsubsection{StreamScope}

In the programming model of StreamScope (StreamS) \cite{LIN:ACM:16}, a data pipeline consists of a sequence of deterministic declarative SQL queries operating on event streams. Queries support relational operators extended with temporal semantics, UDFs, and time-based windows. At the execution level, a data pipeline is represented as a DAG of operators that receive and produce data streams, where each logical operator is translated into multiple physical operators that execute in parallel within a computation stage. 
Each data stream is modeled as an infinite sequence of events, each with a continuously incremented sequence number, which is concurrently accessed (read and write) by operators that receive or produce events in that stream. As in ChronoStream, the events added to a stream are periodically and asynchronously persisted (optimistic logging of events). 

In each execution step, an operator consumes the next event in the input streams, updates its internal state, and possibly produces output events. The execution is tracked using a \emph{snapshot}, which is a triplet containing the current sequence numbers of its input streams, the current sequence numbers of its output streams, and its current internal state. Each operator can independently decide to periodically and asynchronously checkpoint its snapshots. 

When a failure occurs, each failed operator recovers independently. A first strategy, called \emph{checkpoint-based} recovery, consists of loading the most recently saved snapshot and resuming its execution. If the input events required to resume the execution are not available in the input streams, because they have either been garbage collected or not saved due to optimistic logging, then downstream operators must re-generate those events from their most recently saved snapshots (cascaded rollback). Another strategy, called \emph{replay-based} recovery, only applies to operators whose internal state consists of a time window of a certain duration (e.g., last 5 minutes). When a failure occurs, the current event window at the time of failure is reprocessed, which may result in a large number of records being reprocessed in exchange for not incurring the overhead caused by checkpointing during regular execution.
A third strategy is replication-based recovery, in which multiple instances execute the same operator simultaneously. In this strategy, some instances may perform checkpoints, while others do not. Thus, the latency generated by checkpoints is not observed in the final result. In case of a failure, an instance can recover the checkpoint present in another instance to accelerate the recovery process. 

There are differences with LOG.io programming model. First, operators only consist of deterministic SQL queries. Second, SteamS uses the watermark events issued by source operators to force each operator to consume the next events from multiple input streams in a deterministic order, which may cause latency during normal processing. Similarly to LOG.io, StreamS reliably logs events with their data and the use of optimistic logging is done under conditions similar to those pointed out in Section \ref{sec:recovery-with-replay} for LOG.io. However, most of the time, StreamS also stores operator states in snapshot checkpoints (as noted in \cite{LIN:ACM:16}, in a production environment, this happens in 75\% of the cases). StreamS snapshots record processing intervals delimited by the latest input events that updated the state of an operator and the latest output events that were produced. However, as with the lineage graph of Spark Streaming, the snapshots are insufficient to capture the true data lineage of operators. Finally, unlike LOG.io, StreamS does not support dynamic horizontal scaling.

\begin{figure*}[ht]
\begin{center}
    \includegraphics[height=3.16cm, trim={0.5cm 0cm 0.5cm 0cm}]{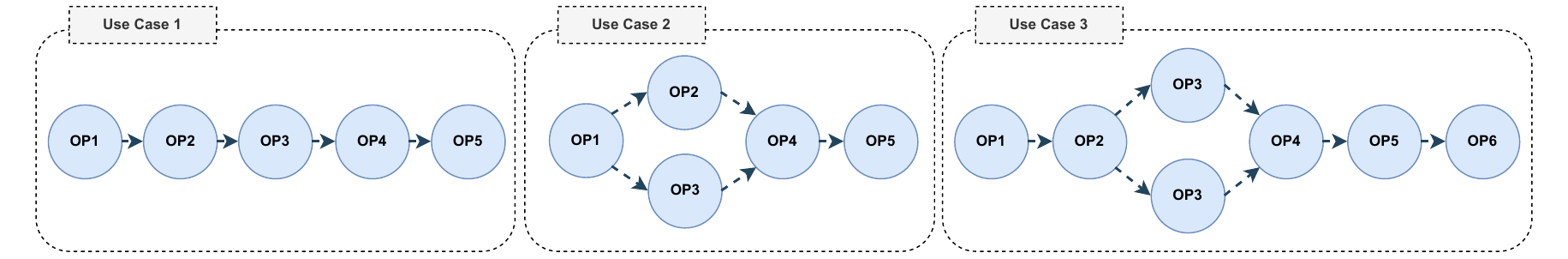}
    \caption{Use Case data pipelines} \label{fig:use-case-graphs}
\end{center}
\end{figure*}


\section{Quantitative Analysis} 
\label{sec:logio-evaluation}

In this section, we conduct a performance evaluation of LOG.io with two goals: (1) compare the performance of the LOG.io and ABS protocols in the context of SAP DI Cloud, and (2) assess the overhead of data lineage capture during a normal execution. 
In Section \ref{sec:guidelines}, we discuss the general guidelines for the design of our data pipelines. Section \ref{sec:experiments} presents the design of our experiments, while the results of the experiments are presented in Section \ref{sec:experiment-results}. Finally, Section \ref{sec:experiments-summary} presents our summary of results. 



\subsection{General design guidelines}
\label{sec:guidelines}

All data pipelines in our experiments are designed to ensure a correct rollback recovery in case of failure. Since ABS has stricter correctness constraints on a data pipeline than LOG.io, we enforced those constraints.
A first constraint of ABS is that Source operators must be replayable, which unlike LOG.io (see Section \ref{sec:correctness-logio}), is not guaranteed by the ABS protocol. Thus, we designed a Source operator that continuously generates output data and is replayable. This operator is configured to generate data at different rates and output events of varying size. 
A second constraint is that, unlike Apache Flink (since version 1.4.0), SAP DI Cloud does not implement a two-step commit protocol between all Writer operators of a data pipeline. Consequently, the conditions for correctness impose that all operators on a path to a Writer operator are deterministic, and write actions are idempotent. The idempotency constraint can be relaxed if write actions are transactional and the Writer operator uses a Write Ahead Log (WAL) to accumulate write actions. This is ensured in the design of our Writer operators. 

There are other considerations to take into account. First, for the reasons explained above, we do not measure the latency incurred by the two-step commit protocol between writers during the normal processing of a data pipeline, as required by the original ABS method. Second, we assume that aligned checkpoints are used by each operator during the alignment phase. As acknowledged in \cite{FlinkAlignment1}, aligned checkpoints yield some latency in data pipelines running under heavy backpressure. Although unaligned checkpoints have been introduced to address this problem in \cite{FlinkAlignment2}, they would not be effective in our data pipelines because the backpressure is due to the high processing time of output event generation. As explained in \cite{FlinkAlignment1}, this is a limitation of the unaligned checkpoints. Third, to facilitate comparisons between different configurations, we ensure that the data pipelines in all our experiments run without any failure in a total time of 5 to 6 minutes. Finally, each operator runs in a separate processing node (Kubernetes pod).

\subsection{Design of experiments}
\label{sec:experiments}

The goal of our \textit{first use case}, illustrated in Figure \ref{fig:use-case-graphs}, is to study the impact of having operators with different processing times on the overhead of LOG.io and ABS. We designed a linear data pipeline consisting of four operators: a (replayable) Source data generator operator OP1, a Middle stateless operator OP2 with a fixed low processing time of 50 ms, a Middle stateful operator OP3 with a varying processing time, a stateful Middle Writer operator OP4 with a low processing time that performs one transactional write action for a set of input events, and a stateful Sink operator OP5 that terminates the execution of the pipeline after receiving a predefined number of input events. 

The controlled processing time of an operator is the time taken during the generation of output events. We vary the processing time of OP3 so that it takes 100, 10, or 2 times the processing time of OP2. To keep the total execution time of the data pipeline without any failure almost constant, we accordingly vary the number and frequency of the events produced by OP1. We also vary the size of the Input Set used to generate output events in OP3. Finally, we vary the operator that fails. 

The goal of our \textit{second use case} is to study the impact of having parallel processing paths on the overhead of rollback recovery with LOG.io and ABS. We designed a data pipeline with two parallel paths. Source operator OP1 and Middle operators OP2 and OP3 are as before. OP4 is a Middle Writer stateful operator in which the two inputs of OP4 are synchronized with the requirement that a number of events must be received on each input port to trigger the generation of output events. The design of OP4 enables us to evaluate the impact of aligned checkpoints in ABS. Finally, OP5 is a Sink operator that terminates the execution as in the previous use case. A failure is artificially generated on OP2. We vary the same parameters as in the first use case. 

The goal of the \textit{third use case} is to study the impact of using data parallelization, as discussed in Section \ref{sec:non-blocking}, on the overhead of LOG.io and ABS. In the data pipeline of Figure \ref{fig:use-case-graphs}, OP1 is a data generator as in the two previous use cases. OP2 is a Dispatcher operator that uses a round-robin strategy to dispatch the events to two replicas of a stateless operator OP3 that has a relatively high processing time with respect to the frequency at which events are generated by OP1 (which is the reason why it was scaled up). OP4 is a Merger operator that collects input events and bundles them into a single output stream. OP5 is a stateful Middle Writer operator with a low processing time, and OP6 is a stateful Sink terminating operator as in the previous use cases. A failure is artificially generated on either one of the replicas of OP3.  


\subsection{Experiment Results}
\label{sec:experiment-results}

\subsubsection{Execution environment and settings}\label{sec:execution-envoriment}

Since the current implementation of LOG.io and ABS are mutually incompatible, the experiments were conducted using two cloud instances of the SAP DI Cloud system, one using the ABS and the other one using LOG.io., hosted over a Kubernetes environment. Both Kubernetes environments are created with GKE (Google Kubernetes Engine) version v1.30.5-gke.1713000, composed of 3 instances n1-standard-8, with 8 vCPUs and 30 GB of memory. A HANA database (version 2414.4.0) is used in each Kubernetes environment to respectively store the asynchronous snapshots and the LOG.io logs.

The experiments compare the total execution time of the data pipelines during a failure-free execution (called \emph{normal} execution) and an execution with failure (called \emph{recovery} execution). 
The execution of a data pipeline without any rollback recovery is called \emph{execution baseline}.
We artificially generate up to three failures for the same execution. 
Each experiment was conducted 10 times for greater precision, and all results were averaged with standard deviation. 

\begin{figure}
    \centering
    \includegraphics[width=0.95\linewidth]{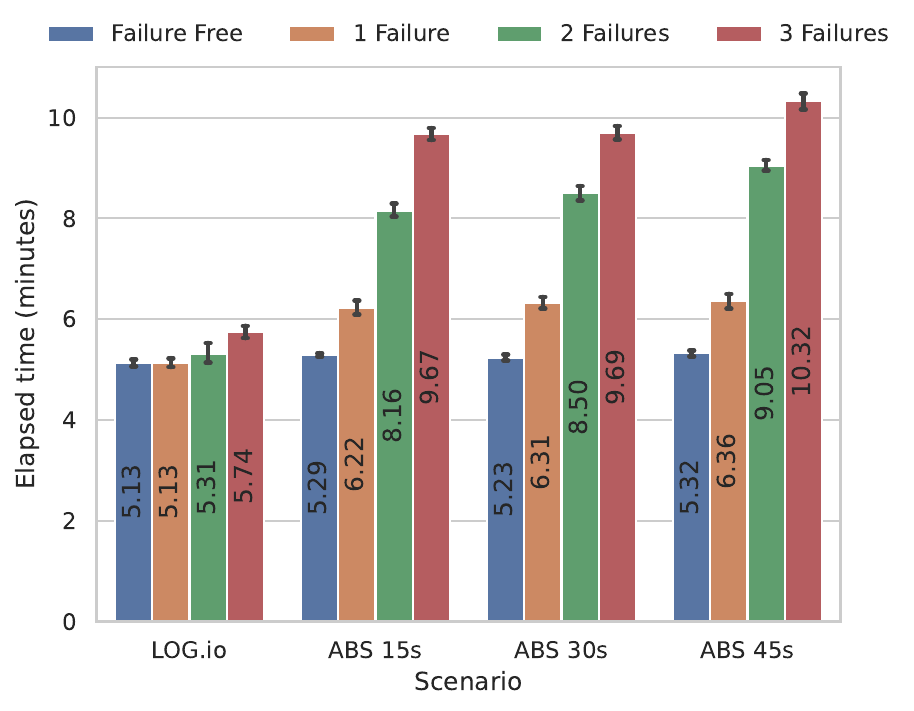}
    \caption{Use Case 1 - 100 events}
    \label{fig:use-case1-execution-time-100-messages}
\end{figure}

\subsubsection{First use case}

In a first series of experiments, OP1 generates a small number of events (100 events) of 10 KB  at a fixed rate of 500 ms. OP2 is a stateless operator that generates 1 output event for each received input event in 50 ms. OP3 accumulates two input events (using an event state of 20 KB) and generates one output event in 5 s (100 times higher than OP2). OP4 accumulates 10 events and, for each set of 10 events, performs 1 write action and generates 1 output event. With ABS, write actions are performed only when an epoch is reached, whereas with LOG.io they are performed as soon as they are produced. OP5 terminates the pipeline after processing 5 events.  For ABS, we use snapshot intervals of 15, 30 (the default recommended parameter in SAP DI), and 45 s.  A failure is artificially generated on OP4 after processing 1, 23, and 45 events, which respectively correspond to the failure points at the beginning, the middle and the beginning of an epoch with ABS.

The processing time of the execution baseline is $\SI{5.14 \pm 0.07}{}$ minutes. The results, depicted in Figure \ref{fig:use-case1-execution-time-100-messages}, show that the overheads of LOG.io and ABS during normal processing with respect to the execution baseline are (considering the standard deviation of the execution baseline): less than 1\% for LOG.io and between 2 and 3\% for ABS, depending on the snapshot interval. For LOG.io, the low logging overhead is explained by the backpressure created by the operator OP3 (the logging of events for OP2 and OP4 occurs in the "shadow" of OP3) and the small number of events output by OP3. For ABS, since there is no alignment phase and snapshotting is done asynchronously, the overhead is expected to be small. There is a small penalty for ABS due to the delay of the write actions done by OP4 until the end of an epoch. Finally, as expected, the overhead for normal processing is roughly the same with different snapshot intervals, the best interval being 15 seconds. 

\begin{figure}
    \centering
    \includegraphics[width=0.95\linewidth]{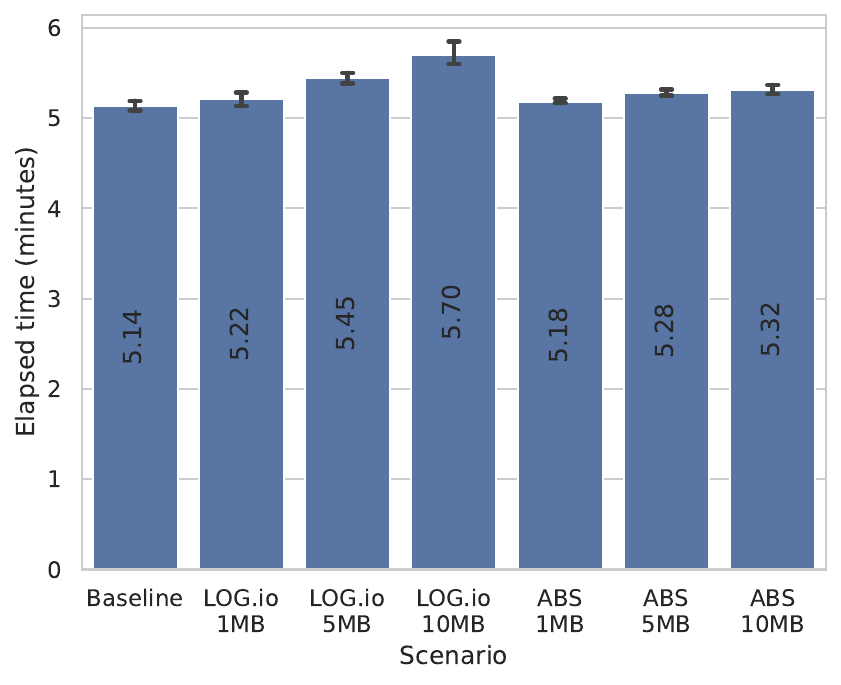}
    \caption{Use Case 1 - Normal processing with different event sizes}
    \label{fig:use-case1-varying-message-size}
\end{figure}

When OP4 fails, the recovery overheads of LOG.io with 1, 2, and 3 failures, with respect to the execution baseline, are: less than 1\%, 3.5\% and 12\% respectively. We see the impact of the non-blocking recovery of LOG.io: when OP4 fails, OP3 keeps its processing and no previous processing effort is lost. Specifically, when the first failure occurred, OP1 produced 10 events. While OP4 is restarted, OP1 produces the remaining events which are then processed by OP2 and OP3. Thus, there is no impact on the overall processing time. With ABS, the overheads are respectively (for a snapshot interval of 15 s): 21.2\%, 59\%, and 88.4\%. We see the impact of the failure point within an epoch. In fact, when OP4 fails just before processing a multiple of 15 input events (end of an epoch for a snapshot interval of 15 s), the processing time lost is: 0.55 s + 15 * 5 s = 75.5 s. The first failure occurs at the beginning of the epoch, and the lost processing time is less than 6 s to which is added the time to restart the data pipeline. The second failure occurs in the middle of an epoch, so lost time is roughly 41 s. 

We then varied the size of the events produced by OP1 and accordingly the size of the Input Set of OP3 which is equal to the size of two input events. We measured the overhead of LOG.io and ABS during a normal execution with respect to the execution baseline. Figure \ref{fig:use-case1-varying-message-size} shows that the overhead of LOG.io for events of 1 MB, 5 MB, and 10 MB is: 1.5\%, 6\%, and 10.6\%, respectively. Due to the low throughput of events in the data pipeline, the advantage of using replay-based recovery is moderate. Unlike ABS, LOG.io is sensitive to the size of events since they are logged.

In a second series of experiments, OP1 generates 1000 events of 10 KB at a fixed rate of 100 ms. OP2 has a processing time of 50 ms as before while OP3 has a ten times higher processing time (500 ms). OP4 performs 1 write action and generates 1 output event, for each accumulated set of 100 events. OP5 terminates the pipeline after processing 5 events.
A failure is artificially generated on OP4 after processing 10, 148, and 375 events, corresponding to the failure points at the beginning, the end, and the middle of an epoch with ABS.

\begin{figure}
    \centering
    \includegraphics[width=0.95\linewidth]{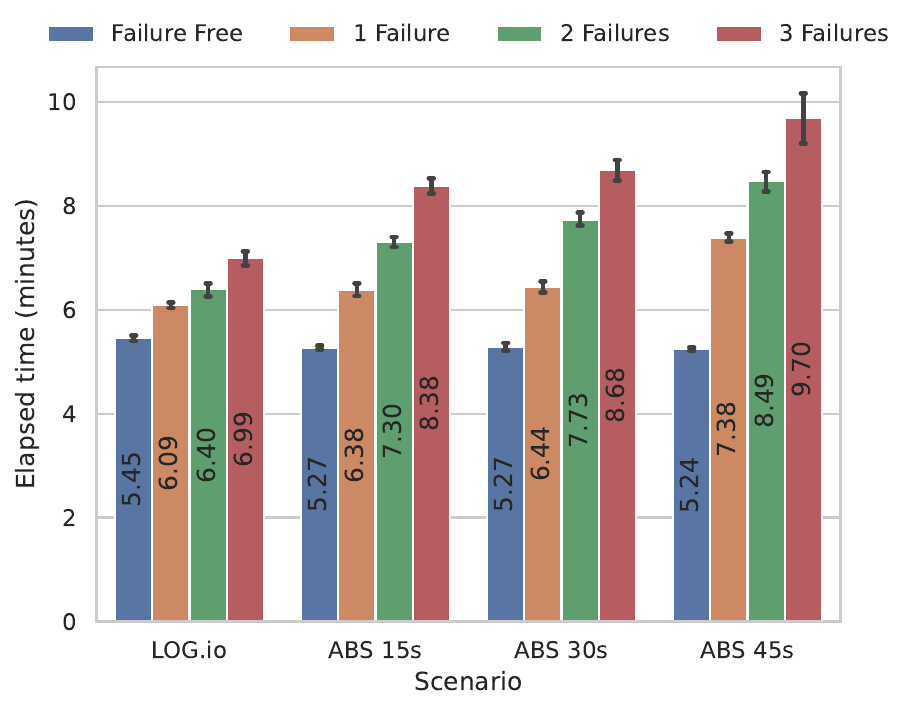}
    \caption{Use Case 1 - 1,000 events and failure in OP4}
    \label{fig:use-case1-execution-time-1000-messages}
\end{figure}

The processing time of the execution baseline is $\SI{5.12 \pm 0.09}{}$ minutes. 
The results in Figure \ref{fig:use-case1-execution-time-1000-messages} show that the overhead during normal processing is small, with an advantage for ABS: 6\% for LOG.io versus 2.8\% for ABS. As before, the logging activity of LOG.io takes advantage of the backpressure created by OP3. However, OP3 generates a bigger number of output events (500), and since LOG.io's logging is pessimistic, it has a higher overhead compared to the previous series of experiments. The overhead of ABS during normal processing is stable around 2.9\%. 


With 1, 2, and 3 failures, the recovery overhead of LOG.io with respect to the execution baseline is: 18.9\%, 25\%, and 36.5\% respectively. It is higher than before because the time to recover the processing of OP4 from the log is also higher. 
With ABS, the overhead is respectively (for the best case of a snapshot interval of 15 s): 24.6\%, 42.5\%, and 63.6\%. Overall, the overhead is smaller than before because the lost processing time after a failure is also smaller. In fact, when OP4 fails just before processing a multiple of 75 events (for a snapshot interval of 15 s), the processing time lost is roughly: 75 * 0.5 s = 37.5 s, to which is added the time to restart the data pipeline.

\begin{figure}
    \centering
    \includegraphics[width=0.95\linewidth]{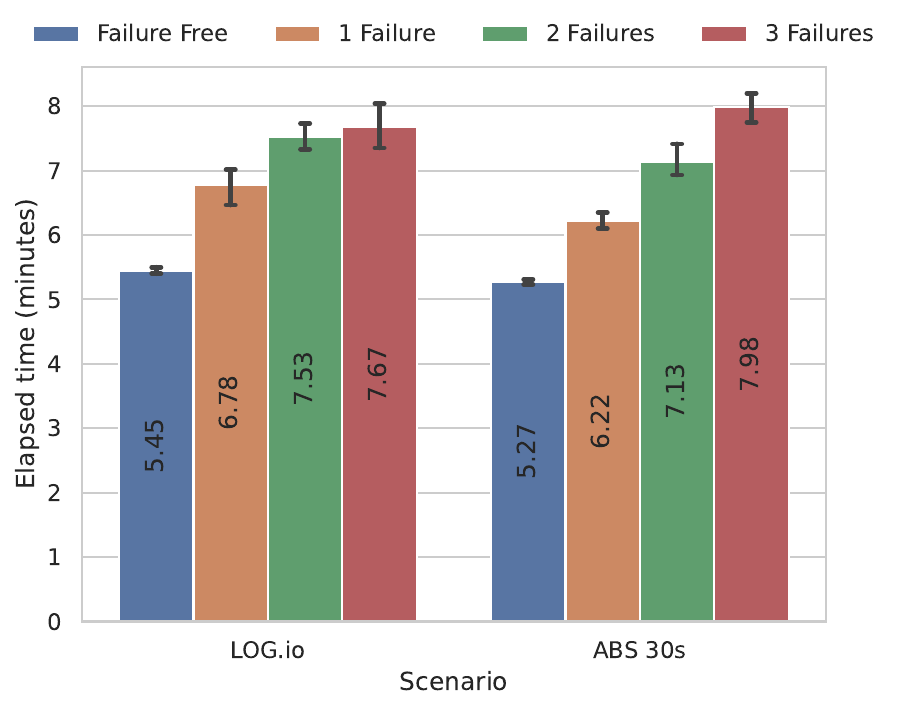}
    \caption{Use Case 1 - 1,000 events and failure in OP3}
    \label{fig:use-case1-1000-events-OP3}
\end{figure}

We now use the same data pipeline configuration as before except that we generate an artificial failure in OP3 after processing 10, 295, and 745 events, which correspond to the beginning, the end, and the middle of an epoch with ABS. We use a snapshot interval of 30 s for ABS. 
The processing time of the execution baseline is as before. The results in Figure \ref{fig:use-case1-1000-events-OP3} show that the overhead of LOG.io during normal processing is 6.4\% and 3\% for ABS. 
When the first failure occurred, OP3 produced 4 events (with a processing time of 2 s) while OP1 produced 21 events. Thus, with LOG.io, while OP3 is restarted, OP1 and OP2 are active, and OP2 continues to output events until the connection is blocked due to backpressure. Afterwards, OP3 recovers its processing from the log, which incurs a delay in the overall processing time. The recovery overhead is: 32\%. The same phenomenon occurs for the other failures, except that OP1 and OP2 are idle because they produced all their events. The overhead with 2 and 3 failures is: 47\% and 50\%. The recovery overhead with ABS for 1, 2 and 3 failures is: 21\%, 39\%, and 56\%. 
With ABS, when OP3 fails just before processing a multiple of 300 events (end of an epoch for a snapshot interval of 30 s), the lost processing time is roughly: 150 * 0.5 s = 75 s. As we conjectured in Section \ref{sec:non-blocking}, the results in Figure \ref{fig:use-case1-execution-time-1000-messages} and \ref{fig:use-case1-1000-events-OP3} show that the overhead of LOG.io varies depending on whether the processing time of the failed operator is relatively high or low with respect to the operators that precede or follow. 

In a third series of experiments, OP1 generates 5000 events of 10 KB at a fixed rate of 30 ms. OP2 keeps a processing time of 50 ms while OP3 has a processing time of 100 ms (4 times higher than OP2). OP4 performs one write action and generates one output event for each accumulated set of 250 input events. OP5 terminates the pipeline after receiving 10 events. For ABS, we used a snapshot interval of 15 s and 30 s. A failure is generated on OP4 after receiving 10, 495, and 1750 events, 
corresponding for ABS to the failure points at the beginning, the end, and the middle of an epoch (for a snapshot interval of 30 s).

\begin{figure}
    \centering
    \includegraphics[width=0.95\linewidth]{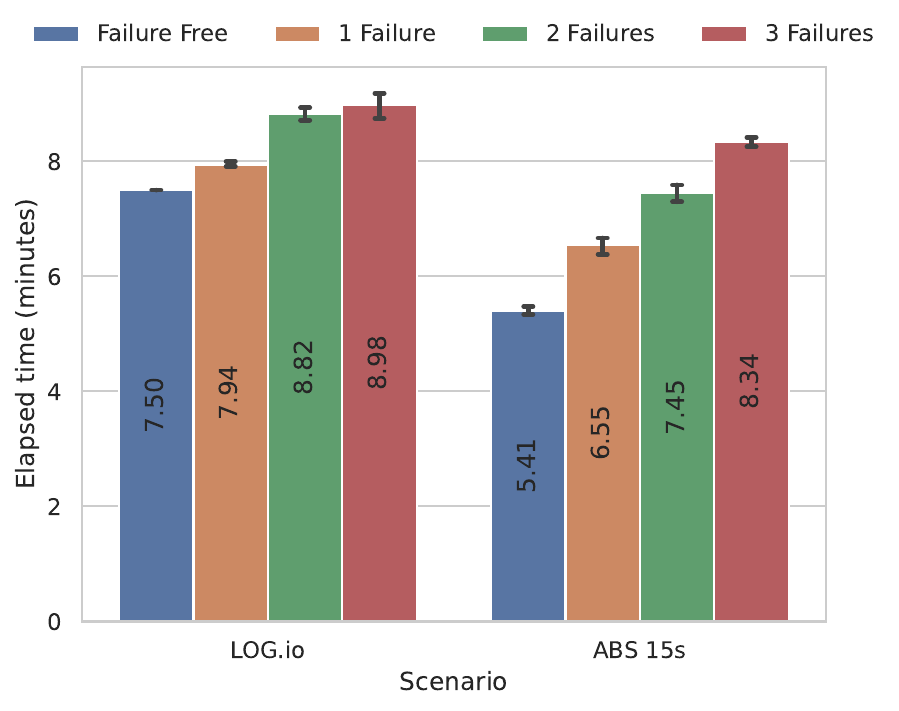}
    \caption{Use Case 1 - 5,000 events and failure in OP4}
    \label{fig:use-case1-execution-time-5000-messages}
\end{figure}

The processing time of the execution baseline is $\SI{5.26 \pm 0.14}{}$ minutes. The results in Figure \ref{fig:use-case1-execution-time-5000-messages} show that the overhead of LOG.io during normal processing is: 42.5\%, a significant increase compared to the previous experiments. Since all operators run with a comparable processing time, LOG.io cannot take advantage of a backpressure created by OP3 to hide the logging activity of OP2. In addition, since OP3 outputs a larger number of events (2,500) than before, the logging effort with pessimistic logging slows down the overall processing of the pipeline. Specifically, we observed that the processing of log updates is relatively slow with the configuration of HANA DB used in our implementation. In fact, when we disable the logging of output event's data, we only obtain a gain of 8\% in the total normal processing time, so the processing cost of every small log update is a limiting factor. However, when the size of the messages increases from 10 KB to 5 MB, the normal processing cost increases by 19\%, which shows that when the throughput of events increases, there is more interest to use replay-based recovery. The overhead of ABS during normal processing is: 3\%, which is stable compared to previous experiments. 

The recovery overhead of LOG.io is: 51\%, 68\%, and 71\%. With ABS, the recovery overhead, for a snapshot interval of 15 s, is: 24.5\%, 42\%, and 58.5\%.  With ABS, when OP4 fails just before processing a multiple of 500 events (end of an epoch for a snapshot interval of 15 s), the lost processing time is roughly 250 * 0.1 s = 25 s. Although LOG.io still recovers faster than ABS (with respect to normal processing), its total processing time is hampered by the overhead of normal processing. This experiment shows that the worst configuration for LOG.io is when there is a little difference between the processing times of operators, and they output a large number of events.  





Finally, we measure the overhead of LOG.io during a failure-free execution, where data lineage is enabled for all operators. Using the previous configurations with 1,000 and 5,000 events, the data lineage overhead with respect to an execution where data lineage is disabled is: 1.2\% and 0.4\%. Thus, data lineage comes almost for free, which is a clear advantage compared to previously published solutions for data streaming systems \cite{glavic:2014, genealog:2018}. 

\begin{figure}
    \centering
    \includegraphics[width=0.95\linewidth]{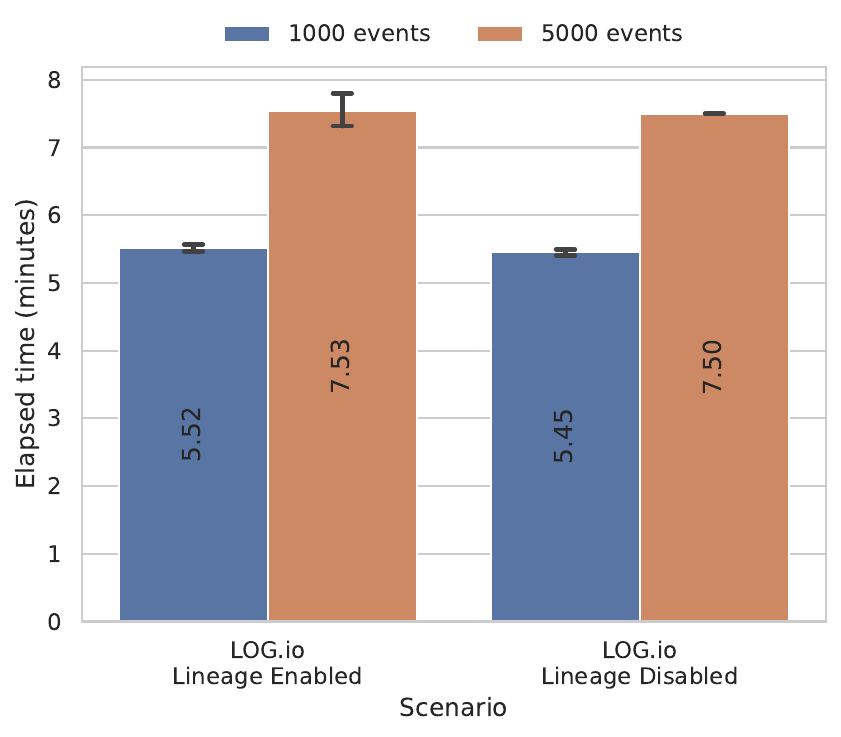}
    \caption{Use Case 1 - data lineage overhead}
    \label{fig:use-case1-lineage-overhead}
\end{figure}

\subsubsection{Second use case}
We run a series of experiments where OP1 generates 1,000 events of 10 KB at a fixed rate of 100 ms and sends them to OP2 and OP3, whose respective processing times are 50 ms and 500 ms, as in the second series of experiments of the first use case.  When the inputs of OP4 are not synchronized, OP4 accumulates 100 events from either OP3 or OP2, performs 1 write action per each set of 100 events, and generates 1 output event. OP5 terminates the pipeline after receiving 15 events. When the two inputs of OP4 are synchronized, OP4 accumulates the events received from OP3 in a chained list of windows of 100 events. When 50 events have been received from OP2, they are grouped with the first window of 100 events in the list, and OP4 performs 1 write action for the group of 150 events and generates 1 output event. Finally, OP5 terminates the pipeline after processing 5 events. We use a snapshot interval of 15 s and 30 s. 
A failure is artificially generated on OP2 corresponding to the failure points at the beginning, the end, and the middle of an epoch, that is, after processing 147, 457, and 825 events for an interval of 15 s, and after processing 295, 605, 750 events for an interval of 30 s. 

\begin{figure}
    \centering
    \includegraphics[width=0.95\linewidth]{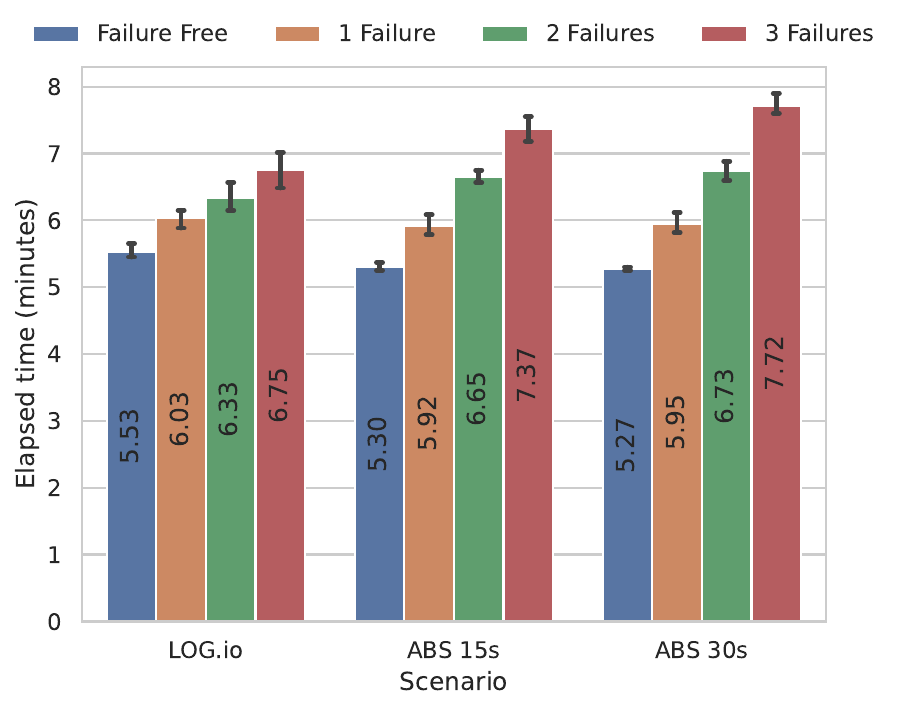}
    \caption{Use Case 2 - 1000 events and failure in OP2}
    \label{fig:use-case2-1000-events}
\end{figure}

The processing time of the execution baseline is $\SI{5.10 \pm 0.09}{}$ minutes. 
The results in Figure \ref{fig:use-case2-1000-events} show that the overhead during normal processing is small, with an advantage for ABS: 8.43\% for LOG.io versus 4\% for ABS. The explanation for the overhead of LOG.io is the same as for the second series of experiments in the first use case with an additional penalty due to the synchronization of input events in OP4. 

The recovery overhead of LOG.io with 1, 2 and 3 failures is: 18.2\%, 24.1\% and 32\%. This is comparable to the results obtained in Figure \ref{fig:use-case1-execution-time-1000-messages}, where OP4 was failing. The reason is that OP3 continues to process its input events while OP2 is restarted. That is, LOG.io exploits the parallelism between OP2 and OP3. Since OP2 is faster than OP3, the impact of the repeated failures of OP2 is limited. With ABS, the recovery overhead for the three failures is: 16.6\%, 32\% and 51.3\%. Here again, we observe the same behavior as in Figure \ref{fig:use-case1-execution-time-1000-messages} because ABS does not exploit the parallelism between OP2 and OP3.




\subsubsection{Third use case}

In a first series of experiments, OP1 sends 1000 events of 10 KB at a fixed rate of 100 ms. OP2 is a stateful Dispatcher operator that uses a round-robin strategy to dispatch its input events to the two replicas of OP3, which is a stateless operator with a processing time of 500ms that produces one output event for each input event. OP4 is a Merger operator that bundles input events into an output stream. The inputs of OP4 are not synchronized. OP4 performs 1 write action and generates 1 output event for each accumulated set of 100 events. OP5 terminates the pipeline after processing 10 events. A failure is artificially generated in one of the two replicas of OP3 (alternatively) after processing 20, 220, and 330 events, which correspond to the failure points at the beginning, the middle, and the end of an epoch for ABS. We use a snapshot interval of 15 s for ABS, which is the best scenario.

\begin{figure}
    \centering
    \includegraphics[width=0.95\linewidth]{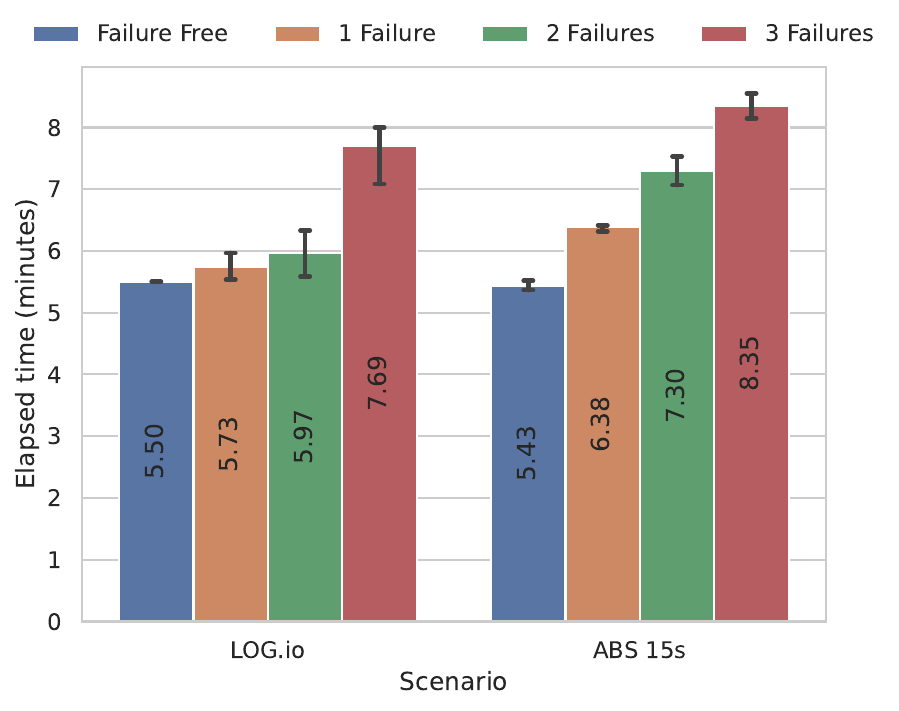}
    \caption{Use Case 3 - 1000 events and failure in one replica of OP3}
    \label{fig:use-case3-1000-events}
\end{figure}

The processing time of the execution baseline is $\SI{5.35 \pm 0.07}{}$ minutes. 
The results in Figure \ref{fig:use-case3-1000-events} show that the overheads during normal processing are small and comparable: 2.48\% for LOG.io versus 1.5\% for ABS. If we compare with the results obtained in Figure \ref{fig:use-case1-execution-time-1000-messages}, the lower overhead is explained by the parallel processing of OP3 (parallel saving of snapshots and parallel logging of events). 

The recovery overhead of LOG.io with 1, 2 and 3 failures is: 7.1\%, 11.6\% and 43.7\%. Here, we clearly see the benefits of the non-blocking behavior of LOG.io that exploits the parallel processing of OP3. When one replica fails, the other continues to process its events. Here, we do not exploit the load balancing strategy described in Section \ref{sec:non-blocking}, so each replica of OP3 eventually processes the same number of events, which explains the overhead when the first replica fails for the second time (third failure).  The recovery overhead for ABS with the three failures is: 19.2\%, 36.4\% and 56\%. We again clearly see the advantage of the non-blocking strategy of LOG.io.

In a second series of experiments, OP1 generates 5000 events of 10 KB at a faster fixed rate of 30 ms, as in the first use case. OP3 has a processing time of 100 ms. OP4 performs 1 write action and generates 1 output event for each accumulated set of 200 events. OP5 terminates the pipeline after receiving 25 events. As before, a failure is alternatively generated in one of OP3's replica after processing 10, 495, and 1620 events, which correspond to the failure points at the beginning, the end, and the middle of an epoch for ABS. We again use a snapshot interval of 15 s for ABS.

\begin{figure}
    \centering
    \includegraphics[width=0.95\linewidth]{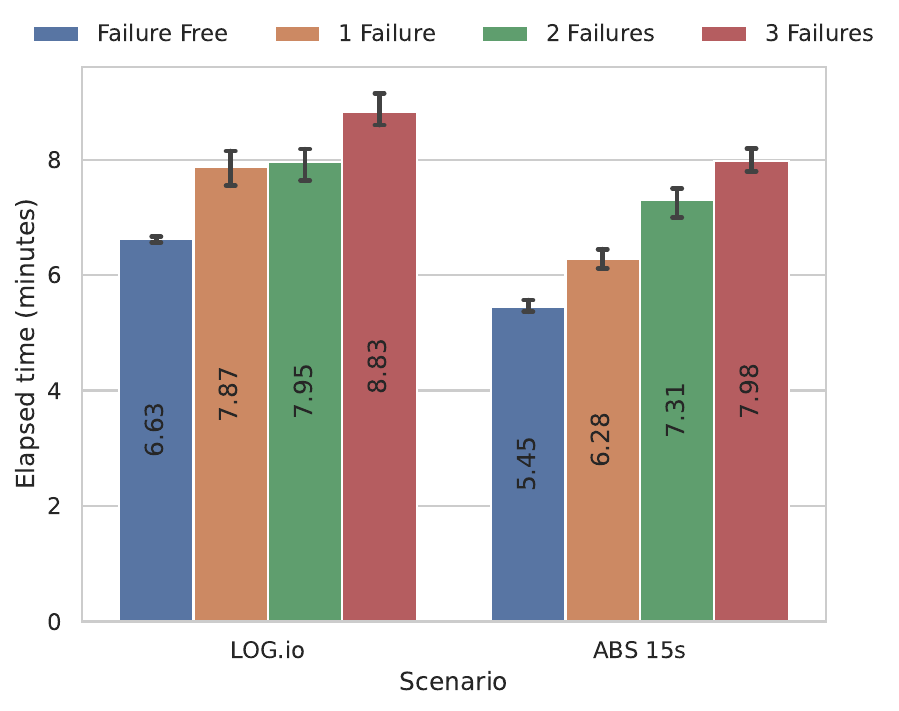}
    \caption{Use Case 3 - 5000 events and failure in one replica of OP3}
    \label{fig:use-case3-5000-events}
\end{figure}

The processing time of the execution baseline is $\SI{5.28 \pm 0.07}{}$ minutes. The results in Figure \ref{fig:use-case3-5000-events} show that the overhead during normal processing is 25\% for LOG.io and 3.2\% for ABS. If we compare with the results of Figure \ref{fig:use-case1-execution-time-5000-messages}, where the overhead of LOG.io was 42.5\%, we see that the parallel execution of OP3 reduces the penalty of pessimistic logging. 

The recovery overhead of LOG.io for the three failures is: 49\%, 50.5\% and 67.2\%, while for ABS, it is: 19\%, 38.4\% and 51.1\%. When the failure occurs at the beginning of the epoch (first failure), ABS is weakly impacted while LOG.io does not benefit yet from the parallel execution of OP3. For the second and third failures, the impact of the non-blocking strategy of LOG.io and the parallel execution of OP3, is visible when we compare with the numbers obtained in Figure \ref{fig:use-case1-execution-time-5000-messages}. This is not the case for ABS and therefore the numbers of LOG.io and ABS get closer.



\subsection{Summary of experiments}\label{sec:experiments-summary}

\paragraph{Normal processing}
The overhead of ABS during a normal processing is quite stable and remains below 3\% in our experiments. This is due to the asynchronous snapshotting of the states of stateful operators. It does not change much with the variation of the snapshot interval. Note that our experiments are still optimistic because we did not stress the checkpointing alignment and stateful operators do not use large event states.  

When a data pipeline contains straggler operators, that is, operators that execute slower than other operators located on the same path of a data pipeline (e.g., at least 10 times slower), and the throughput of events is moderate (e.g., 1 event every 100 ms or more), LOG.io performs almost as well as ABS during normal processing. The bigger the difference between the processing times of the operators, and the smaller the throughput, the better LOG.io is during normal processing. 
We also observe that LOG.io is more sensitive to the size of events than ABS, which is expected due to the logging of events done by LOG.io. 
Alternatively, when the throughput of events is high (e.g., 1 event every 30 ms) and the processing times of the slowest operators are close to the faster ones (e.g., four times slower),
the logging effort required by LOG.io causes a significant overhead (e.g., 42\%). Data parallelization can largely reduce this overhead (e.g.,  25\% with 2 replicas), while it has very little impact on the processing time of ABS. 

\paragraph{Recovery}
When there are operators that execute much slower than others (e.g., 100 times slower) and the throughput of events is low (e.g., 1 event every 500 ms), LOG.io clearly outperforms ABS during recovery (e.g., 1\% overhead for LOG.io vs 21\% for ABS, with 1 failure). The difference between LOG.io and ABS decreases as the difference of processing times between operators decreases and the throughput of events increases.  We also observe that LOG.io performs better when the failing operator has a short processing time compared to other operators on the same processing path, while ABS is not affected. At this point, ABS can perform slightly better than LOG.io during recovery. However, the situation can be reversed if data parallelization is used on the failing operator having a relatively higher processing time. 

When there are no straggler operators and the throughput of events is high (e.g. 1 event every 30 ms), ABS outperforms LOG.io during recovery (e.g., 51\% for LOG.io vs 24.5\% for ABS, with one failure).  However, as for normal processing, data parallelization largely reduces the overhead of LOG.io (e.g., 50.5\% for LOG.io vs 38.4\% for ABS with 2 failures). 




\paragraph{Data lineage capture}
LOG.io captures a fine-grain data lineage at the granularity of events between any two operators of a data pipeline, including custom and possibly non-deterministic operators. This is a notable feature of LOG.io compared to previous solutions. Our experiments show that the overhead of data lineage capture is marginal, with less than 1.5\% in all of our experiments. This is again a strong differentiator with respect to previous solutions. Indeed, data lineage capture capitalizes on the logging mechanism of LOG.io, which simultaneously log output events and marks the input events used to generate them. The only additional effort required is to store the relationship between input and output events in a dedicated EVENT\_LINEAGE table with a small footprint.

\section{Conclusion and future work}
\label{sec:conclusions}
We presented the design of LOG.io, a novel log-based rollback recovery protocol that is unified with a fine-grain data lineage capture at the granularity of events. LOG.io supports a general programming model, accommodating non-deterministic operators, interactions with external systems, and arbitrary custom code. It is non-blocking, allowing failed operators to recover independently without interrupting other active operators, thereby leveraging data parallelization. LOG.io also enables the dynamic scaling (up and down) of operators during pipeline execution.

We compared the performance of LOG.io with the Asynchronous Barrier Snapshotting (ABS) protocol, within the SAP Data Intelligence system. Our experiments show that when there are straggler operators in a data pipeline and the throughput of events is moderate (e.g., 1 event every 100 ms), LOG.io performs as well as ABS during normal processing and outperforms ABS during recovery. Otherwise, ABS performs better than LOG.io for both normal processing and recovery. However, we showed that in these cases, data parallelization can largely reduce the overhead of LOG.io while ABS does not improve. Finally, we showed that the overhead of data lineage capture is marginal, with less than 1.5\% in all our experiments.

Our experiments also highlighted several directions of improvement. First, we should determine when recovery with replay can be exploited to enable an optimistic logging of events' data, particularly in cases where the throughput of events is high and the size of events is large. Another direction is the use of faster and more scalable persistent data structures for logging that could replace the use of the HANA DB in our current implementation. 

\section{Acknowledgments}
We wish to warmly thank our colleagues of SAP involved in the implementation of the data pipeline engine of DI who greatly helped us during the implementation of LOG.io: Rafael Bortolone, Gustavo Marques Netto, and Gabriel Fazenda. We also thank former colleagues of PUCRS, Junior Löff and Claudio Scheer, who contributed to preliminary prototypes of LOG.io. 

\bibliographystyle{elsarticle-num.bst}
\bibliography{myreferences}


\end{document}